\documentclass[aps,prd,twocolumn,superscriptaddress,preprintnumbers,nofootinbib]{revtex4-2}
\usepackage{showyourwork}
\usepackage[nolist,nohyperlinks]{acronym}
\usepackage{amsmath}
\usepackage{amssymb}
\usepackage{booktabs}
\usepackage{xspace}
\usepackage[caption=false]{subfig}
\usepackage{stackengine}

\newcommand{\beq}{\begin{equation}}
\newcommand{\eeq}{\end{equation}}

\newcommand{\dcc}{LIGO-P2300088}

\newcommand{\Nevents}{0}
\newcommand{\varimpNx}{}
\newcommand{\varimpNy}{}
\newcommand{\varimpNz}{}
\newcommand{\varimpJx}{}
\newcommand{\varimpJy}{}
\newcommand{\varimpJz}{}
\newcommand{\clJ}{}
\newcommand{\clN}{}

\newcommand{\Nitersel}{}
\newcommand{\Nstartsel}{}
\newcommand{\Nmaxsel}{}
\IfFileExists{output/macros.tex}{\renewcommand{\Nevents}{57\xspace}
\renewcommand{\varimpNx}{25\%\xspace}
\renewcommand{\varimpNy}{20\%\xspace}
\renewcommand{\varimpNz}{15\%\xspace}
\renewcommand{\varimpJx}{15\%\xspace}
\renewcommand{\varimpJy}{10\%\xspace}
\renewcommand{\varimpJz}{9\%\xspace}
\renewcommand{\clN}{1\%\xspace}
\renewcommand{\clJ}{49\%\xspace}
\renewcommand{\Nitersel}{11\xspace}
\renewcommand{\Nstartsel}{4\xspace}
\renewcommand{\Nmaxsel}{8192\xspace}}{}

\begin{document}

\title{The directional isotropy of LIGO-Virgo binaries}

\newcommand{\CCA}{\affiliation{Center for Computational Astrophysics, Flatiron Institute, 162 5th Ave, New York, NY 10010}}
\newcommand{\STBR}{\affiliation{Department of Physics and Astronomy, Stony Brook University, Stony Brook NY 11794, USA}}
\newcommand{\AEI}{\affiliation{Max Planck Institute for Gravitational Physics
    (Albert Einstein Institute), Am M\"uhlenberg 1, Potsdam 14476, Germany}}
\newcommand{\UMassD}{\affiliation{Department of Mathematics,
    Center for Scientific Computing and Data Science Research,
    University of Massachusetts, Dartmouth, MA 02747, USA}}

\author{Maximiliano Isi}
\email{misi@flatironinstitute.org}
\CCA

\author{Will M. Farr}
\email{will.farr@stonybrook.edu}
\CCA
\STBR

\author{Vijay Varma}
\email{vijay.varma@aei.mpg.de}
\thanks{Marie Curie fellow}
\AEI
\UMassD

\hypersetup{pdfauthor={Isi,Farr,Varma}}

\date{\today}

\begin{abstract}
We demonstrate how to constrain the degree of absolute alignment of the total angular momenta of LIGO-Virgo
binary black holes, looking for a special direction in space that
would break isotropy. We also allow for inhomogeneities in the
distribution of black holes over the sky. Making use of dipolar models for
the spatial distribution and orientation of the sources, we analyze \Nevents signals with false-alarm rates $\leq 1 / \mathrm{yr}$ from
the third LIGO-Virgo observing run. Accounting for selection biases, we find
the population of LIGO-Virgo black holes to be fully consistent with both homogeneity and isotropy.
We additionally find the data to constrain some directions of alignment more than others, and produce posteriors for the directions of total angular momentum of all binaries in our set.
All code and data are made publicly available in \url{https://github.com/maxisi/gwisotropy/}.
\end{abstract}

\maketitle

\begin{acronym}
\acro{GW}{gravitational wave}
\acro{GR}{general relativity}
\acro{CBC}{compact-binary coalescence}
\acro{BH}{black hole}
\acro{BBH}{binary black hole}
\acro{FITS}{Flexible Image Transport System}
\end{acronym}

\section{Introduction}
\label{sec:intro}

With the increasing number of \acp{BBH} detected by LIGO \cite{TheLIGOScientific:2014jea} and Virgo \cite{TheVirgo:2014hva}, it has become possible to study the distribution of such \ac{GW} sources over time and space \cite{LIGOScientific:2021psn,Fishbach:2018edt,Fishbach:2021yvy,Stiskalek:2020wbj,Payne:2020pmc,Cavaglia:2020fnc,Essick:2022slj}.
Since \acp{BBH} can be detected up to nonnegligible redshifts (currently, $\lesssim 1$), we expect their distribution at large scales to reflect the homogeneity and isotropy that characterize the universe cosmologically---a departure from that expectation would reveal a major shortcoming in our understanding of the detection biases affecting the LIGO-Virgo instruments, or, more tantalizingly, point to fundamentally new physics or astrophysics.

The homogeneity \cite{Stiskalek:2020wbj,Payne:2020pmc,Cavaglia:2020fnc,Essick:2022slj} and isotropy \cite{Vitale:2022pmu} of \acp{BBH} have been studied before under different frameworks.
In this work, we reconsider the problem from a new point of view by quantifying the degree of alignment of the \ac{BBH} orbits; in other words, we ask the question: could the total angular momentum vectors of LIGO-Virgo \acp{BBH} be preferentially aligned with a special direction in space?
We consider this possibility as we simultaneously search for angular inhomogeneities in the spatial distribution of sources, thus constraining the existence of special directions controlling both the alignment and location of \acp{BBH}.

\begin{figure*}
\includegraphics[width=0.3\textwidth]{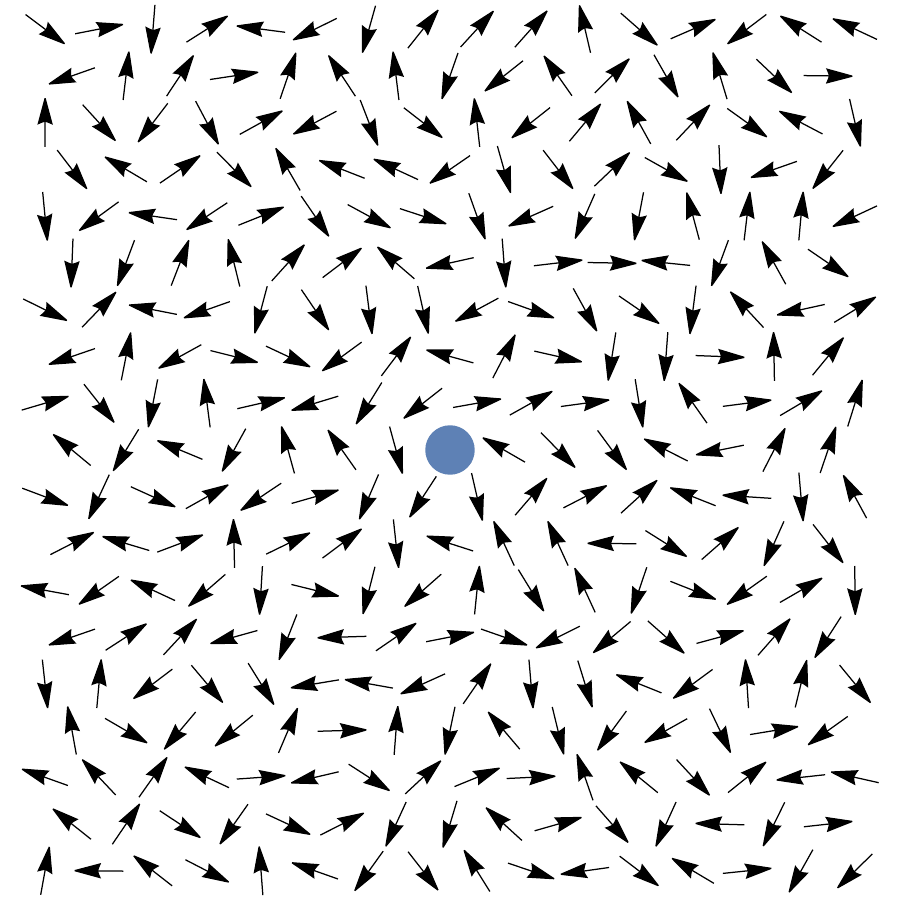}
\includegraphics[width=0.3\textwidth]{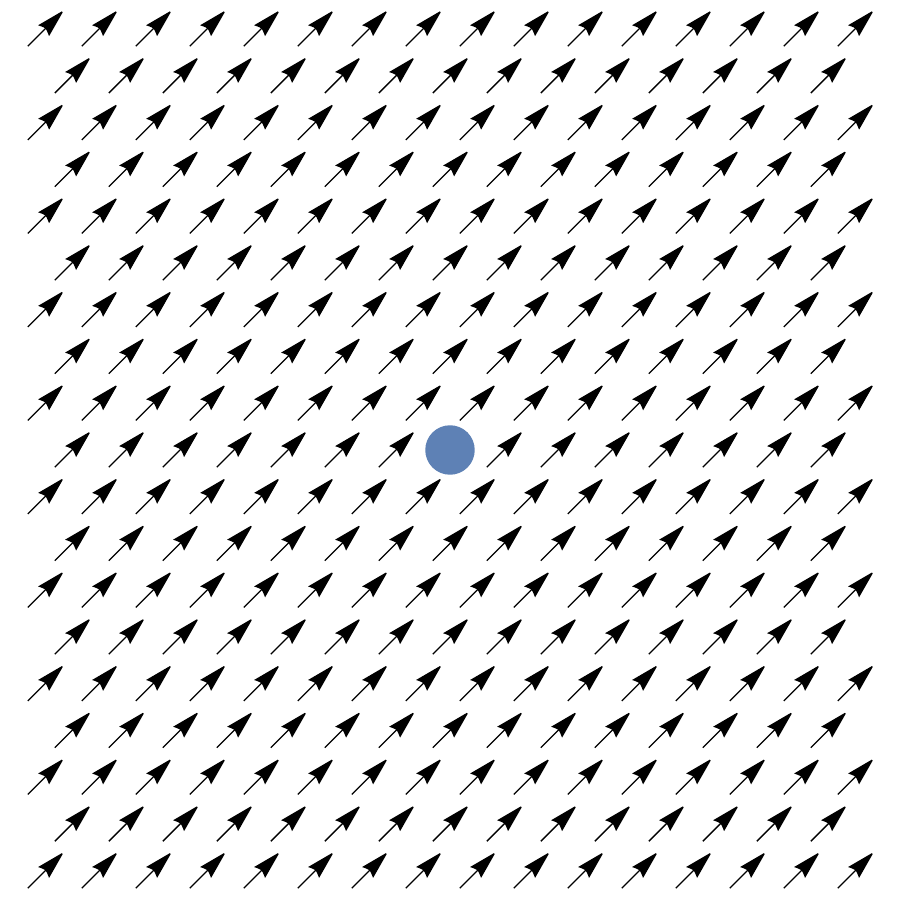}
\includegraphics[width=0.3\textwidth]{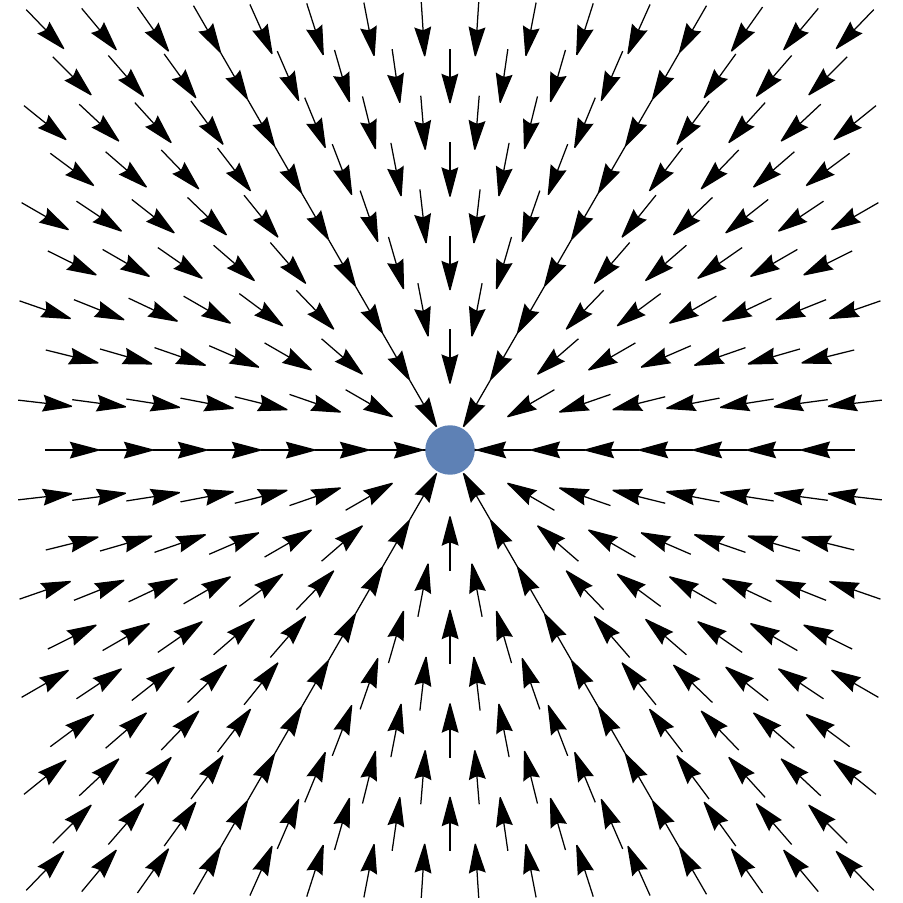}
\caption{\emph{\ac{BBH} orientation models.} By default, we expect \ac{BBH} angular momenta (arrows) to be oriented randomly with respect to Earth (circle) or each other, reflecting isotropy (first panel).
In this study, we consider the possibility that \ac{BBH} orbits follow a special direction in space, the extreme of which is full alignment (second panel).
Previous studies, like Ref.~\cite{Vitale:2022pmu}, have considered models in which binaries are (or are perceived to be) aligned anomalously with respect to Earth, e.g., pointing preferentially towards it (third panel).
The first two panels both have a distribution of inclinations that looks isotropic to analyses like Ref.~\cite{Vitale:2022pmu}.
}
\label{fig:vectors}
\end{figure*}

Unlike previous studies, we look for a breaking of isotropy in the angular momenta through the existence of a special direction in space with reference to some absolute frame, like the cosmic microwave background or far away stars (Fig.~\ref{fig:vectors}, second panel), and not with respect to Earth.
The discovery of such a special direction could reveal the presence of a vector field breaking Lorentz symmetry.
This differs from previous works like Ref.~\cite{Vitale:2022pmu}, which checked for anomalies in the alignment of sources with respect to Earth, as reflected in the distribution of \ac{BBH} inclinations relative to the line of sight (Fig.~\ref{fig:vectors}, third panel).
Such studies are not sensitive to the kind of overall alignment in absolute space that we look for here.

In Sec.~\ref{sec:method} we describe the population model we use to constrain anisotropies, as well as our assumptions about the astrophysical distribution of \ac{BBH} properties like masses and redshifts, and our treatment of selection biases.
In Sec.~\ref{sec:data} we outline the data products used in our analysis.
We present our results in Sec.~\ref{sec:results}, including constraints on the degree of orientation and location inisotropies, as well as maps of possible preferred directions.
In Sec.~\ref{sec:validation}, we summarize validation tests of our infrastructure to contextualize our measurement.
We conclude in Sec.~\ref{sec:conclusion}.
Appendices show how to obtain posteriors on the angular-momentum direction, discuss hyperparameter prior choices and display posteriors for the angular-momentum direction for all events in our set.

\section{Method}
\label{sec:method}

\subsection{Isotropy modeling}

\newcommand*{\dip}[1]{\vec{v}_{#1}}
\newcommand*{\dipraw}[1]{\vec{u}_{#1}}

In order to study the spatial distribution and orientation of \acp{BBH}, we must analyze the collection of detections under a hierarchical population model that allows for variable degrees of spatial and directional correlations (see, e.g., Ref.~\cite{Essick:2022slj}).
This requires modeling the distribution of source locations, $\hat{N}$, and total-angular-momentum directions, $\hat{J}$.
For modeling purposes, the $\hat{N}$ and $\hat{J}$ vectors must be expressed through their Cartesian components in some absolute reference frame.

We choose to work in geocentric, equatorial celestial coordinates, with the vernal equinox as the $x$ axis.
In that frame, the sky-location vector is, for each \ac{BBH},
\begin{equation} \label{eq:n}
\hat{N} = \left( \cos\alpha \cos\delta, \sin\alpha \cos\delta, \sin\delta \right) ,
\end{equation}
for the right ascension $\alpha$ and declination $\delta$; meanwhile, the orientation vector is $\hat{J} \equiv \vec{J} / |\vec{J}|$ where
\begin{equation} \label{eq:j}
\vec{J} = \vec{L} + \vec{S}_1 + \vec{S}_2 \, ,
\end{equation}
for the orbital angular momentum $\vec{L}$ and individual (dimensionful) \ac{BH} spin angular momenta $\vec{S}_{1/2}$.
The Cartesian components of $\hat{J}$ can be computed in the above frame as a function of $\alpha$, $\delta$, and the polarization angle $\psi$ \cite{Isi:2022mbx}, as well as the component masses $m_{1/2}$, the dimensionless spin vectors $\vec{\chi}_{1/2}$ and the orbital phase $\phi_{\rm ref}$ at some reference frequency $f_{\rm ref}$; we outline this calculation in Appendix \ref{app:j} and release relevant code in \cite{repo}.
We prefer to work with $\hat{J}$ rather than $\hat{L}$ because the former is conserved over the coalescence to a high degree, even for precessing systems \cite{poisson2014gravity}.

Our goal is to quantify the degree of isotropy in $\hat{N}$ and $\hat{J}$.
To this end, we model each of those vectors as drawn from an isotropic distribution with a dipolar correction of variable magnitude.
This \emph{ad hoc} modification may be generally seen as the first term in a harmonic expansion around isotropy, and is specifically well suited to capture the existence of a preferred direction in space.
The dipolar components in the location and orientation distributions are defined by dipole vectors $\dip{N/J}$ whose magnitudes control the degree of deviation from isotropy for $\hat{N}$ and $\hat{J}$ respectively. 

Concretely, we implement the following population-level likelihood for each event (indexed by $i$)
\begin{equation}
\label{eq:lnlike}
p(\hat{N}_i,\hat{J}_i \mid \dip{N}, \dip{J}) \sim \frac{1}{16 \pi^2}  \hspace{-3pt}\left(1 + \dip{N} \cdot \hat{N}_i\right) \hspace{-3pt} \left(1 + \dip{J} \cdot \hat{J}_i\right) \hspace{-2pt} ,
\end{equation}
for some special-direction vectors $\dip{N/J}$, with $0 \leq \left| \dip{N/J} \right| < 1$, to be inferred as hyperparameters from the collection of detections.
The fully isotropic case is recovered for $\dip{N} = \dip{J} = 0$.
On the other hand, setting $\dip{J} = 0$ alone allows for a nonhomogenous distribution of sources in the sky while assuming isotropic source orientations; this reduces to the ``dipole'' model in \cite{Essick:2022slj}.
Modeling the likelihood as in Eq.~\eqref{eq:lnlike}, with $\left| \dip{N/J} \right| < 1$, ensures that the likelihood itself remains positive everywhere.

It is convenient to rephrase the constraint that $0 \leq \left| \dip{N/J} \right| <
1$ by re-parameterizing the population in Eq.~\eqref{eq:lnlike}, through two
corresponding, auxiliary vectors $\dipraw{N/J}$ defined such that
\begin{equation}
\dip{N/J} = \frac{\dipraw{N/J} }{\sqrt{1 + |\dipraw{N/J}|^2}} .
\end{equation}
In this way, we ensure $0 \leq \left| \dip{N/J} \right| \leq 1$ for $-\infty <
\dipraw{N/J} < \infty$; $\dipraw{N/J}$ are unconstrained.  This ``decompactifying'' transformation
is more effective for sampling purposes than enforcing a sharp constraint on the
magnitudes of $\dip{N/J}$. For small $\left| \dip{N/J} \right| \ll 1$, $\dip{N/J}
\simeq \dipraw{N/J}$ to second order in $\left| \dip{N/J} \right|$. As a prior, we choose a three-dimensional Gaussian distribution on the components of
each $\dipraw{N}$ and $\dipraw{J}$, with zero mean and standard deviation
$\sigma = 0.4$. This choice of prior is designed to be fairly uninformative about
$\dip{N/J}$ (i.e., lacking a strong gradient) while still peaking at $\dip{N/J} = \vec{0}$; see Appendix
\ref{app:prior} for a description of this feature.

\subsection{Reweighting to an astrophysical population}
\label{sec:reweight}

Besides the location and orientation modeling described above, we need to ensure that our assumptions about the parameters of each individual \ac{BBH}, like masses and redshift, are astrophysically sensible.
To that end, we assume a redshift distribution that follows the Madau-Dickinson star formation rate \cite{Madau:2014bja} in the comoving frame.
For the masses, we assume a prior distribution inversely proportional to the heaviest component mass ($\propto 1/m_1$) and uniform in the mass ratio (constant in $q = m_2/m_1$), and restrict our sample to \acp{BH} with (posterior-median) masses in the range $5 \, M_\odot < m_2 \leq m_1 < 150 \, M_\odot$; within that range, this choice is a simple approximation to the measurement in \cite{LIGOScientific:2021psn}.
We assume the population of component spins are isotropically oriented with respect to the orbital angular momentum, with a uniform distribution over their dimensionless magnitude.%
\footnote{This implies that our modeling of $\hat{J}$ following Eq.~\eqref{eq:lnlike} can be reinterpreted as a nontrivial modeling of $\vec{L}$ through Eq.~\eqref{eq:j}.}
A future analysis may fit the astrophysical distribution of these parameters jointly with the orientation and location of the binaries.

\subsection{Selection biases}
\label{sec:selection}

Since we are attempting to model the \emph{intrinsic} distribution of all \ac{BBH} sources, not merely those that were detected, we must account for the difference in LIGO-Virgo's sensitivity to various sources.
This is true for both intrinsic parameters, like \ac{BH} masses and spin magnitudes, as well as the location and orientation parameters in which we are interested for this work (namely, $\hat{N}$ and $\hat{J}$).
With knowledge of the instruments' sensitivity over parameter space, we use the measured selection function to obtain a measurement of the intrinsic distribution of parameters out of the distribution of detected sources \cite{Loredo:2004nn,Mandel:2018mve}.

Evaluating the detectors' sensitivity over parameter space requires large simulation campaigns that quantify the end-to-end performance of LIGO-Virgo detection pipelines by injecting and recovering synthetic signals.
As in \cite{Essick:2022slj}, we take advantage of the \ac{BBH} dataset in \cite{o3-selection} for this purpose.%
\footnote{Specifically, the \texttt{endo3\_bbhpop-LIGO-T2100113-v12} injection set.}
Since this injection campaign only covered LIGO-Virgo's third observing run, we only consider events detected during that period; together with the mass constraints cited above, this means there are \Nevents \ac{BBH} events to be included in our analysis (listed in Table \ref{tab:events}). 

\section{Data}
\label{sec:data}

Our analysis starts from posterior samples for individual events reported by LIGO-Virgo in \cite{LIGOScientific:2021djp,LIGOScientific:2021usb} and publicly released in \cite{zenodo:GWTC-2.1,zenodo:GWTC-3} through the Gravitational Wave Open Science Center \cite{GWOSC:GWTC-2.1,GWOSC:GWTC-3,LIGOScientific:2019lzm}.
Specifically, we make use of results obtained with the \textsc{IMRPhenomXPHM} waveform \cite{Pratten:2020ceb,Pratten:2020fqn,Garcia-Quiros:2020qpx,Garcia-Quiros:2020qlt} that have been already reweighted to a distance prior uniform in comoving volume.
The single-event inference was carried out by the LIGO-Virgo collaborations using the \textsc{Bilby} parameter estimation pipeline \cite{Romero-Shaw:2020owr,Ashton:2018jfp}, as detailed in \cite{LIGOScientific:2021djp,LIGOScientific:2021usb}.
We reweight those samples following the astrophysical population described above (see, e.g., \cite{Miller:2020zox,Essick:2022ojx} for methodology) and then use them to produce distributions for the components of $\hat{N}$ and $\hat{J}$, which we use as input for our hierarchical analysis based on Eq.~\eqref{eq:lnlike}.

\setlength{\tabcolsep}{4pt}
  \begin{table}
\caption{Events considered.}
\label{tab:events}
\resizebox{\columnwidth}{!}{\begin{tabular}{lcr}
\toprule
\midrule
GW190408\_181802 & GW190708\_232457 & GW191129\_134029 \\
GW190412\_053044 & GW190719\_215514 & GW191204\_171526 \\
GW190413\_052954 & GW190720\_000836 & GW191215\_223052 \\
GW190413\_134308 & GW190727\_060333 & GW191216\_213338 \\
GW190421\_213856 & GW190728\_064510 & GW191222\_033537 \\
GW190503\_185404 & GW190731\_140936 & GW191230\_180458 \\
GW190512\_180714 & GW190803\_022701 & GW200112\_155838 \\
GW190513\_205428 & GW190805\_211137 & GW200128\_022011 \\
GW190517\_055101 & GW190828\_063405 & GW200129\_065458 \\
GW190519\_153544 & GW190828\_065509 & GW200202\_154313 \\
GW190521\_030229 & GW190910\_112807 & GW200208\_130117 \\
GW190521\_074359 & GW190915\_235702 & GW200209\_085452 \\
GW190527\_092055 & GW190925\_232845 & GW200216\_220804 \\
GW190602\_175927 & GW190929\_012149 & GW200219\_094415 \\
GW190620\_030421 & GW190930\_133541 & GW200224\_222234 \\
GW190630\_185205 & GW191103\_012549 & GW200225\_060421 \\
GW190701\_203306 & GW191105\_143521 & GW200302\_015811 \\
GW190706\_222641 & GW191109\_010717 & GW200311\_115853 \\
GW190707\_093326 & GW191127\_050227 & GW200316\_215756 \\
\bottomrule
\end{tabular}}
\end{table}
\label{output/events.tex}\unskip%

The six-dimensional posteriors for the components of $\hat{J}$ and $\hat{N}$ for each event are the primary input for our hierarchical isotropy analysis.
We show the posteriors for $\hat{J}$ in Fig.~\ref{fig:skymaps-1} as skymaps for all events in our set, produced using standard LIGO-Virgo tools for representing probability densities over the sky \cite{skymap,Singer:2016eax,Singer:2016erz}; we make these posteriors, resulting from the calculation detailed in Appendix \ref{app:j}, available in our data release \cite{repo}.
The equivalent figures for $\hat{N}$ are nothing but the sky-localization maps already made available by LIGO-Virgo \cite{zenodo:GWTC-2.1,zenodo:GWTC-3}.

\section{Results}
\label{sec:results}

\begin{figure*}[hbtp]
\includegraphics[width=\textwidth]{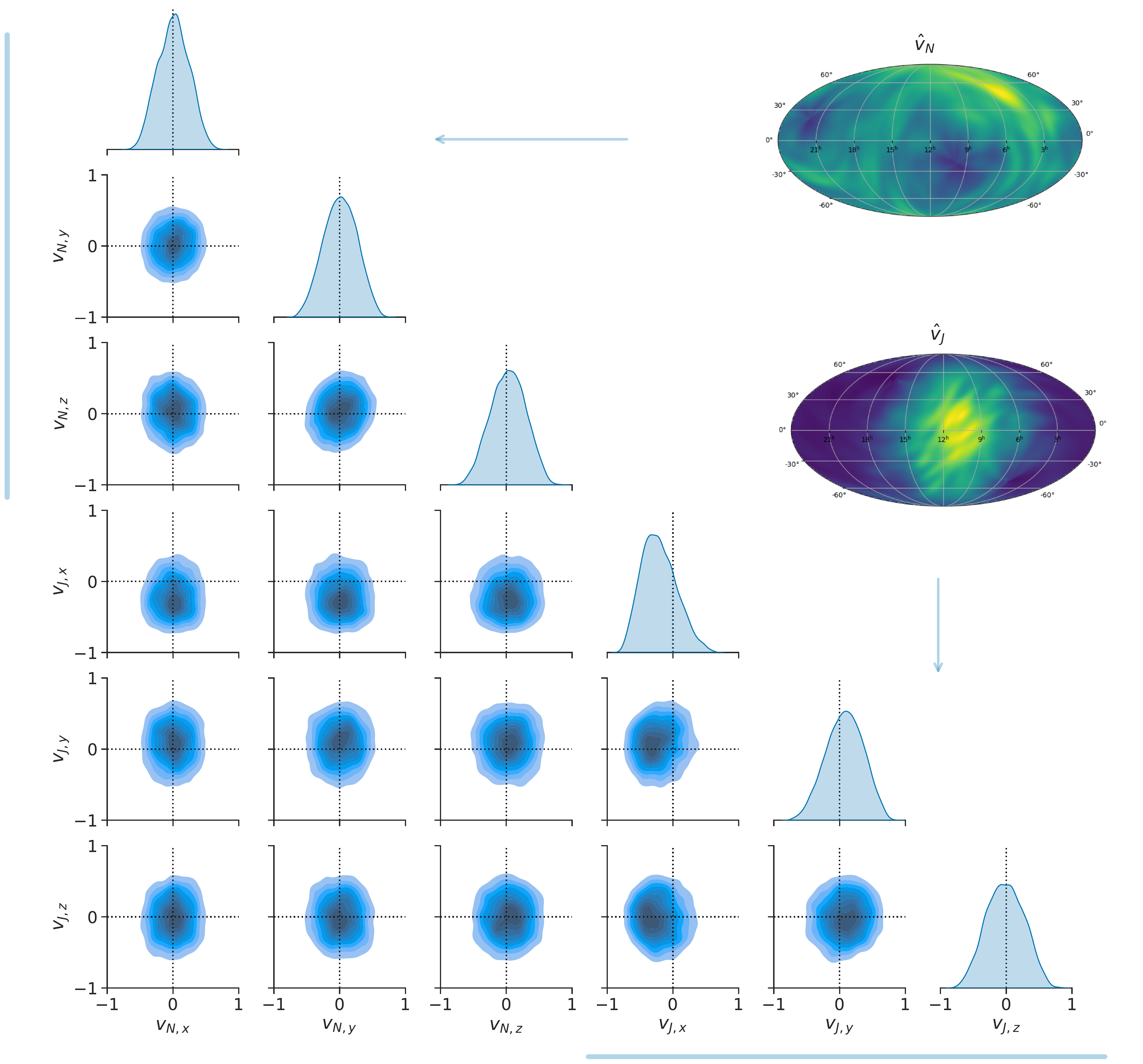}
\caption{\emph{Isotropy measurement.} Result of the simultaneous measurement of location and orientation isotropy through the model in Eq.~\eqref{eq:lnlike}, as represented by the posterior distribution on the dipole vectors $\dip{N/J}$ (corner plot), and the corresponding projections over the sky (Mollweide insets).
The six-dimensional posterior distribution is represented through credible levels over two-dimensional slices (blue contours, spaced at intervals corresponding to 10\% increments in probability mass, with the outer contour enclosing 90\% of the probability), and one-dimensional marginals (diagonal).
The upper-left and lower-right sub-corners encode constraints on the individual components of each $\dip{N}$ and $\dip{J}$ respectively (highlighted with vertical and horizontal lines in the margin), while the other panels encode potential correlations between the location and orientation inisotropies.
The measurements for $\dip{N/J}$ can be projected into distribution over the sky as in the top-right insets, which show the allowed dipole orientations for $\hat{v}_N \equiv \vec{v}_N / |\vec{v}_N|$ (top) or $\hat{v}_J \equiv \vec{v}_J / |\vec{v}_J|$ (bottom), with lighter colors encoding higher probability density over the celestial sphere \cite{skymap,Singer:2016eax,Singer:2016erz}; inhomogeneities in these sky-maps do not constitute evidence for anisotropies.
Isotropy is recovered for $\dip{N} = \dip{J} = 0$ (dotted lines), which is well supported by the 6D posterior.
}
\label{fig:jn_corner}
\script{corner_plot.py}
\end{figure*}

\begin{figure*}
\subfloat[][$\dip{J}$ posterior]{\includegraphics[width=0.325\textwidth]{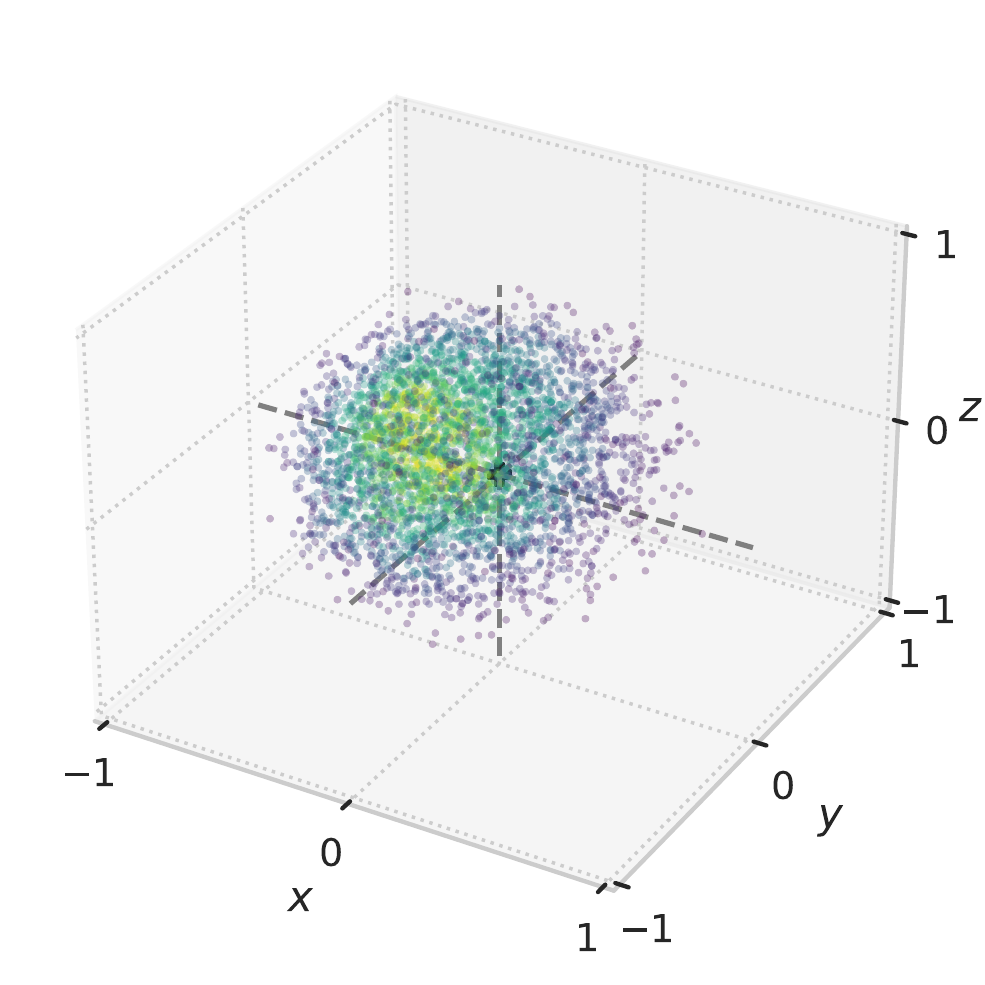}}
\subfloat[][$\dip{N}$ posterior]{\includegraphics[width=0.325\textwidth]{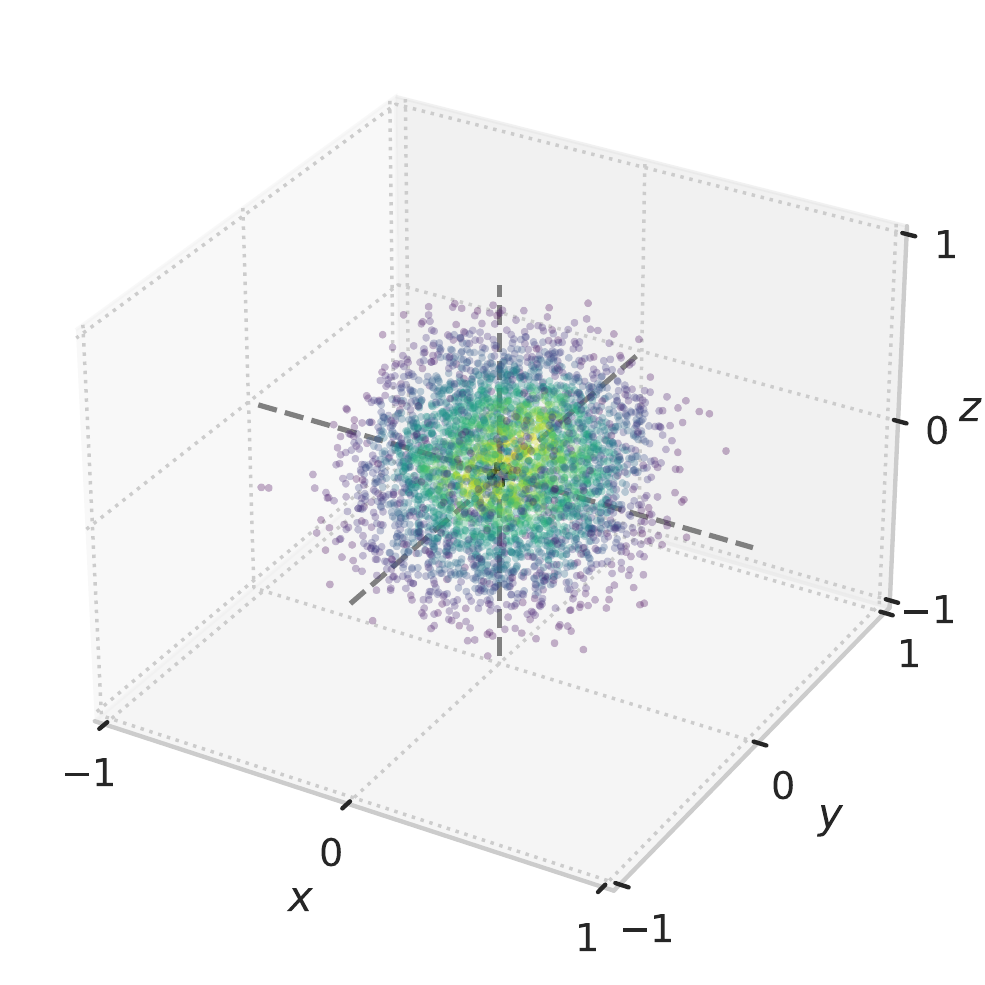}}
\subfloat[][prior]{\includegraphics[width=0.325\textwidth]{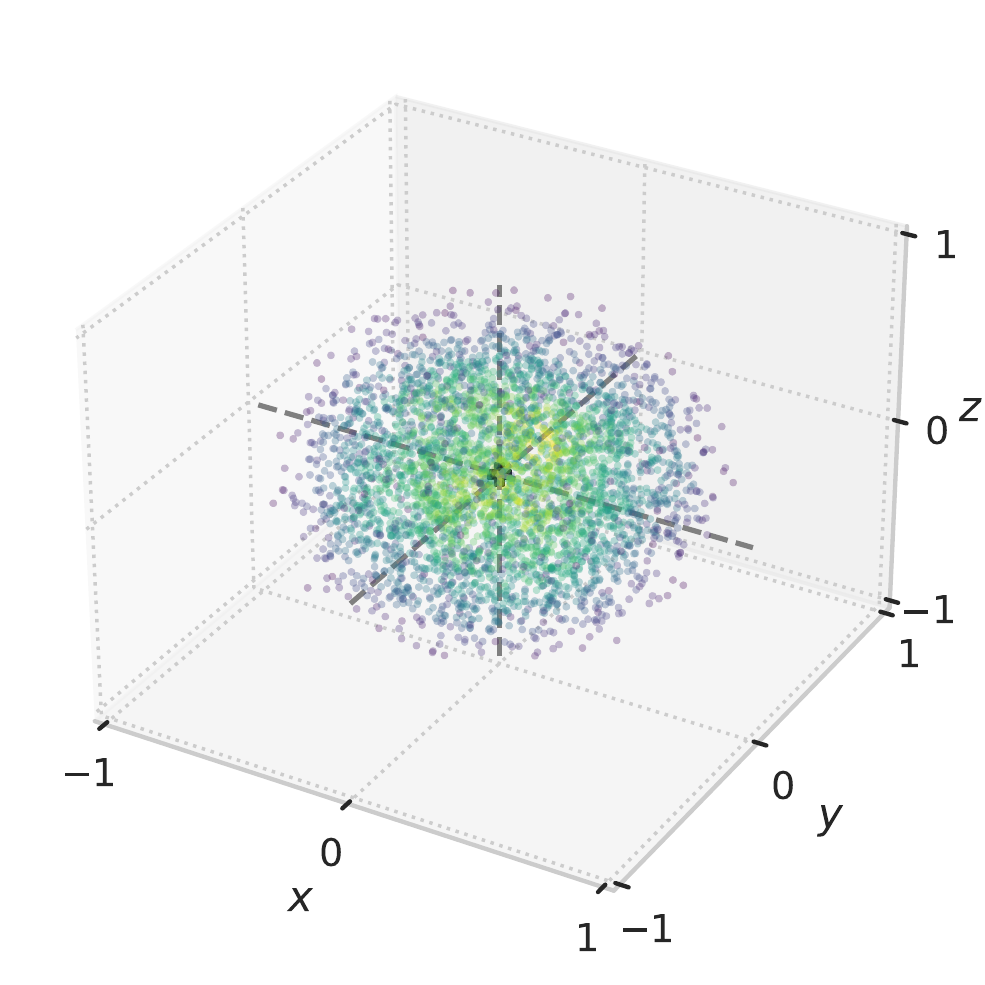}}
\caption{\emph{3D distributions.} Three-dimensional representation of the $\dip{J/N}$ measurement in Fig.~\ref{fig:jn_corner} (first two panels), in comparison to the prior (last panel).
Each point is drawn from the corresponding three-dimensional distribution, with color proportional to the probability density (lighter colors for higher density).
The origin, representing isotropy, is well favored in all cases (intersection of gray dashed lines).
The posteriors are tighter with respect to the prior, as is also seen in Fig.~\ref{fig:jn_norm}.
}
\label{fig:density_3d}
\script{density_3d_plot.py}
\end{figure*}

We showcase the full result of our analysis in Fig.~\ref{fig:jn_corner}, which represents the simultaneous measurement of location and orientation anisotropies through the six-dimensional posterior on the components of $\dip{N}$ and $\dip{J}$.
The result is fully consistent with both kinds of isotropy, with $\dip{N} = 0$ and $\dip{J} = 0$ supported with high credibility, falling close to the peak of the marginal distributions on the \clN and \clJ quantiles respectively;%
\footnote{These three-dimensional quantiles correspond to the fraction of $\dip{J/N}$ samples with higher probability density than the origin.}
 this feature is also reflected in the fact that the origin is well supported in all the panels of the corner plot in Fig.~\ref{fig:jn_corner}.
The result for $\dip{N}$ is consistent with previous studies \cite{Essick:2022slj}, which did not find evidence against isotropy in the location of LIGO-Virgo sources.

To the extent that there is any support for nonzero dipolar contributions to the location or orientation densities (namely, for $|\dip{J/N}| > 0$), their possible directions in the sky are represented by the insets on the top right of Fig.~\ref{fig:jn_corner}: a higher density indicates a potentially allowed orientation for the $\dip{N}$ (top) or $\dip{J}$ (bottom) dipole vectors.
These skymaps reveal that the data are not in conflict with the existence of a weak dipole along the vernal equinox in the celestial equatorial plane (the direction of the $x$ axis in our Cartesian coordinate system), as implied by the marginal on $\dip{J,x}$ in Fig.~\ref{fig:jn_corner}; this dipole is allowed, but not required, by the data, since the posterior is fully consistent with $\dip{J} = 0$.

Indeed, although inhomogeneities appear in these Mollweide projections, this should not be interpreted as evidence for anisotropies: the density of points in those maps only encodes \emph{permissible} directions for the dipole, without implications for its magnitude.
In fact, inhomogeneities will appear in such plots any time the posterior does not happen to peak exactly at the three-dimensional origin of $\dip{J/N}$, as we expect to be commonly the case even if $\dip{J/N}=\vec{0}$ is the underlying truth.%
\footnote{With a finite number of events in the catalog, even if $\dip{J/N}$ is zero in truth, the maximum-likelihood estimate will not, in general, be zero. The posterior will therefore peak away from zero (but be fully consistent with zero). In this situation inhomogeneity can appear as in Fig.\ \ref{fig:jn_corner}.}

\begin{figure}
\includegraphics[width=0.9\columnwidth]{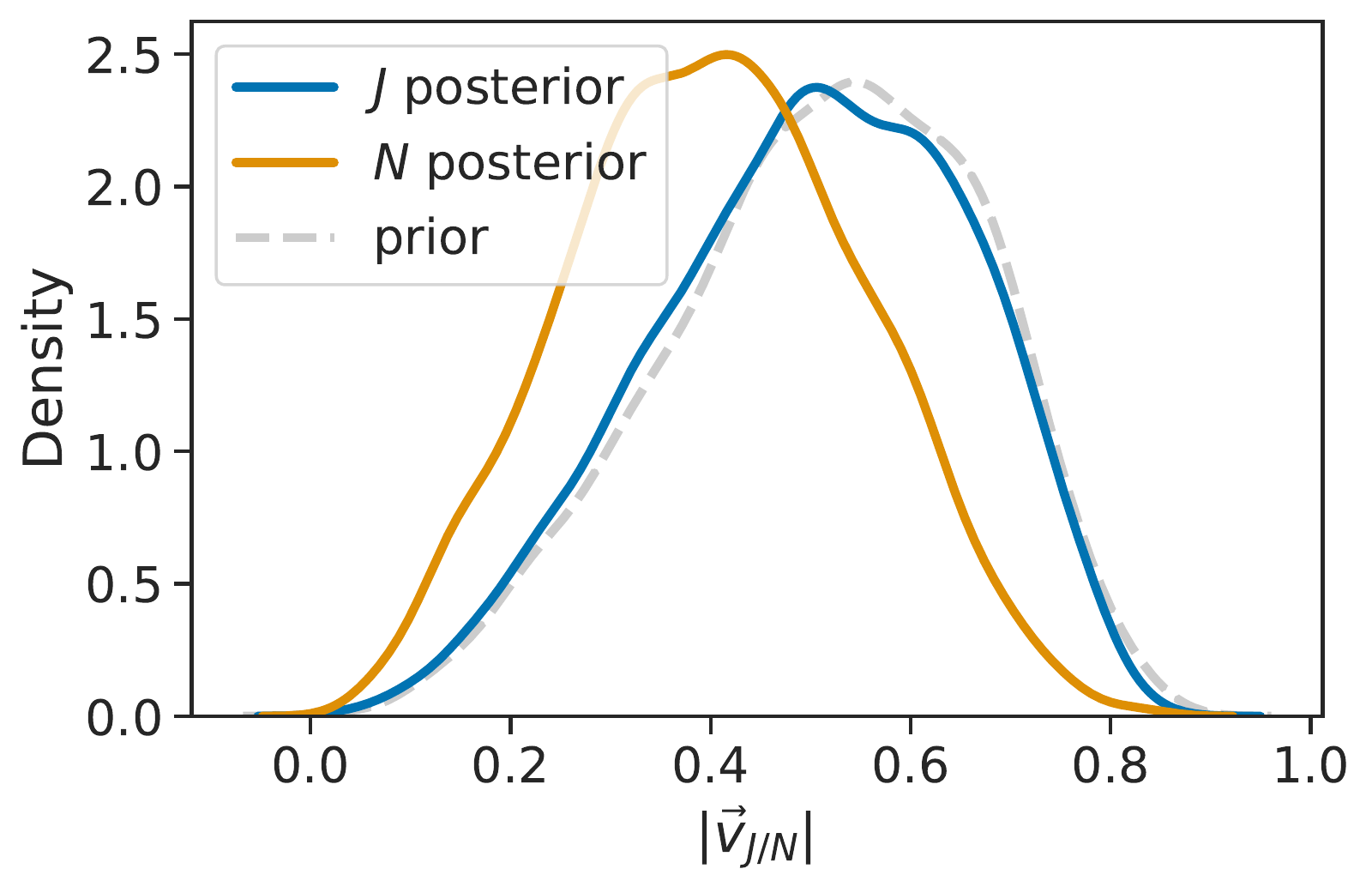}
\caption{\emph{Posterior on dipole magnitudes.} The measurement of Fig.~\ref{fig:jn_corner} translated into one-dimensional posteriors on the $|\dip{J}|$ (blue) and $|\dip{N}|$ (orange) magnitudes.
These are shifted rightward with respect to the implied prior (gray), which itself heavily disfavors $|\dip{J/N}|=0$ due to the reduced phase space near the origin in a three-dimensional space.
}
\label{fig:jn_norm}
\script{norm_plot.py}
\end{figure}

In Fig.~\ref{fig:density_3d}, we provide an additional representation of this posterior projected onto the three-dimensional spaces of $\dip{J}$ and $\dip{N}$; we also present the prior for comparison.
Large dipolar contributions are disfavored by the posterior in all cases, and the posterior distributions are more concentrated than the prior, indicating that the data are informative.
The posterior standard deviations for the three Cartesian components of $\dip{J}$ and $\dip{N}$ are smaller by $\{\varimpJx, \varimpJy, \varimpJz\}$ and $\{\varimpNx, \varimpNy, \varimpNz\}$ respectively with respect to the prior.
The stronger tightening in the $\dip{N}$ distribution is a consequence of $\hat{N}$ being better constrained than $\hat{J}$ in individual events.
Figure \ref{fig:density_3d} again makes it apparent that data disfavor certain directions for the $\dip{J}$ dipole (towards the positive-$x$ quadrants, away from the vernal equinox).

We may translate the result in Fig.~\ref{fig:jn_corner} into a posterior on the magnitudes $|\dip{J/N}|$ of the dipole components, as we do in Fig.~\ref{fig:jn_norm}.
However, the one-dimensional posteriors on these quantities is heavily dominated by the dimensionality of the problem, which results in a Jacobian disfavoring small values of $|\dip{J/N}|$ due to the limited phase space near $\dip{J}=0$ or $\dip{N} = 0$.%
\footnote{In other words, although the probability density is high near the origin, this is outweighed by the greater volume found far away, so that the overall probability of finding $\dip{J/N}=0$ is small.  Any density in $\dip{J/N}$ that is finite at the origin will produce a density on $\left| \dip{J/N} \right|$ that behaves as $\left| \dip{J/N} \right|^2$ near the origin.}
Therefore, even though our three-dimensional prior does not treat $\dip{J/N} = 0$ as a special point (Appendix \ref{app:prior}), the effective prior induced on the magnitudes heavily disfavors the origin in $\left| \dip{J/N} \right|$ due to the reduced available prior volume (gray curve in Fig.~\ref{fig:jn_norm}); that explains why the posteriors in Fig.~\ref{fig:jn_norm} themselves appear to disfavor $|\dip{J/N}| = 0$.
With that in mind, the influence of the data can be seen in the leftward shift of the $|\dip{J/N}|$ distributions with respect to the prior; this is effect is more pronounced for $\dip{N}$, which is a consequence of the fact that $\vec{N}$ is generally better measured than $\vec{J}$.
Although the shift in $|\dip{J}|$ is slight, the data \emph{are} informative about $\dip{J}$---constraining some of its possible orientations (Figs.~\ref{fig:jn_corner} and \ref{fig:density_3d}), if not its overall magnitude.

\section{Validation}
\label{sec:validation}

In this section, we validate our setup by studying simulated datasets in which $\hat{J}$ and $\hat{N}$ are isotropically distributed (\ref{sec:validation:selection}).
We also revisit our assumptions about the astrophysical distribution of \ac{BBH} parameters, described in Sec.~\ref{sec:reweight}, and show that they are robust.

\subsection{Injections from selection set}
\label{sec:validation:selection}

We validate our infrastructure on the set of injections used to evaluate the selection function of the instruments, which was drawn from an intrinsically isotropic distribution \cite{o3-selection}.
This provides an end-to-end test of our setup, including the computation of location and orientation vectors, the selection function and the inference process.

Concretely, we simulate catalogs of detections drawn from the injection set used to evaluate the selection function as described in Sec.~\ref{sec:selection} \cite{o3-selection}.
For simplicity, we treat the injection parameters as a single sample of a fictitious posterior: the input to our hierarchical analysis is one sample per synthetic event, drawn from the distribution in Sec.~\ref{sec:selection}.
At each iteration, we double the size of the simulated catalog.
Since the injection distribution was constructed to be isotropic \cite{o3-selection}, we expect this experiment to indicate $\dip{J/N} = \vec{0}$, with certainty growing with catalog size.

\begin{figure}
\includegraphics[width=\columnwidth]{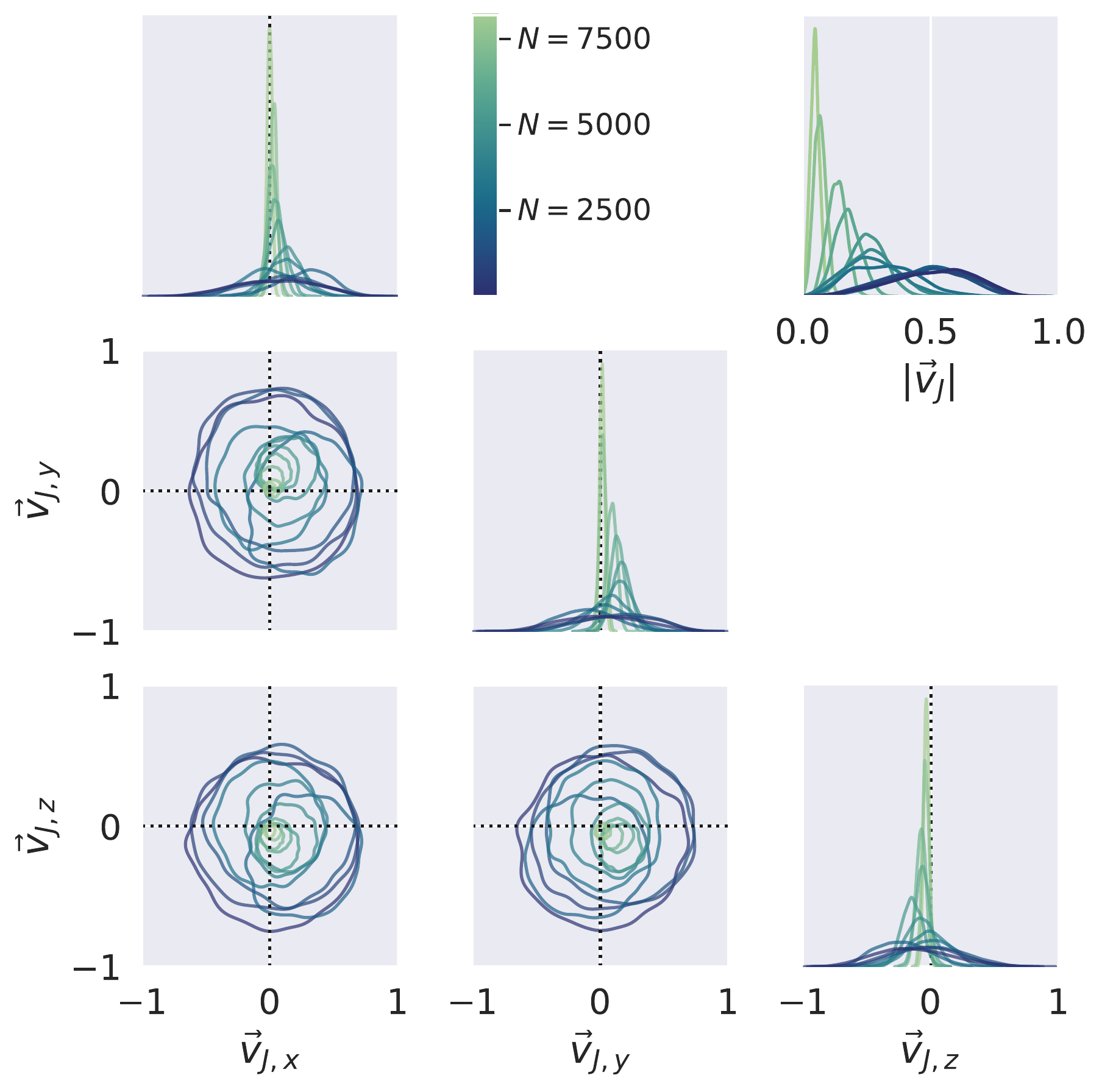}
\caption{\emph{Validation on synthetic catalogs drawn from selection injection set.}
Result of hierarchical analyses on synthetic catalogs of increasing size $N$ (color), obtained by taking draws from the injection set used to evaluate the selection function (Sec.~\ref{sec:selection}) and using them as single samples from the $\hat{J}$ and $\hat{N}$ posterior of synthetic events.
We show the posterior on the components of $\dip{J}$ (main corner), and the implied posterior on the magnitude $|\dip{J}|$ (upper right).
A lighter color corresponds to a larger catalog: starting with $N = \Nstartsel$ events for the darkest color and progressively doubling the catalog size \Nitersel times to reach $N = \Nmaxsel$ events for the lightest color.
The hierarchical analysis measures $\dip{J} = \vec{0}$ with growing precision as the size of the catalog increases.
}
\label{fig:control-sel}
\script{plot_control_selection.py}
\end{figure}

We show the result in Fig.~\ref{fig:control-sel} for $\dip{J}$ (the result for $\dip{N}$ is similar).
As expected, $\dip{J/N} = \vec{0}$ is supported with increasing certainty as the synthetic catalog grows.
This is what we expect if our infrastructure is working as designed.

\subsection{Astrophysical distributions}
\label{sec:validation:rates}

As described, in Sec.~\ref{sec:reweight}, analysis above (Figs.~\ref{fig:jn_corner}--\ref{fig:jn_norm}) hinged on simplified assumptions about the astrophysical distribution of \ac{BH} masses, spins and redshifts.
To evaluate the impact of this simplification, we repeat our analysis but now reweighting to a different astrophysical population based on the measurements in \cite{LIGOScientific:2021psn}.
Specifically, we make use of the highest-probability instantiation of the \textsc{Power Law + Peak} parametric mass model and the \textsc{Default} spin model, with parameters obtained from the posterior samples released in \cite{rpdata}.
This model treats the astrophysical distribution of \ac{BH} masses as a power law plus a Gaussian peak, with density evolving over comoving volume as a power law; the spins are isotropic with a possible Gaussian overdensity in alignment with the orbital angular momentum, and with magnitudes following a Beta distribution (see \cite{LIGOScientific:2021psn} for details).
To implement this new astrophysical prior, both individual detections and the selection injections are reweighted accordingly.

\begin{figure}
\includegraphics[width=\columnwidth]{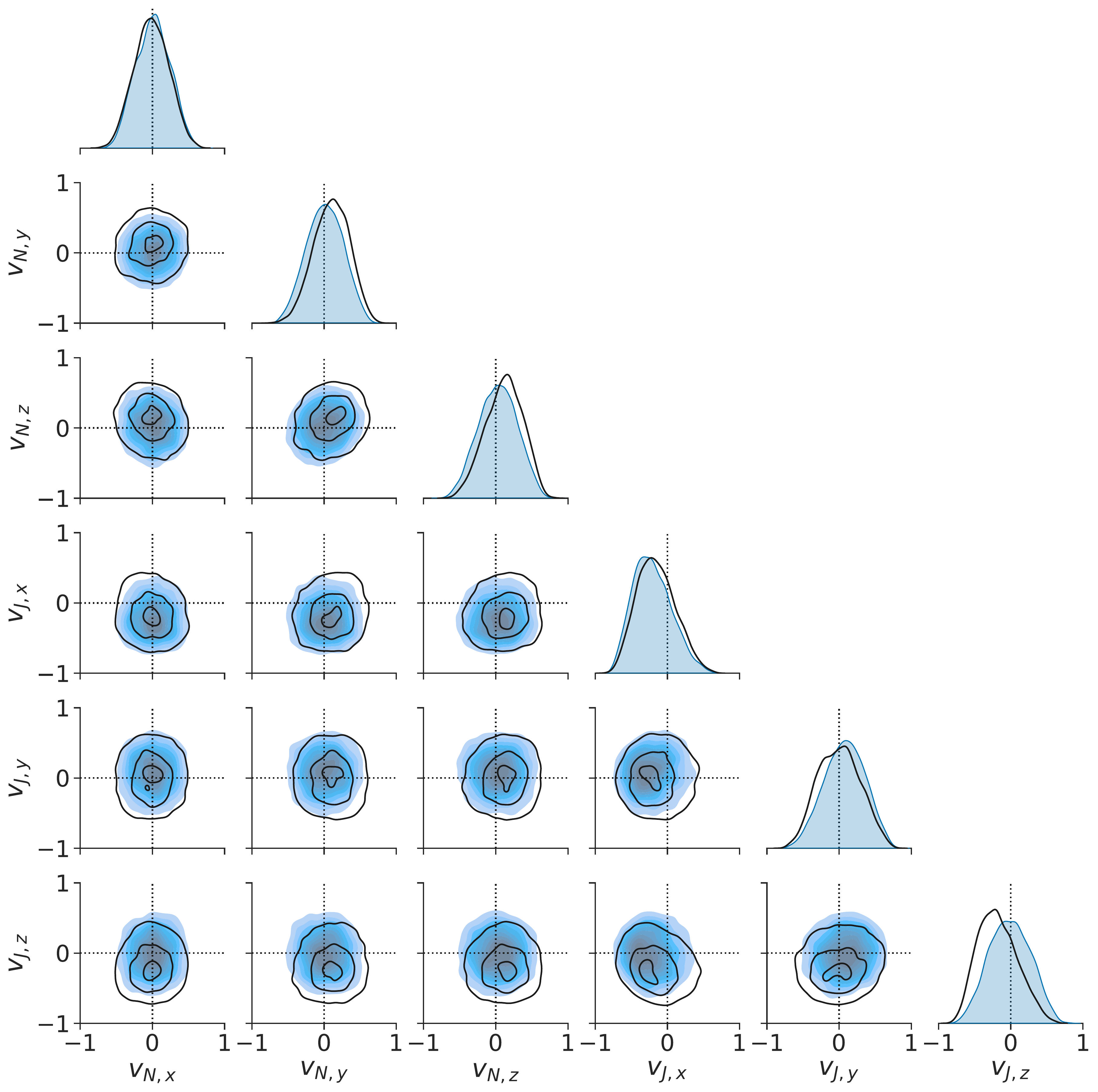}
\caption{\emph{Effect of astrophysical population.} We repeat the isotropy analysis with a different assumption about the underlying distribution of \ac{BBH} parameters based on the measurement in \cite{LIGOScientific:2021psn}.
This yields the result shown in black, as opposed to the result in Fig.~\ref{fig:jn_corner}, which is reproduced here in blue for comparison.
The black contours enclose 90\%, 50\% and 10\% of the marginal probability mass.
}
\label{fig:control-spins}
\script{control_rates_plot.py}
\end{figure}

The result of assuming this different astrophysical distribution is shown in Fig.~\ref{fig:control-spins}, compared to the main result above.
Although the new posterior differs slightly from our primary one in Fig.~\ref{fig:jn_corner}, as we might expect given the difference in models, the change is quite limited and does not qualitatively impact the discussion above.
The discrepancy is somewhat more pronounced for the components of $\dip{J}$, as we might expect from the fact that the reconstruction of $\hat{J}$ must factor our inference on the masses and three dimensional spins of each \ac{BBH}.
In the future, a more comprehensive analysis might simultaneously measure $\dip{J/N}$ and the distribution of astrophysical properties.

\section{Conclusion}
\label{sec:conclusion}

We have demonstrated a measurement constraining the potential alignment of the total angular momenta of \acp{BBH} detected by LIGO and Virgo, using \Nevents detections from their third observing run and duly accounting for selection effects.
In addition to alignment of momenta, we simultaneously looked for inhomogeneities in the distribution of sources over the sky.
We found no evidence against isotropy in either the orientation or, consistent with previous works, the location of LIGO-Virgo \acp{BBH}.
Additionally, we determined that the GWTC-3 data disfavor certain orientations of the potential preferred alignment of angular momenta more than others.

Future measurements will improve as the LIGO, Virgo and KAGRA detectors grow in sensitivity, resulting in many more detections at higher signal-to-noise ratios, which will result in more precise isotropy constraints.
The advent of next generation detectors, like Cosmic Explorer \cite{Dwyer:2014fpa,Evans:2016mbw,Reitze:2019iox} or the Einstein Telescope \cite{Punturo:2010zz}, will enable the exploration of higher order anisotropies and other interesting effects, like correlations with redshift or, potentially, local correlations in the directions $\hat{N}$ and $\hat{J}$ on the sky.

\begin{acknowledgments}
We thank Reed Essick for valuable comments.
The Flatiron Institute is a division of the Simons Foundation.
V.V.~acknowledges funding from the European Union's Horizon 2020 research and
innovation program under the Marie Skłodowska-Curie grant agreement No.~896869.
This material is based upon work supported by NSF's LIGO Laboratory which is a major facility fully funded by the National Science Foundation.
This research has made use of data or software obtained from the Gravitational Wave Open Science Center (gw-openscience.org), a service of LIGO Laboratory, the LIGO Scientific Collaboration, the Virgo Collaboration, and KAGRA. LIGO Laboratory and Advanced LIGO are funded by the United States National Science Foundation (NSF) as well as the Science and Technology Facilities Council (STFC) of the United Kingdom, the Max-Planck-Society (MPS), and the State of Niedersachsen/Germany for support of the construction of Advanced LIGO and construction and operation of the GEO600 detector. Additional support for Advanced LIGO was provided by the Australian Research Council. Virgo is funded, through the European Gravitational Observatory (EGO), by the French Centre National de Recherche Scientifique (CNRS), the Italian Istituto Nazionale di Fisica Nucleare (INFN) and the Dutch Nikhef, with contributions by institutions from Belgium, Germany, Greece, Hungary, Ireland, Japan, Monaco, Poland, Portugal, Spain. The construction and operation of KAGRA are funded by Ministry of Education, Culture, Sports, Science and Technology (MEXT), and Japan Society for the Promotion of Science (JSPS), National Research Foundation (NRF) and Ministry of Science and ICT (MSIT) in Korea, Academia Sinica (AS) and the Ministry of Science and Technology (MoST) in Taiwan.
This is a reproducible article compiled with \textsc{showyourwork} \cite{Luger2021}.
This paper carries LIGO document number \dcc{}.
\end{acknowledgments}

\appendix

\section{Computing the total angular momentum}
\label{app:j}

Here we describe how to compute the total angular momentum vector $\vec{J}$ in the celestial coordinate frame of the main text, starting from \ac{BBH} parameters measured in LIGO-Virgo analyses.
The total angular momentum is defined, as in Eq.~\eqref{eq:j}, to be the vector sum of the orbital angular momentum $\vec{L}$ and the dimensionful \ac{BH} spins $\vec{S}_{1/2}$.
We can thus compute $\vec{J}$ by summing the components of $\vec{L}$ and $\vec{S}_{1/2}$ in some common frame with known orientation relative to our target.

The Cartesian components of the \ac{BH} dimensionless spins, $\vec{\chi}_{1/2} \equiv \vec{S}_{1/2} / m_{1/2}^2$ in units where $G=c=1$, can be readily obtained from the LIGO-Virgo samples, in a frame in which the $z$-axis points along (the Newtonian) $\vec{L}$ and the $x$-axis points along the orbital vector from the lighter to the heavier body \cite{LALSuite:spins} (see, e.g., Fig.~18 in \cite{Isi:2022mbx}); this specification is established in reference to some specific point in the evolution of the system, e.g., when the \ac{GW} signal at the detector reaches some frequency $f_\mathrm{ref}$.
With that information, all we need to compute the components of $\vec{J}$ in that same frame is the magnitude $|L|$ of the orbital angular momentum; then, the $\vec{J}$ vector will be specified by
\begin{subequations}
\label{eq:jcomp}
\begin{align}
J_x &= S_{1,x} + S_{2,x}, \\
J_y &= S_{1,y} + S_{2,y}, \\
J_z &= S_{1,z} + S_{2,z} + |L|,
\end{align}
\end{subequations}
and $\vec{J} = J_x \hat{x} + J_y \hat{y} + J_z \hat{L}$, where $\hat{L} \equiv \vec{L}/|L|$ is the direction of the orbital angular momentum, $\hat{x}$ is the orbital vector at the reference time, and $\hat{y}$ completes the triad.

In the above equations, $(S_{1/2,x},S_{1/2,y},S_{1/2,z})$ are the components of the dimensionful spins, which we can derive from the respective components of the dimensionless spins $\vec{\chi}_{1/2}$ using the masses $m_{1/2}$ and the definition above. 
Following the standard LIGO-Virgo calculation \cite{LALSuite:spins},%
\footnote{Here we are following the definition of $|L|$ within the LIGO-Virgo software; this is adopted in defining the inclination parameters $\theta_{JN}$ and $\iota$. The effect of truncating the series at the first post-Newtonian order is expected to be small, but could be revisited in future work.}
we can approximate the magnitude of the orbital angular momentum as \cite{Kidder:1995zr,Bohe:2012mr}
\begin{equation}
|L| = L_N \left(1 + \ell_{\rm 1PN}\right),
\end{equation}
where $L_N = m_1 m_2 / v$ is the Newtonian angular momentum, $v = \left( \pi M f_{\rm ref} \right)^{1/3}$ is the post-Newtonian expansion parameter at the reference frequency, and $\ell_{\rm 1PN} = v^2 \left( 3 + \eta /3 \right)/2$ is the first-order correction to $|L|$, for the symmetric mass ratio $\eta \equiv m_1 m_2/M^2$ and the total mass $M \equiv m_1 + m_2$.
We can then get the components of the orientation vector $\hat{J} \equiv \vec{J} /|J|$ by normalizing Eq.~\eqref{eq:jcomp}.

The components of $\vec{J}$ in Eq.~\eqref{eq:jcomp} are defined in a binary-specific frame which doest not facilitate comparisons across different systems.
To express $\vec{J}$ in a common frame for all binaries, all we need is to express the $(\hat{x}, \hat{y}, \hat{L})$ coordinate basis of Eq.~\eqref{eq:jcomp} in a Celestial coordinate frame.
We can achieve this by first obtaining the Cartesian components for $\vec{L}$ in this frame using the measured right ascension $\alpha$, declination $\delta$, inclination $\iota$, polarization angle $\psi$, and the reference orbital phase $\phi_\mathrm{ref}$ (see, e.g., Figs.~6 and 8 in \cite{Isi:2022mbx}).
In doing so, however, it is important to keep in mind that, since the polarization angle only enters the waveform as $2\psi$ (i.e., $\psi$ and $\psi + \pi$ are degenerate), the LIGO-Virgo analyses usually only allow $0 \leq \psi < \pi$; yet, $\psi$ and $\psi + \pi$ are two physically distinct configurations, so we must for our purposes mirror the samples to ensure they span the full range of polarization angles, $0 \leq \psi < 2\pi$ (or, equivalently, randomly add $0$ or $\pi$ to $\psi$ for each sample).

Having properly accounted for this degeneracy, $\hat{L}$ can be obtained from the source location $\hat{N} \equiv - \hat{k}$ in Eq.~\eqref{eq:n}, $\iota$ and $\psi$ via rotations and geometric products as
\begin{equation}
\hat{L} = R_{\hat{w}_y}(\iota)\, \hat{k} \, ,
\end{equation}
where $R_{\hat{v}}(\theta)$ is a right-handed rotation by an angle $\theta$ around some direction $\hat{v}$, and $\hat{w}_y \propto \hat{k} \times \hat{w}_x$ with $\hat{w}_x = R_{\hat{k}}(\psi) \hat{u}$, for $\hat{u}$ due west (see \cite{Anderson:T010110} for details).
We can similarly obtain $\hat{x}$ by rotating $\hat{w}_y$ around $\vec{L}$ by $\phi_\mathrm{ref}$, namely
\begin{equation}
\hat{x} = R_{\hat{L}}(\phi_{\rm ref})\, \hat{w}_y \, ,
\end{equation}
and $\hat{y} \propto \hat{L} \times \hat{x}$ completes the triad.
This provides expressions for the components of the $(\hat{x}, \hat{y}, \hat{L})$ basis for each binary in a common reference frame, from which we can obtain the components of all $\hat{J}$ vectors in that same frame via Eq.~\eqref{eq:jcomp}.
Those components are the input for the hierarchical analysis described in the main text.

\section{Prior in Unconstrained Parameters}
\label{app:prior}

\begin{figure}
\includegraphics[width=\columnwidth]{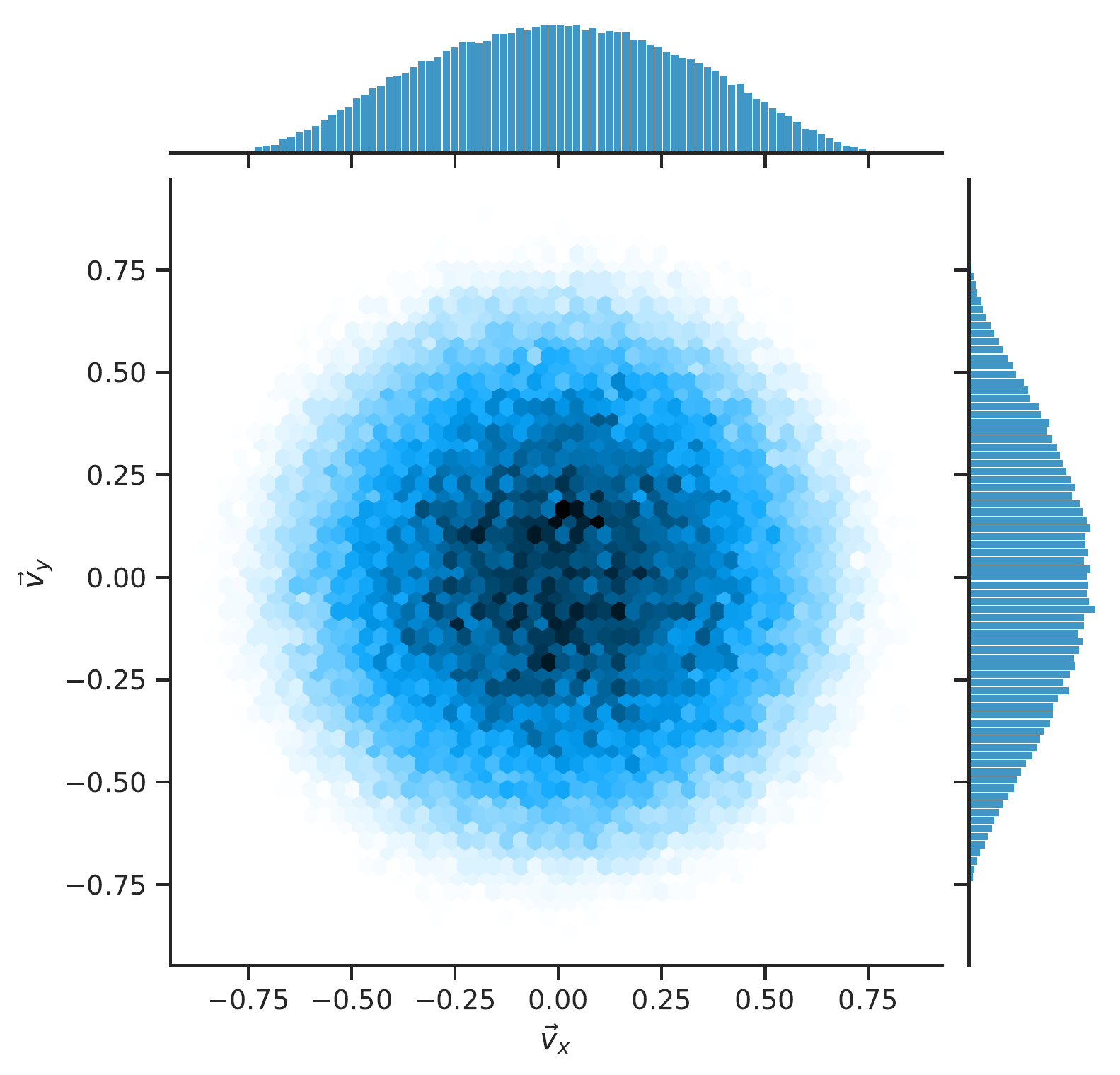}
\caption{\emph{Prior}. Two-dimensional slice of the Gaussian prior on the components of $\dip{J/N}$ for $\sigma = 0.4$, as used for the analysis in the main paper.
With this choice, the origin is not disfavored by the three-dimensional prior probability density.
}
\label{fig:prior}
\script{prior_plot.py}
\end{figure}

Recall that we sample in parameters $\dipraw{N/J}$ defined by 
\begin{equation}
  \dipraw{N/J} = \frac{\dip{N/J}}{\sqrt{1 - \dip{N/J} \cdot \dip{N/J}}}
\end{equation}
that map the unit ball $\mathbb{B}^3$ to $\mathbb{R}^3$.  A Gaussian prior with
mean $\vec{0}$ and standard deviation in each component of $\sigma$ on
$\dipraw{N/J}$ induces a prior on $\dip{N/J}$ that is 
\begin{multline}
  \log p\left( \dip{N/J} \right) = \mathrm{const} \\ -\frac{\left| \dip{N/J} \right|^2}{2 \sigma^2 \left(1 - \left| \dip{N/J} \right|^2 \right)} - \frac{5}{2} \log \left( 1 - \left| \dip{N/J} \right|^2 \right).
\end{multline}
The derivative of the density with respect to $\left| \dip{N/J} \right|$
vanishes at the origin, as it must by symmetry.  But the second derivative does
not, and is 
\begin{equation}
  \frac{\partial^2 \log p}{\partial \left| \dip{N/J} \right|^2} = 5 - \frac{1}{\sigma^2}.
\end{equation}
Thus, such a Gaussian prior will have a \emph{maximum} at $\dip{N/J} = \vec{0}$
only if $\sigma < 1/\sqrt{5} \simeq 0.45$; otherwise it places too much prior
mass on large $\left|\dipraw{N/J}\right|$ which generates a ring-shaped maximum
in the prior for some $\left| \dip{N/J} \right| > 0$ and a \emph{minimum} at the
origin.  With the desire to be uninformative about the typical scale of the components of $\dip{N/J}$
while keeping the maximum prior density at the origin, we choose $\sigma = 0.4 <
1/\sqrt{5}$ for the analysis in this paper.
A two-dimensional slice of the prior on the components of $\dip{N/J}$ for this choice is illustrated in Fig.~\ref{fig:prior}.

\section{Direction of total angular momentum for individual events}
\label{app:skymaps}

In Fig.~\ref{fig:skymaps-1}, we present posteriors on the direction of the total angular momentum $\hat{J}$ for all \Nevents events in our set.
We produced these posteriors by applying the calculation described in Appendix \ref{app:j} to the samples released by LIGO-Virgo \cite{zenodo:GWTC-2.1,zenodo:GWTC-3}, after reweighting as outlined in Sec.~\ref{sec:reweight}.

The skymaps in the figure provide a Molleweide projection of probability density over the celestial sphere, in the standard equatorial, geocentric coordinates used in the main text.
A darker color corresponds to a direction in the sky with more probability density.
From these maps, it is clear that not all events are equally informative about $\hat{J}$; better constrained events will tend to dominate our hierarchical measurement.

The skymaps were produced using the \textsc{ligo.skymap} package \cite{skymap,Singer:2016eax,Singer:2016erz}, a set of standard LIGO-Virgo tools for the processing of probability densities over the sphere.
We make these figures, corresponding \ac{FITS} files, and code used to generate them, available in our data release \cite{repo}.

\bibliography{bib}

{\marginicon{} %
  \begin{figure*}
\centering
\subfloat{\centering
    \stackunder{\includegraphics[width=0.20\textwidth]{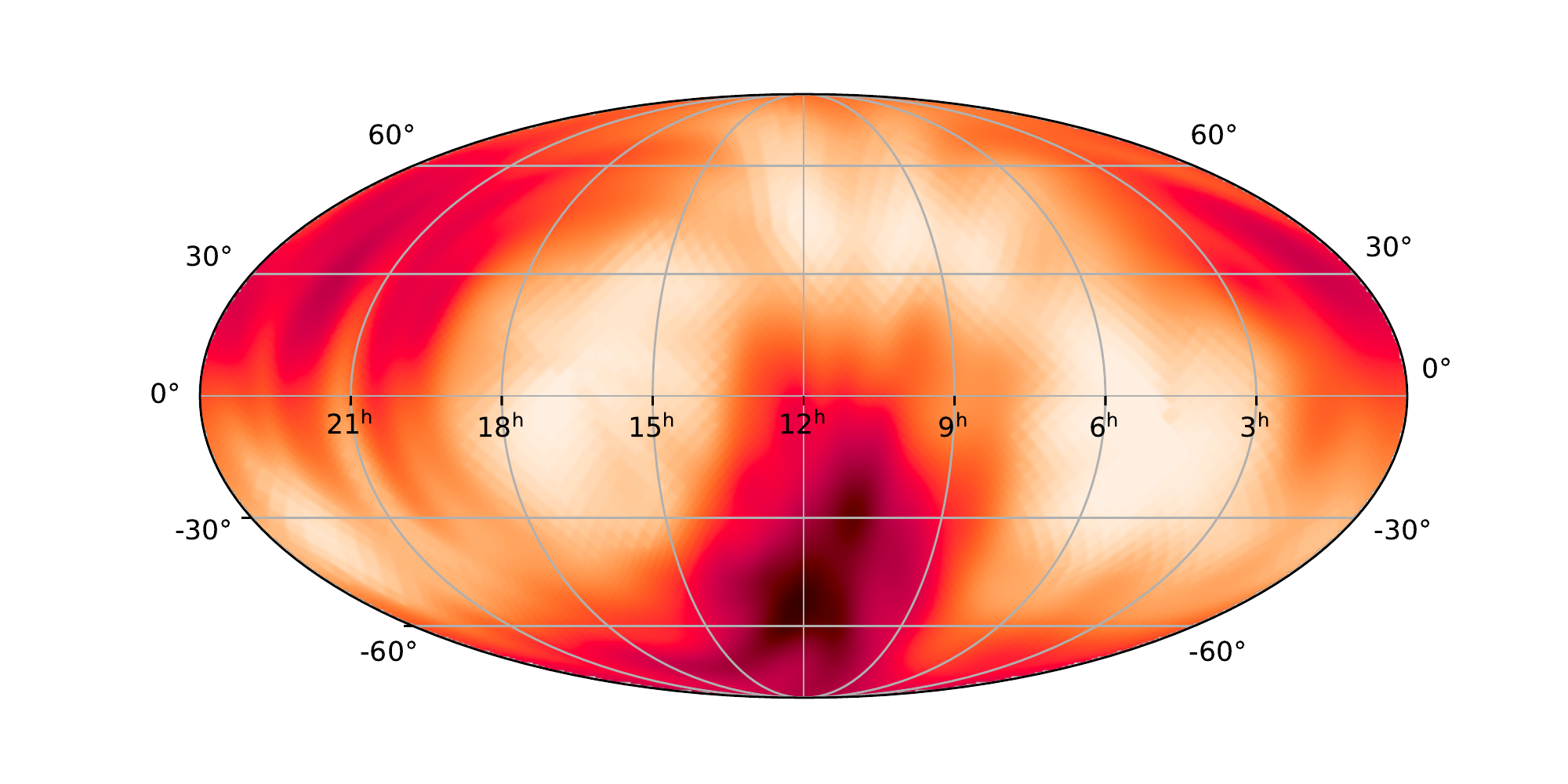}}{GW190408\_181802}
}%
\subfloat{\centering
    \stackunder{\includegraphics[width=0.20\textwidth]{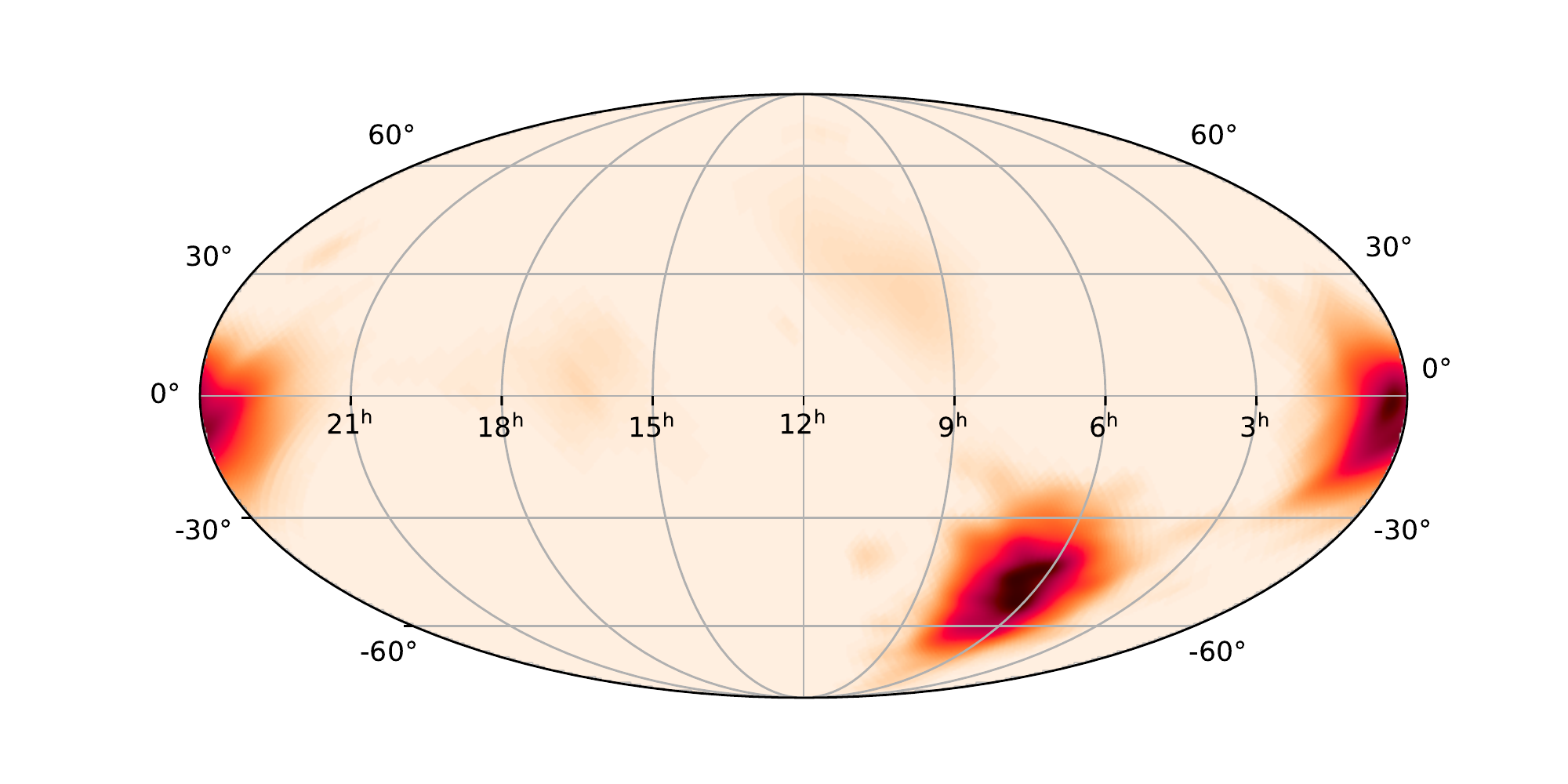}}{GW190412\_053044}
}%
\subfloat{\centering
    \stackunder{\includegraphics[width=0.20\textwidth]{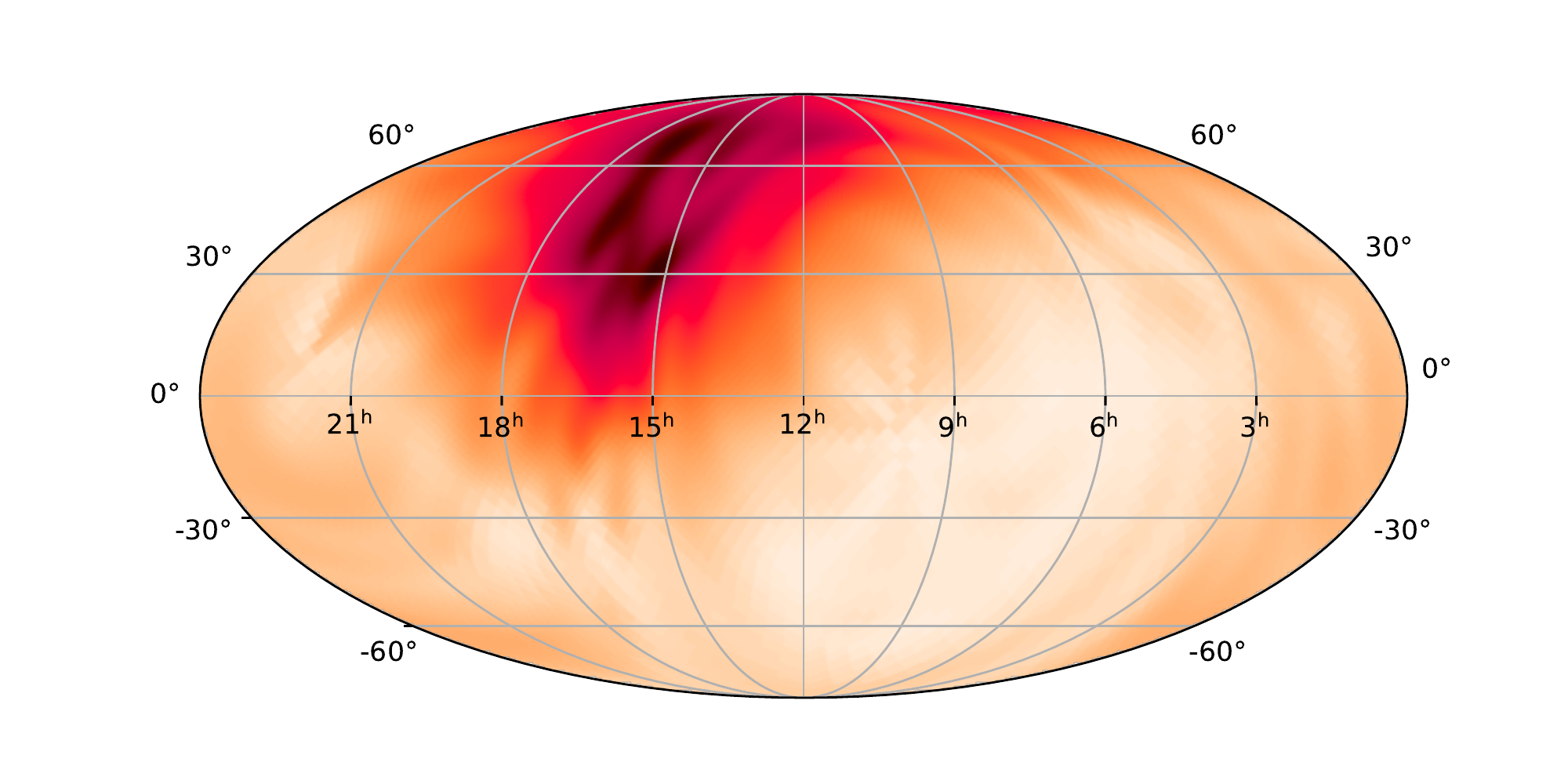}}{GW190413\_052954}
}%
\subfloat{\centering
    \stackunder{\includegraphics[width=0.20\textwidth]{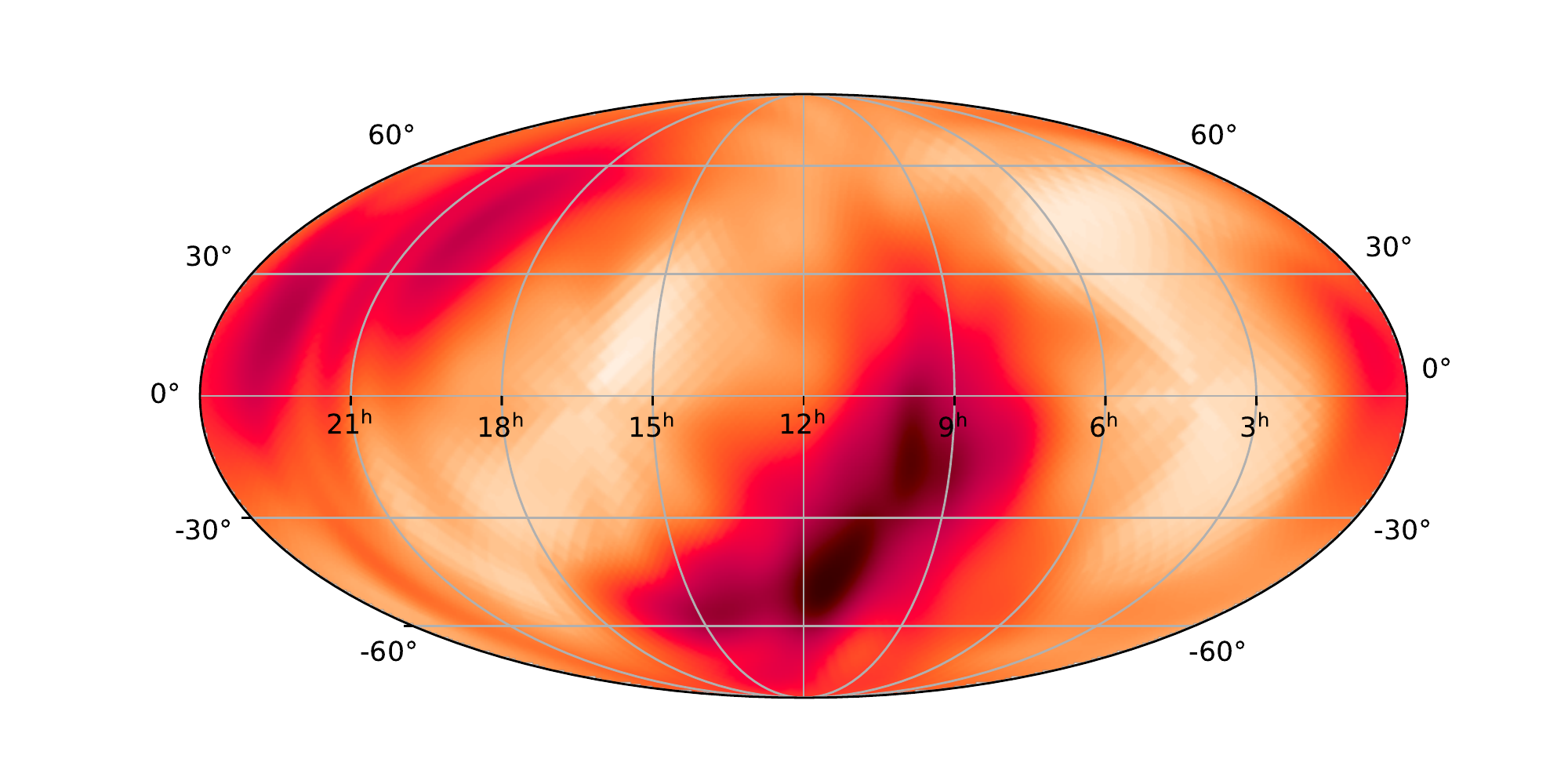}}{GW190413\_134308}
}%
\subfloat{\centering
    \stackunder{\includegraphics[width=0.20\textwidth]{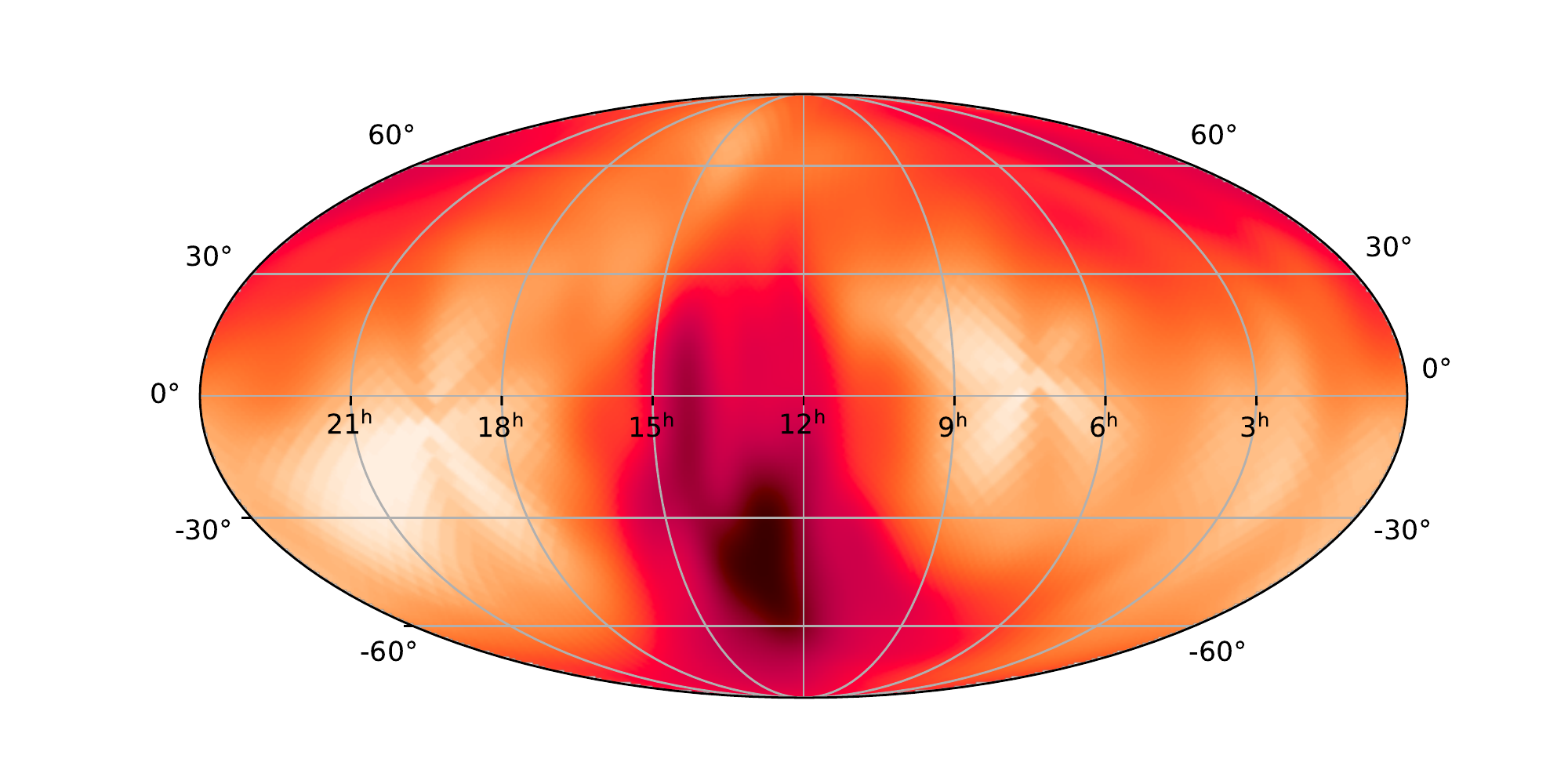}}{GW190421\_213856}
}\\
\subfloat{\centering
    \stackunder{\includegraphics[width=0.20\textwidth]{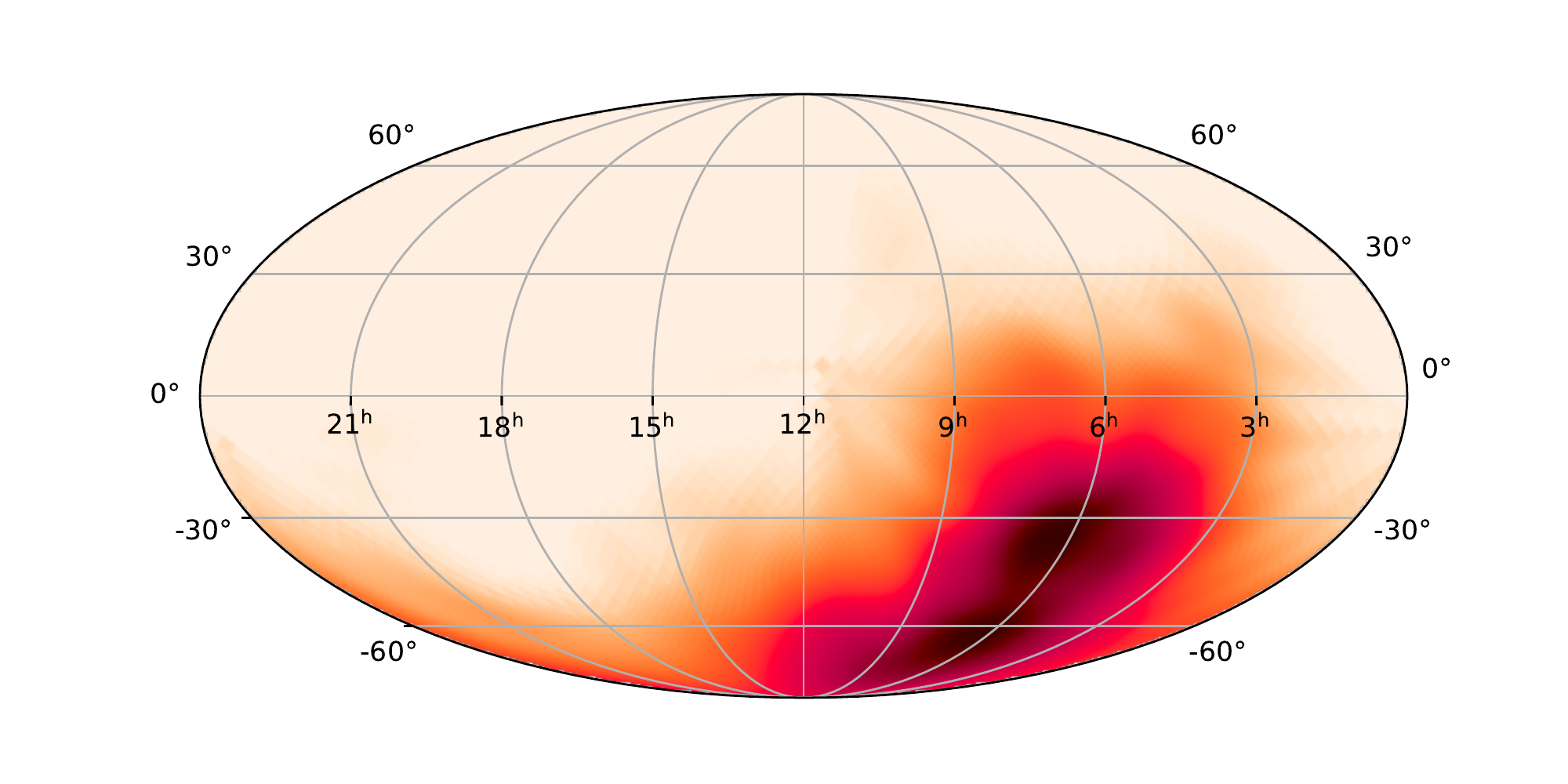}}{GW190503\_185404}
}%
\subfloat{\centering
    \stackunder{\includegraphics[width=0.20\textwidth]{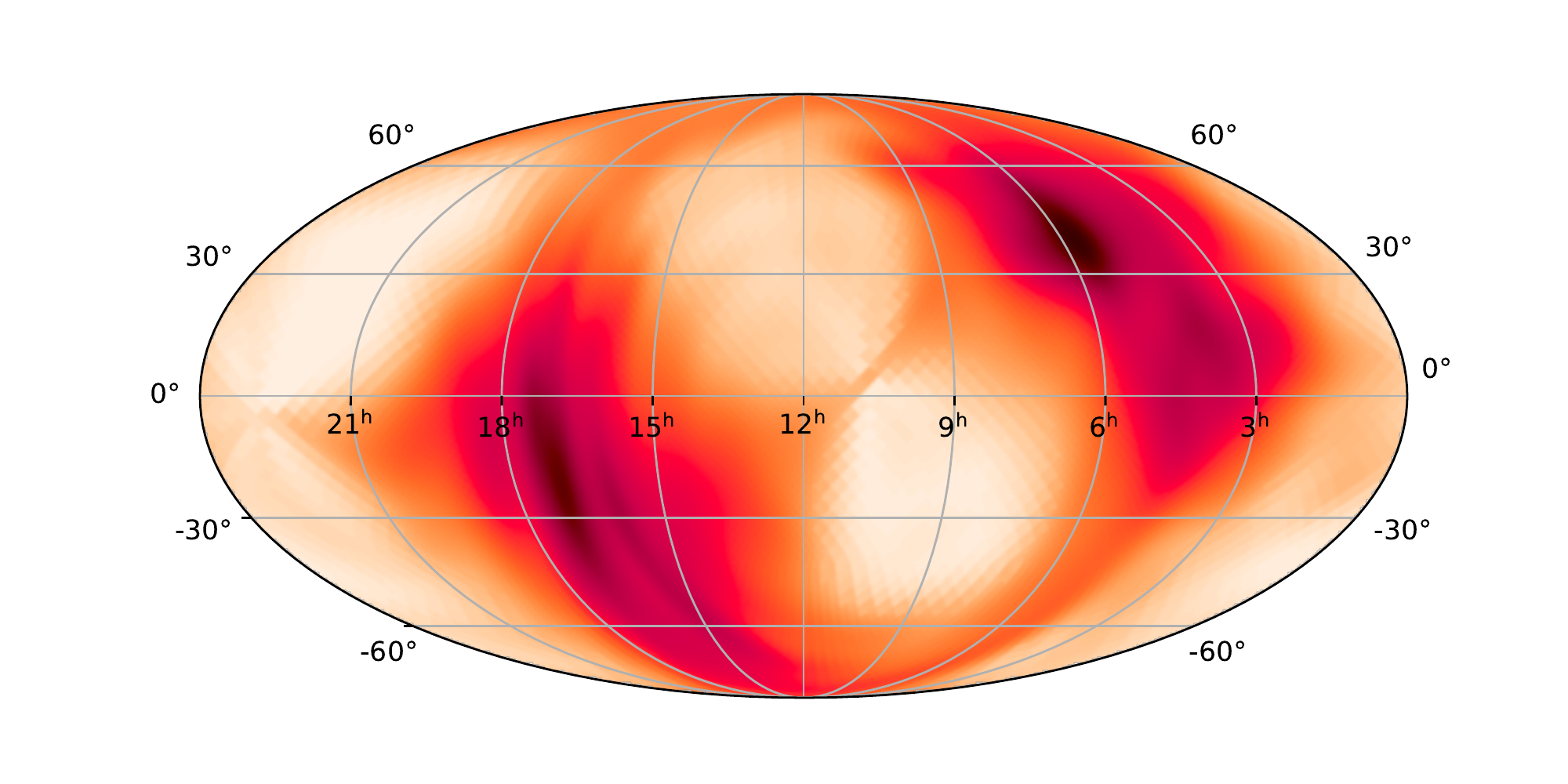}}{GW190512\_180714}
}%
\subfloat{\centering
    \stackunder{\includegraphics[width=0.20\textwidth]{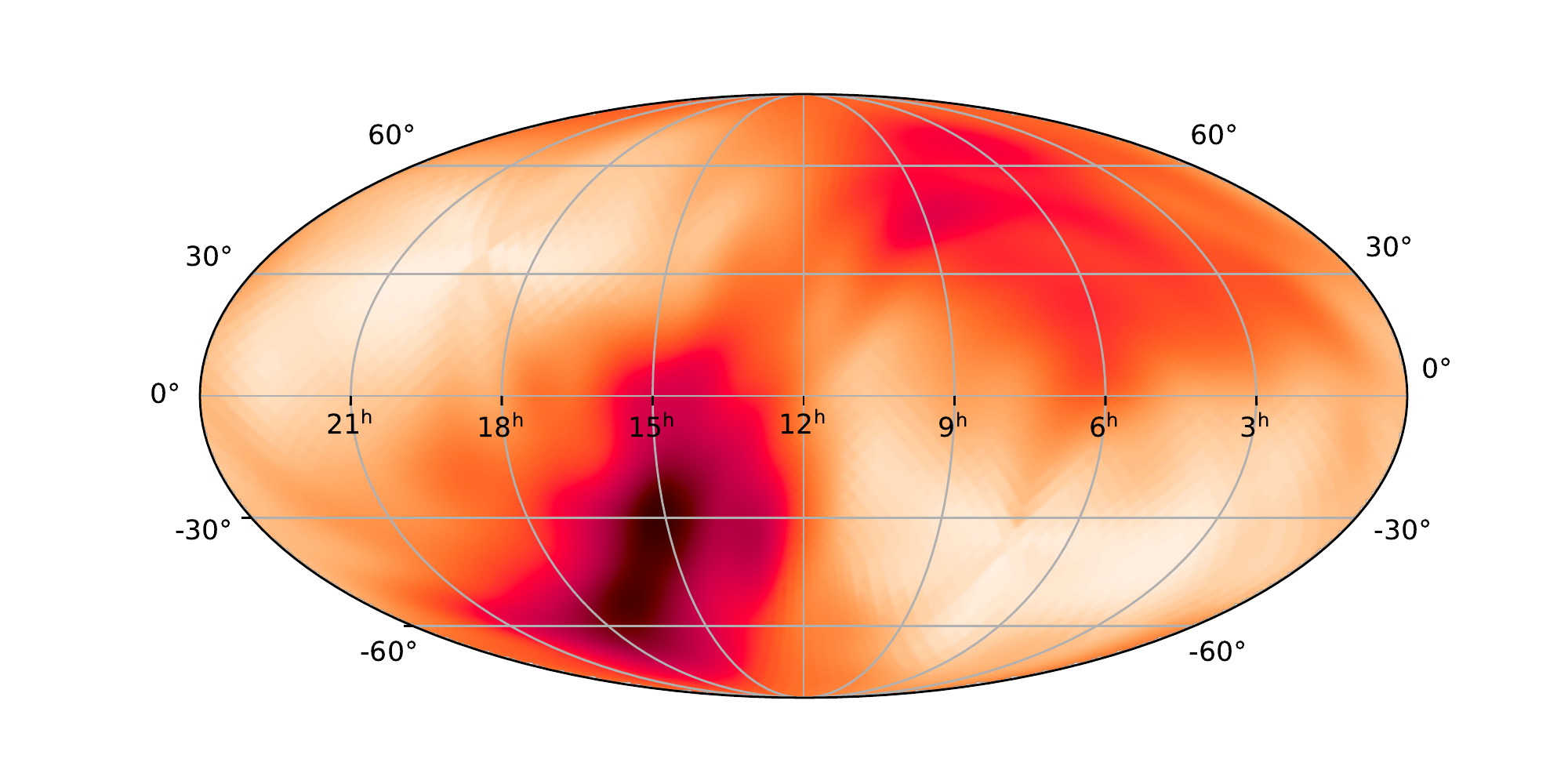}}{GW190513\_205428}
}%
\subfloat{\centering
    \stackunder{\includegraphics[width=0.20\textwidth]{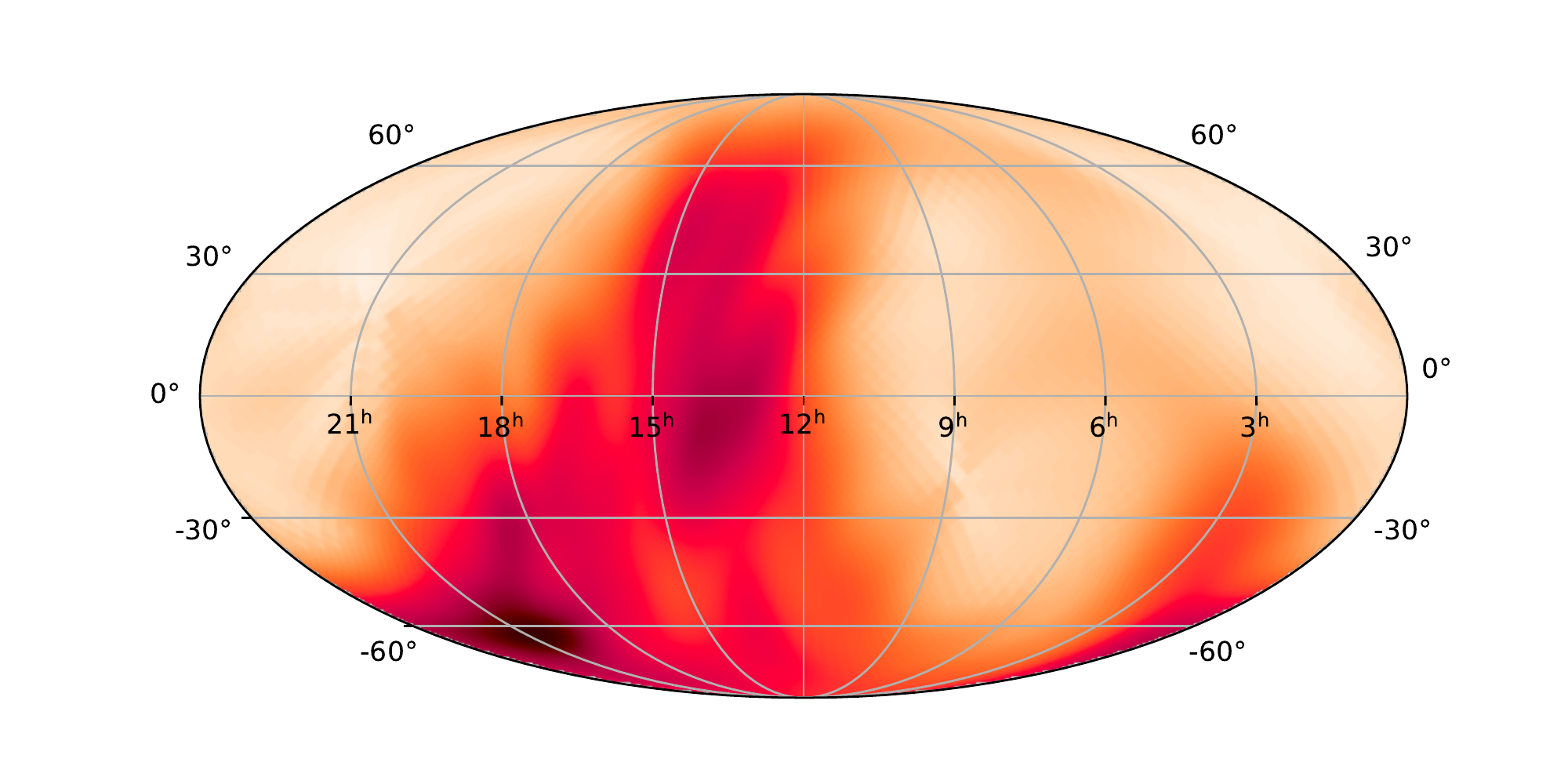}}{GW190517\_055101}
}%
\subfloat{\centering
    \stackunder{\includegraphics[width=0.20\textwidth]{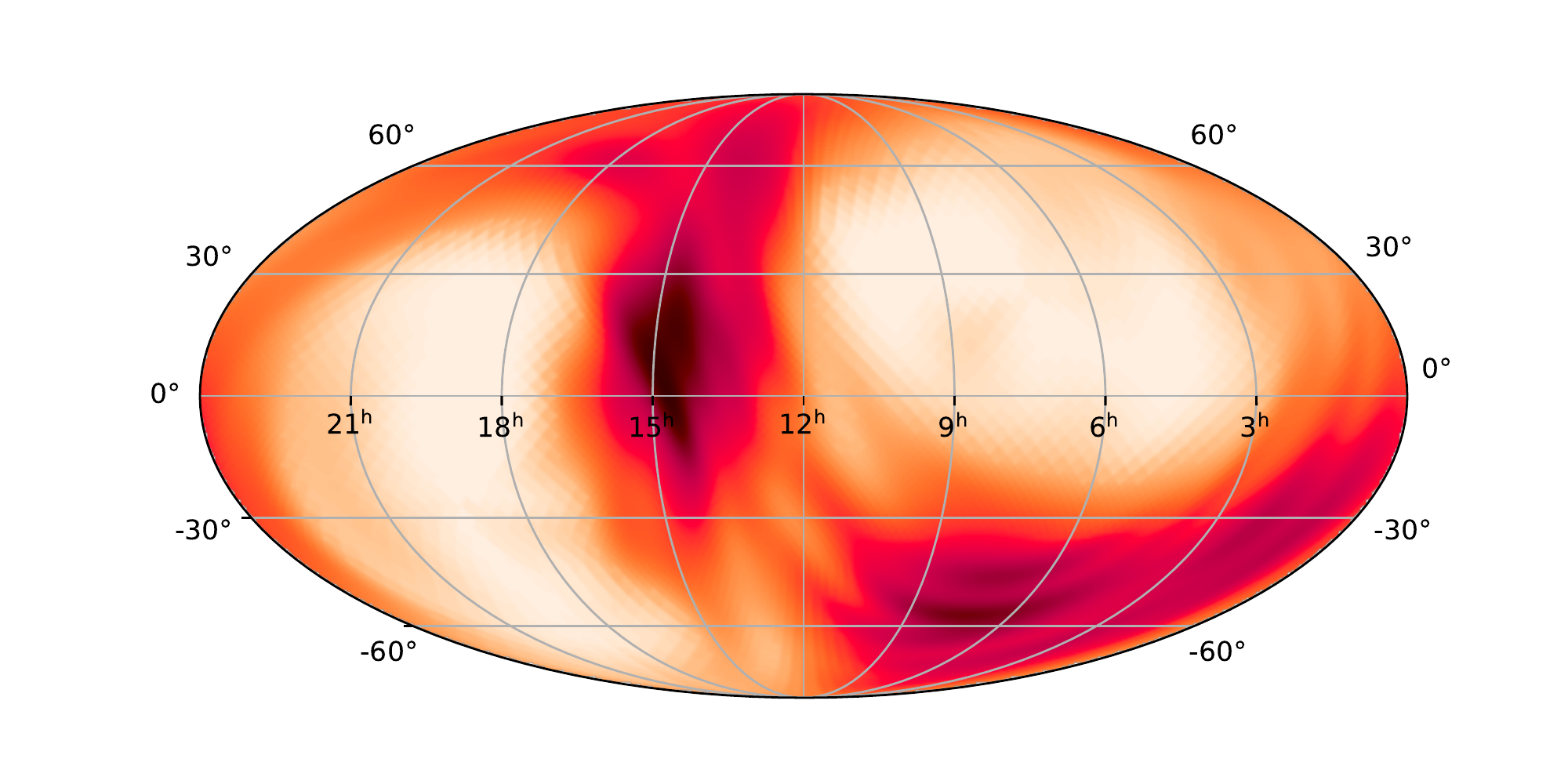}}{GW190519\_153544}
}\\
\subfloat{\centering
    \stackunder{\includegraphics[width=0.20\textwidth]{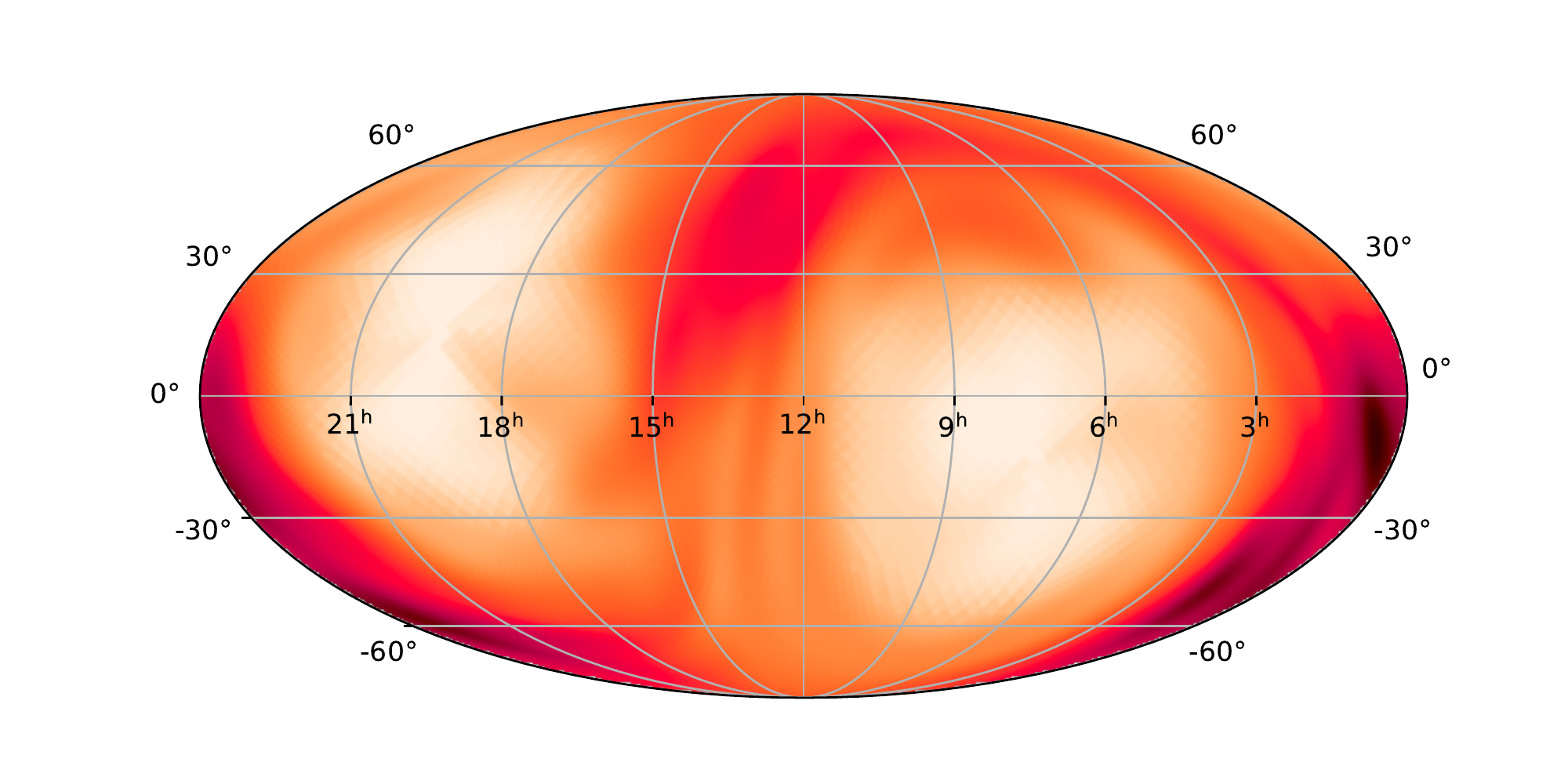}}{GW190521\_030229}
}%
\subfloat{\centering
    \stackunder{\includegraphics[width=0.20\textwidth]{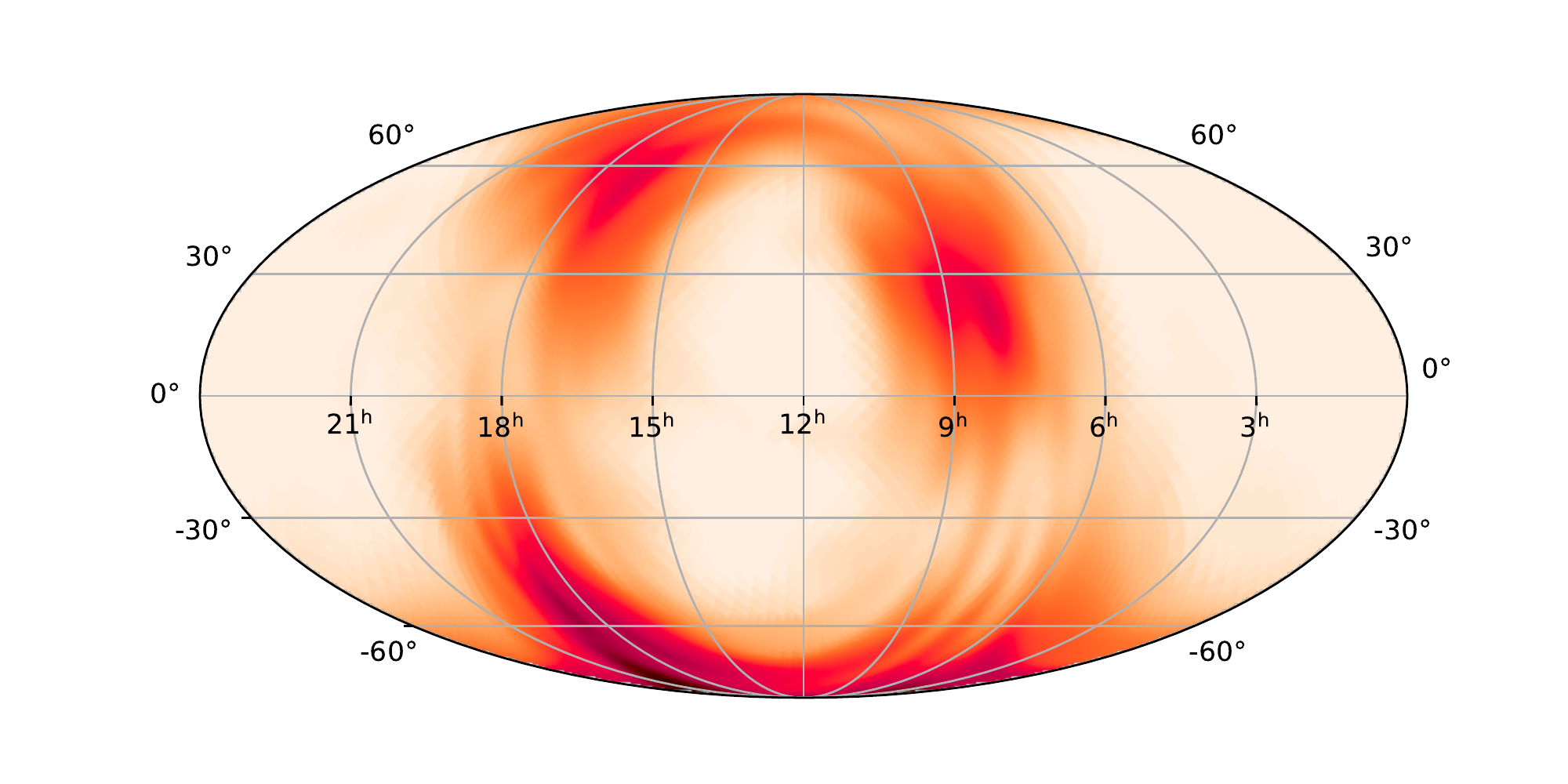}}{GW190521\_074359}
}%
\subfloat{\centering
    \stackunder{\includegraphics[width=0.20\textwidth]{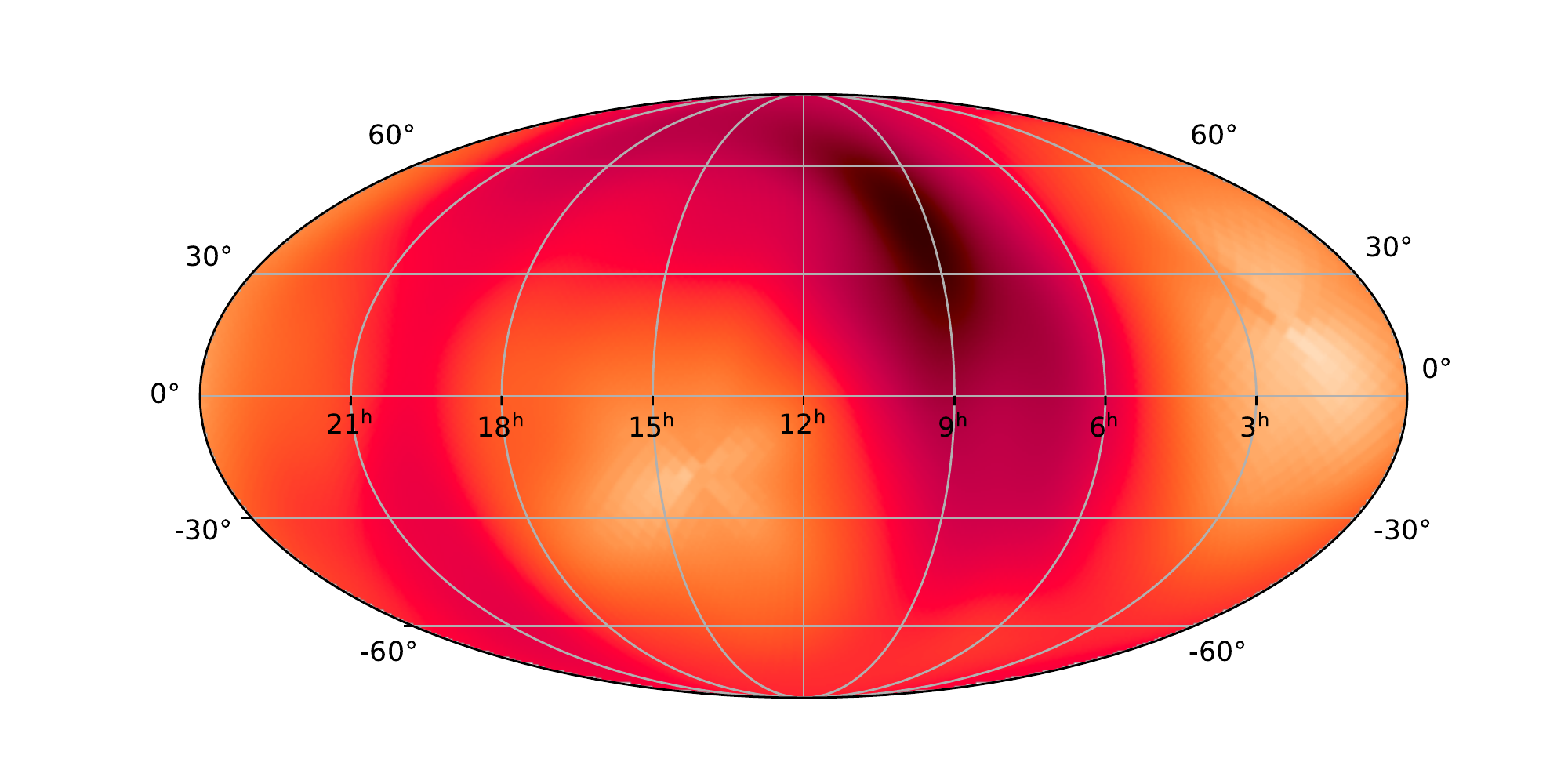}}{GW190527\_092055}
}%
\subfloat{\centering
    \stackunder{\includegraphics[width=0.20\textwidth]{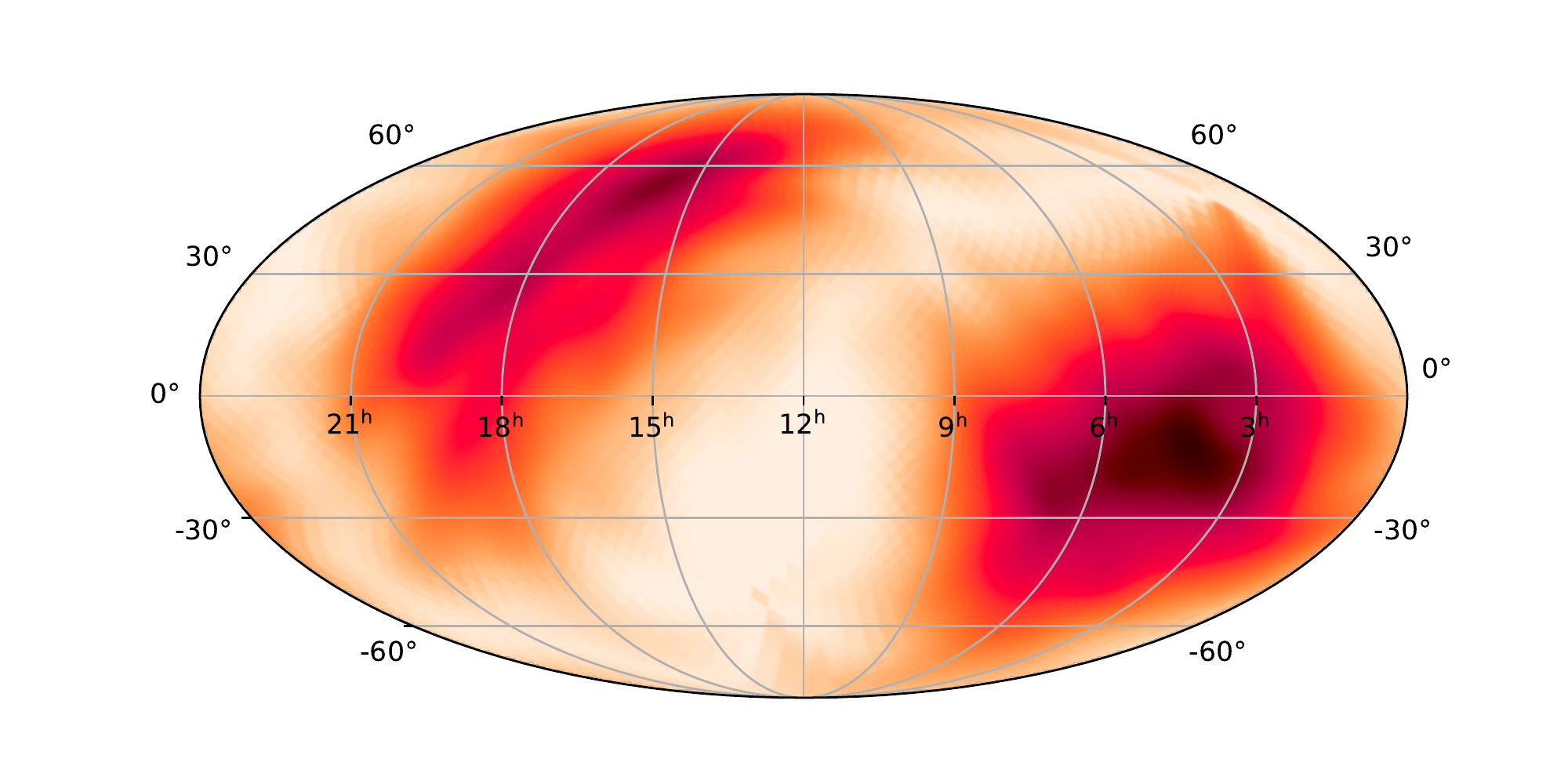}}{GW190602\_175927}
}%
\subfloat{\centering
    \stackunder{\includegraphics[width=0.20\textwidth]{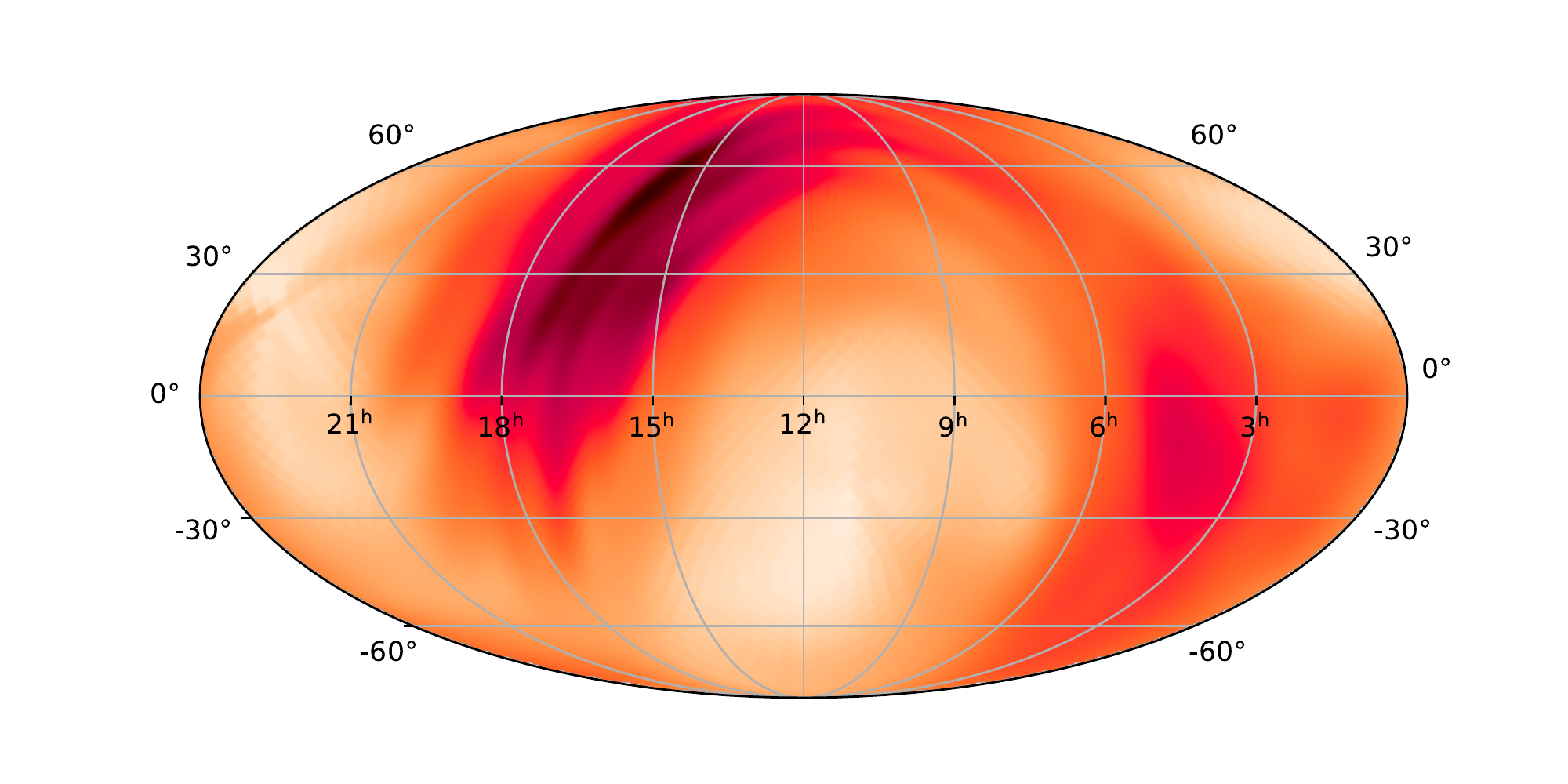}}{GW190620\_030421}
}\\
\subfloat{\centering
    \stackunder{\includegraphics[width=0.20\textwidth]{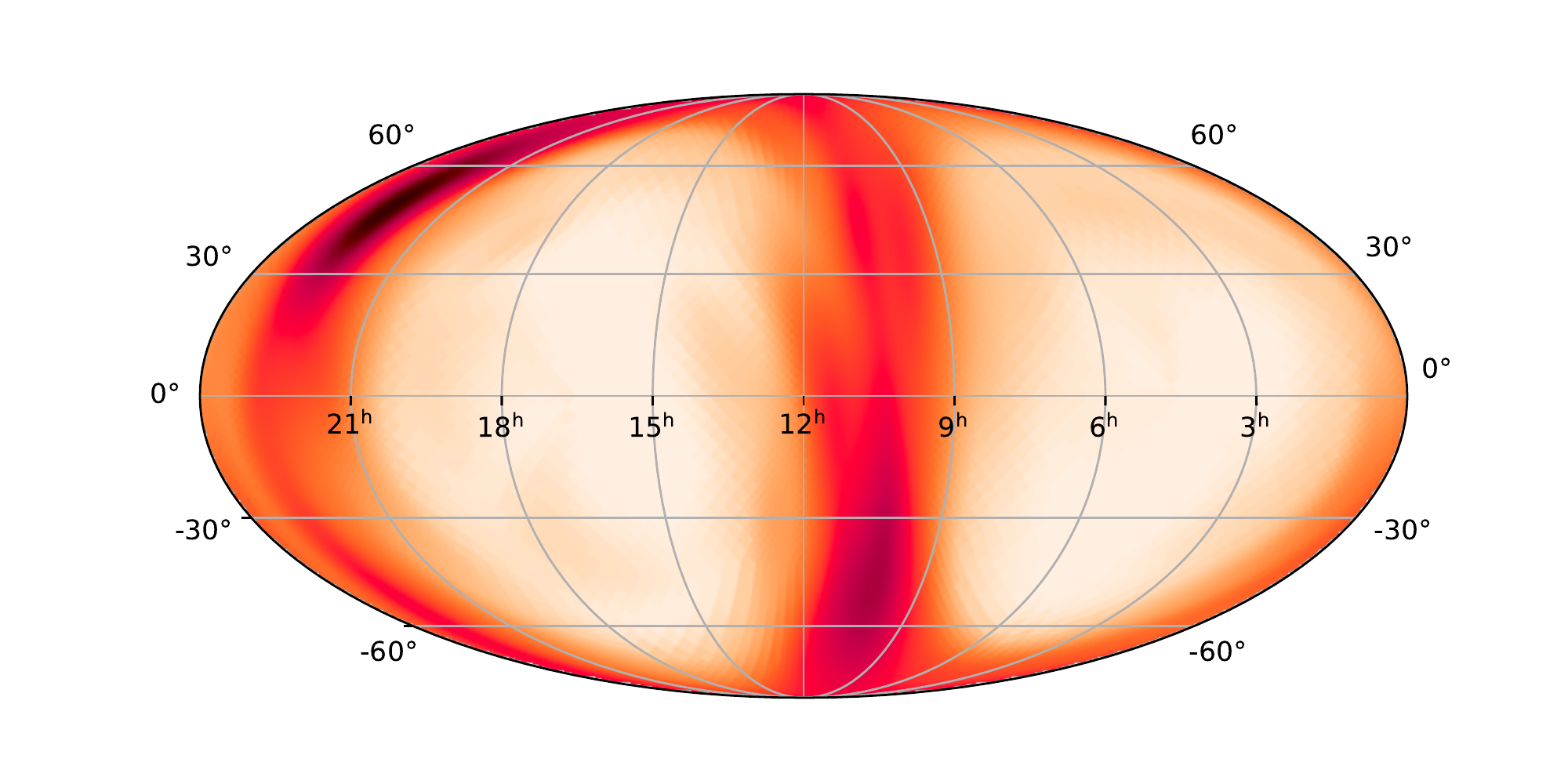}}{GW190630\_185205}
}%
\subfloat{\centering
    \stackunder{\includegraphics[width=0.20\textwidth]{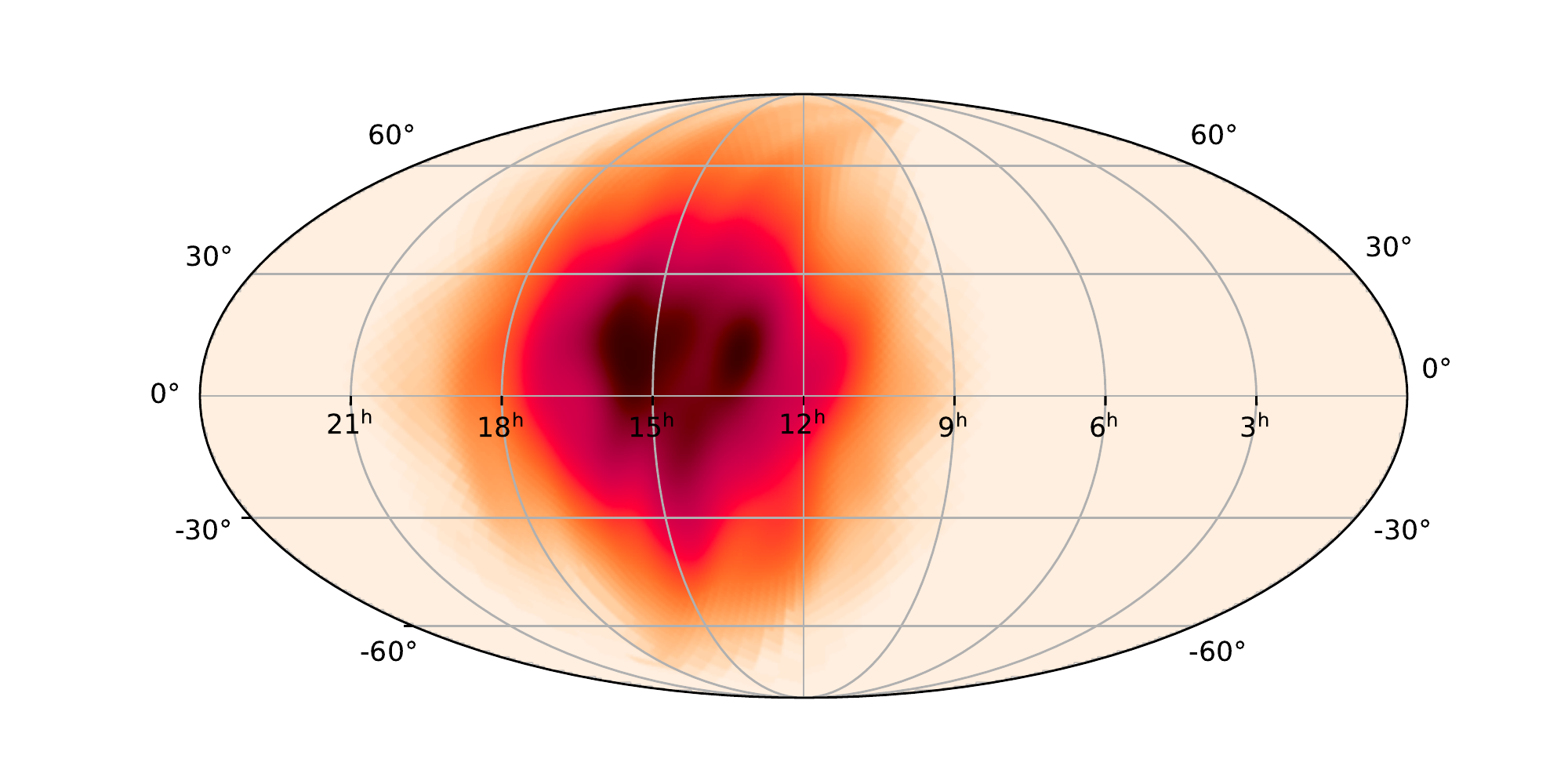}}{GW190701\_203306}
}%
\subfloat{\centering
    \stackunder{\includegraphics[width=0.20\textwidth]{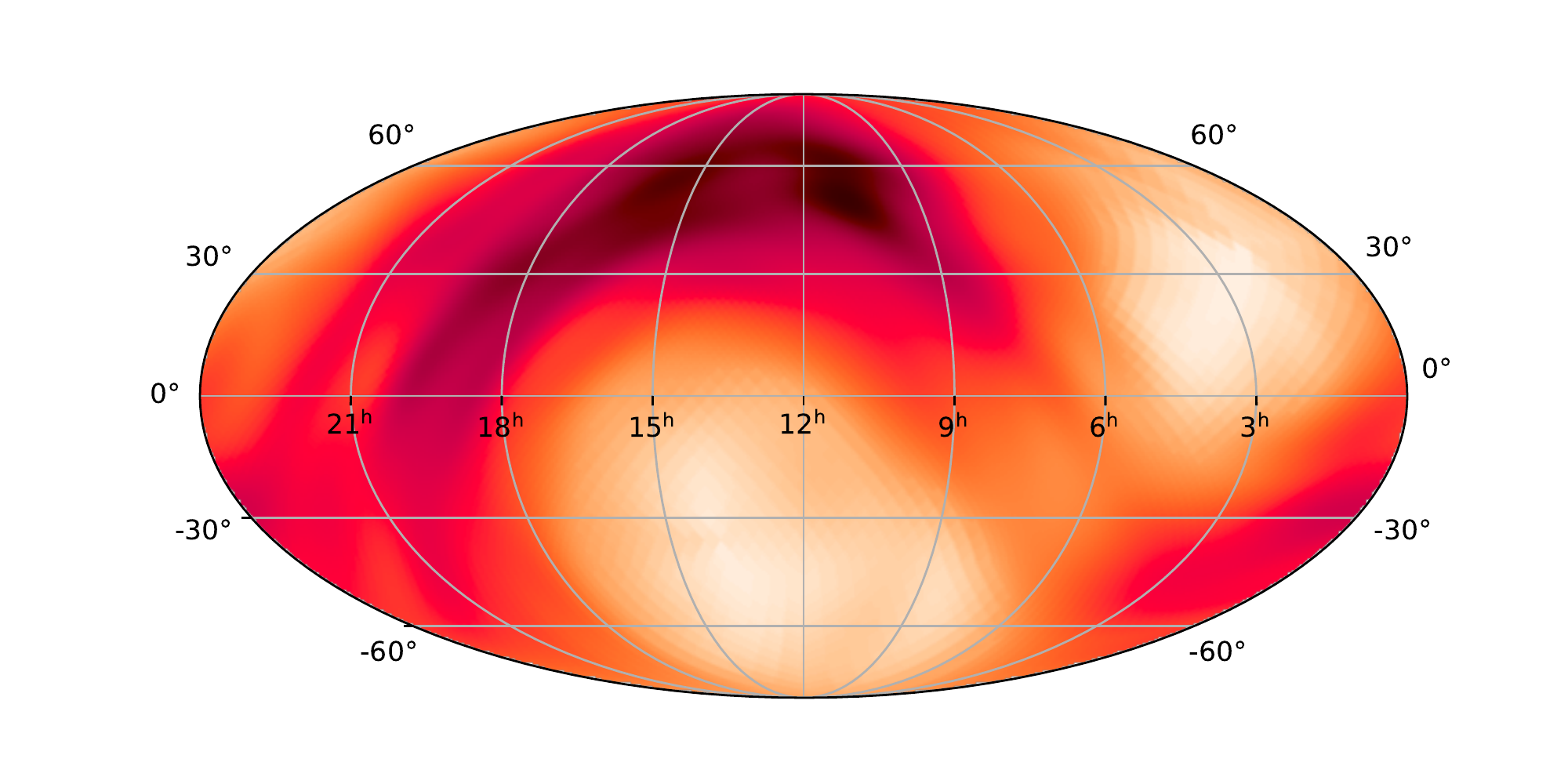}}{GW190706\_222641}
}%
\subfloat{\centering
    \stackunder{\includegraphics[width=0.20\textwidth]{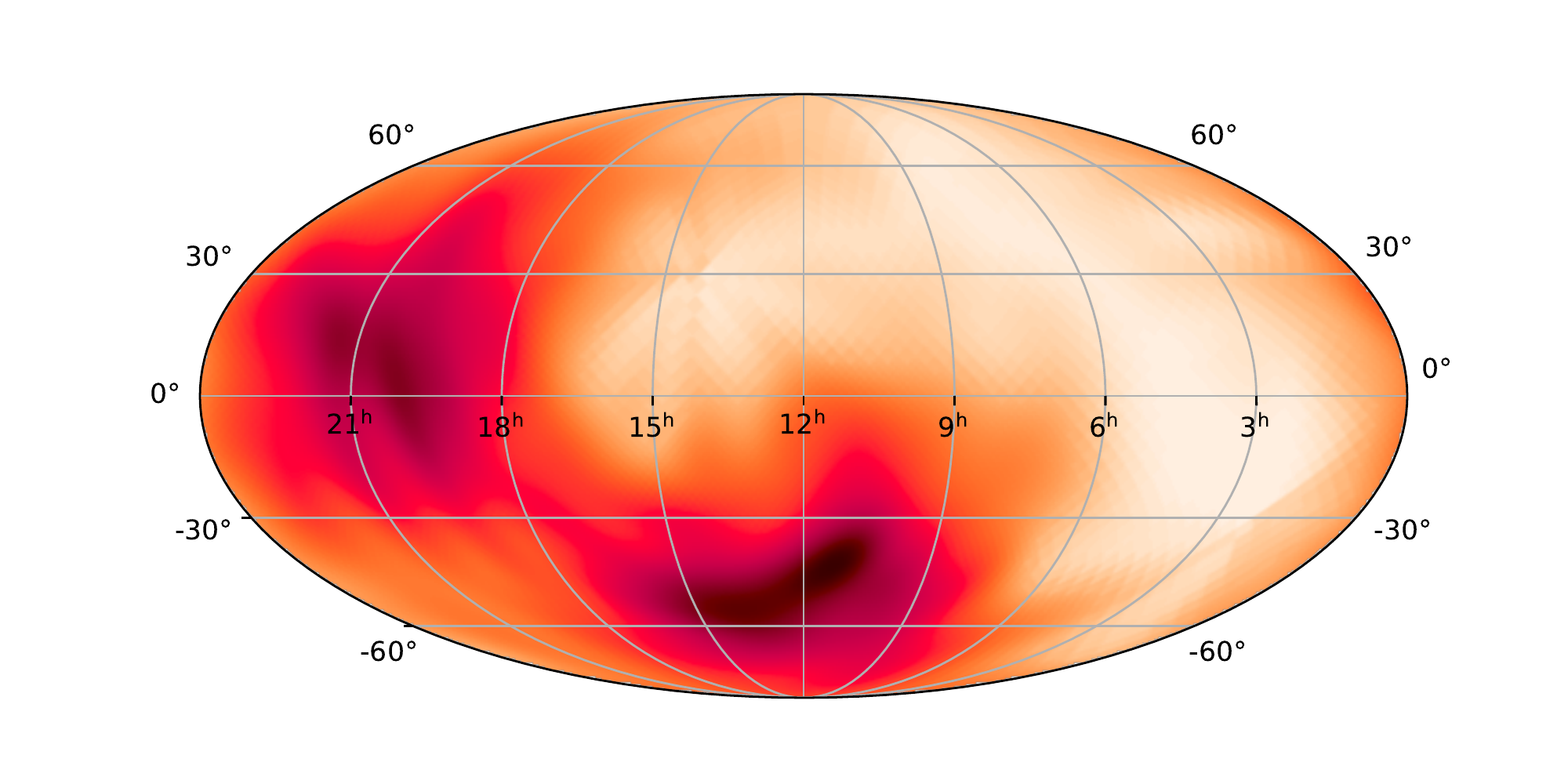}}{GW190707\_093326}
}%
\subfloat{\centering
    \stackunder{\includegraphics[width=0.20\textwidth]{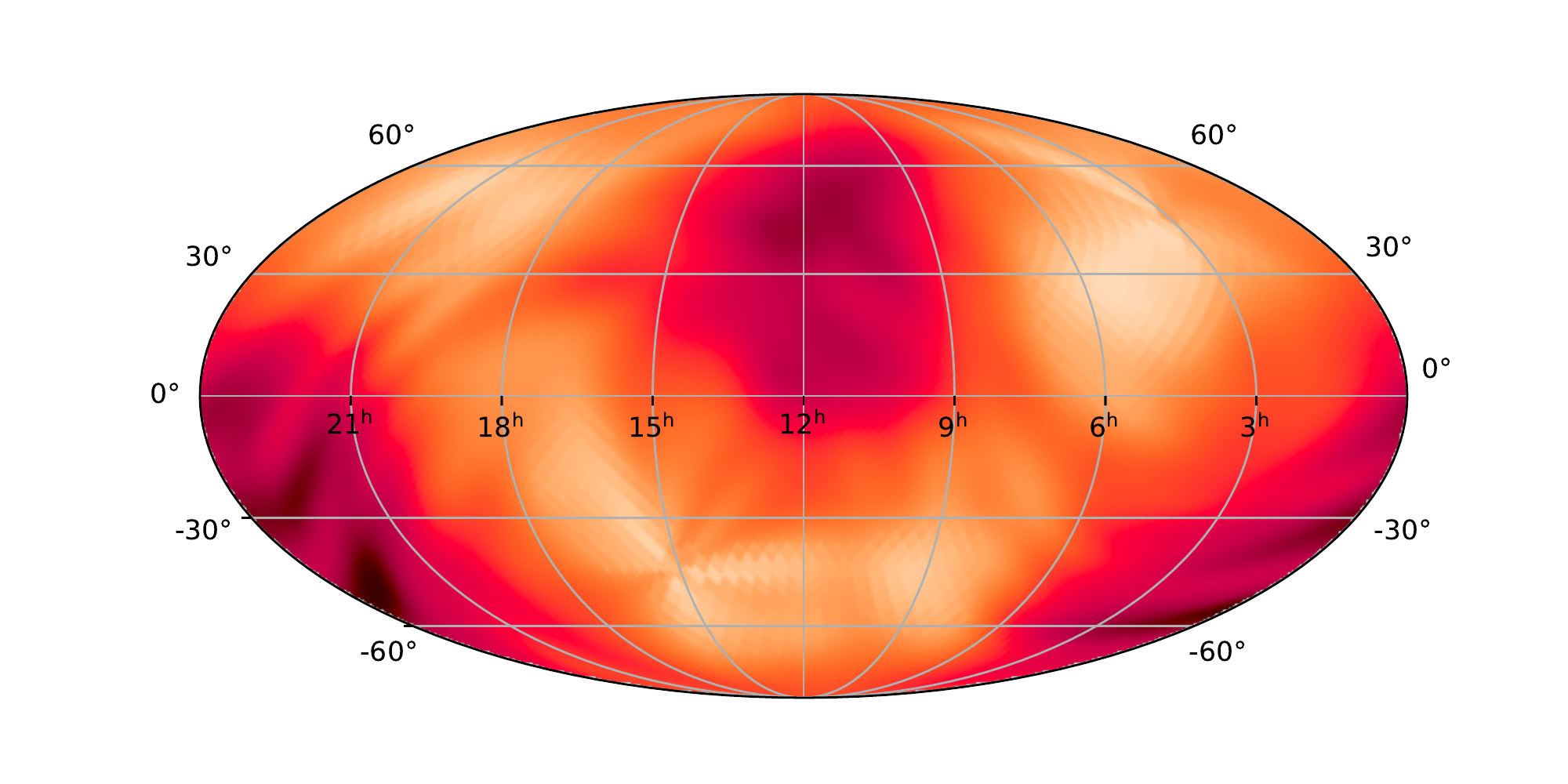}}{GW190708\_232457}
}\\
\subfloat{\centering
    \stackunder{\includegraphics[width=0.20\textwidth]{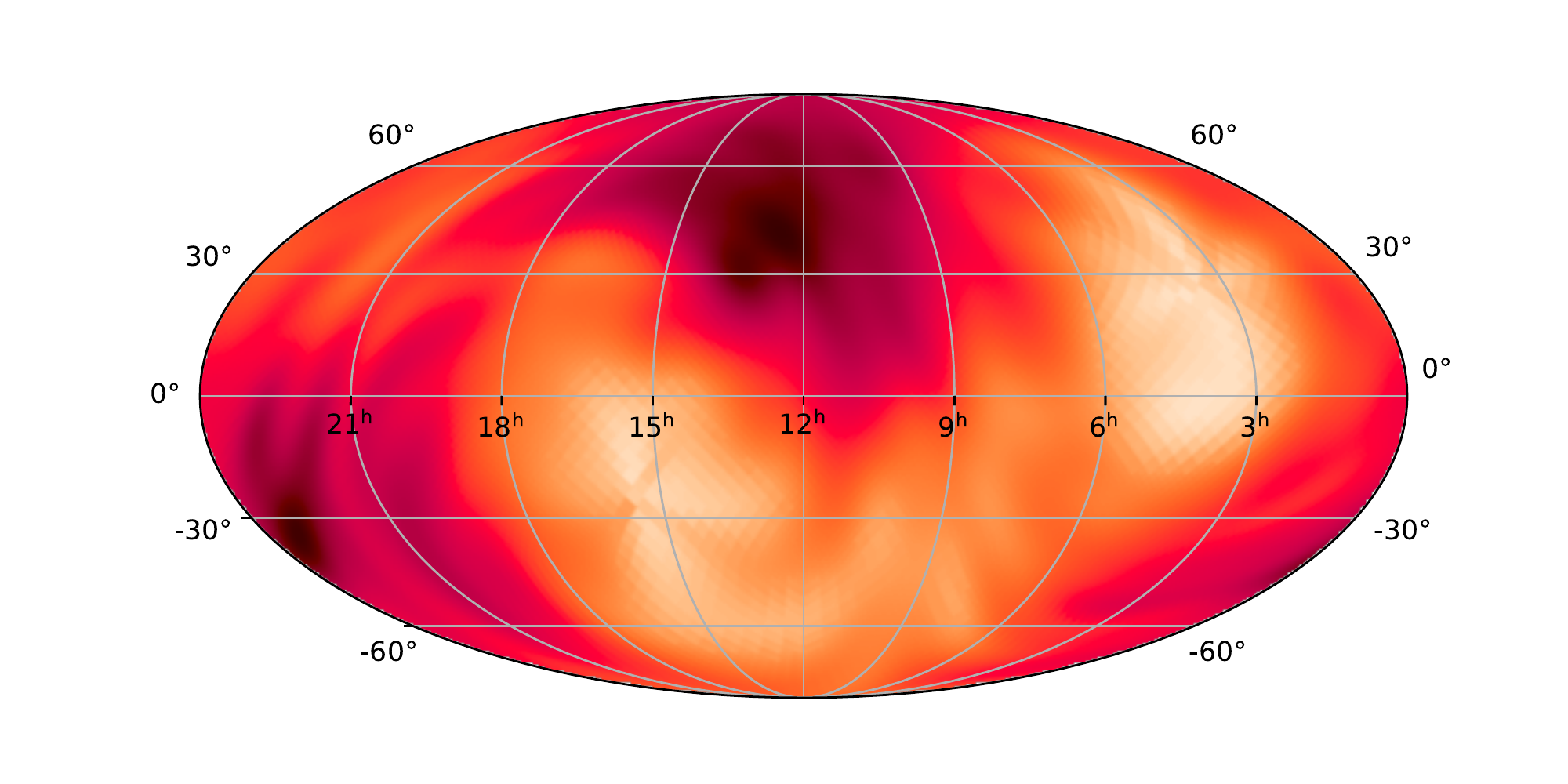}}{GW190719\_215514}
}%
\subfloat{\centering
    \stackunder{\includegraphics[width=0.20\textwidth]{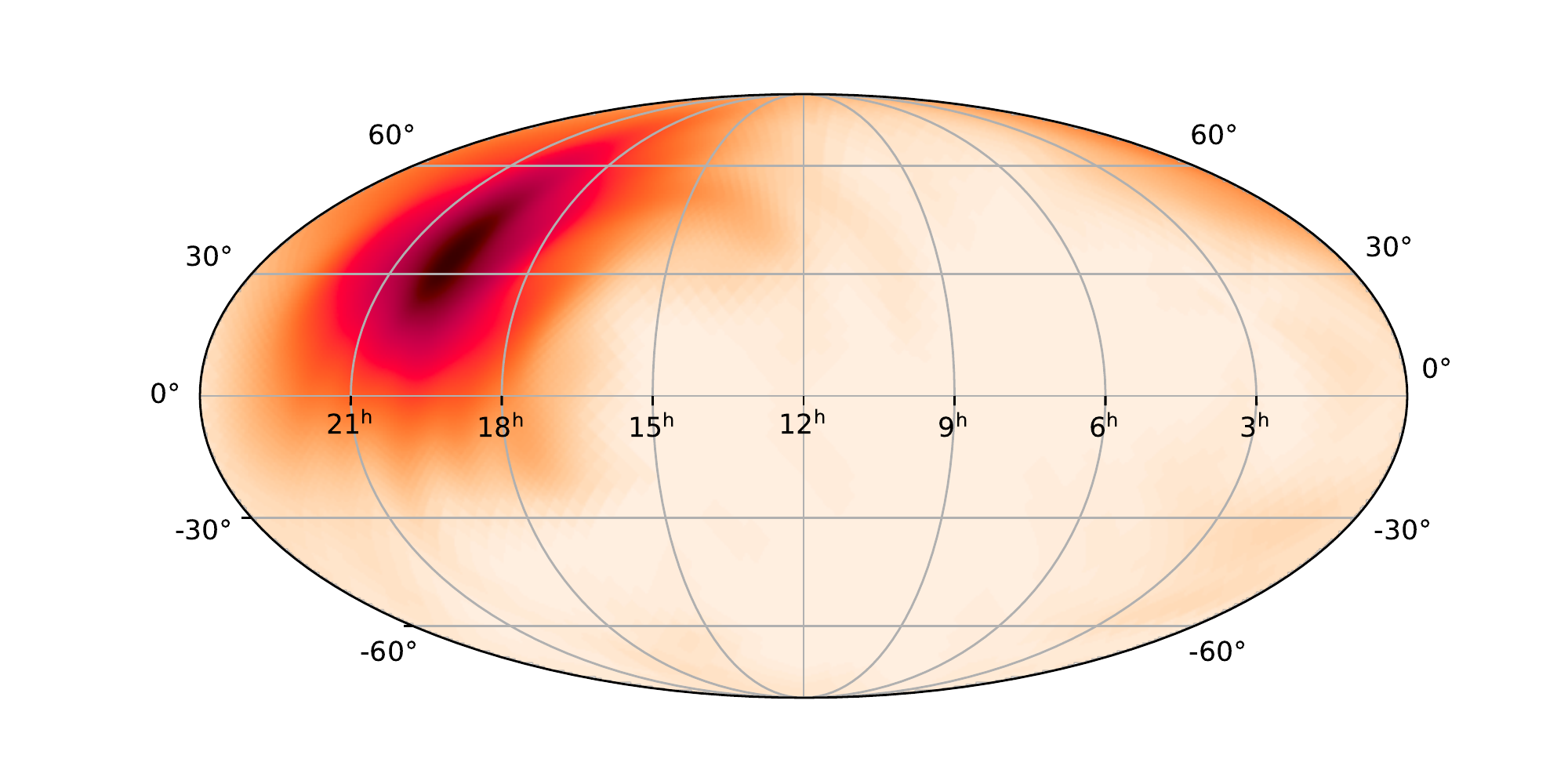}}{GW190720\_000836}
}%
\subfloat{\centering
    \stackunder{\includegraphics[width=0.20\textwidth]{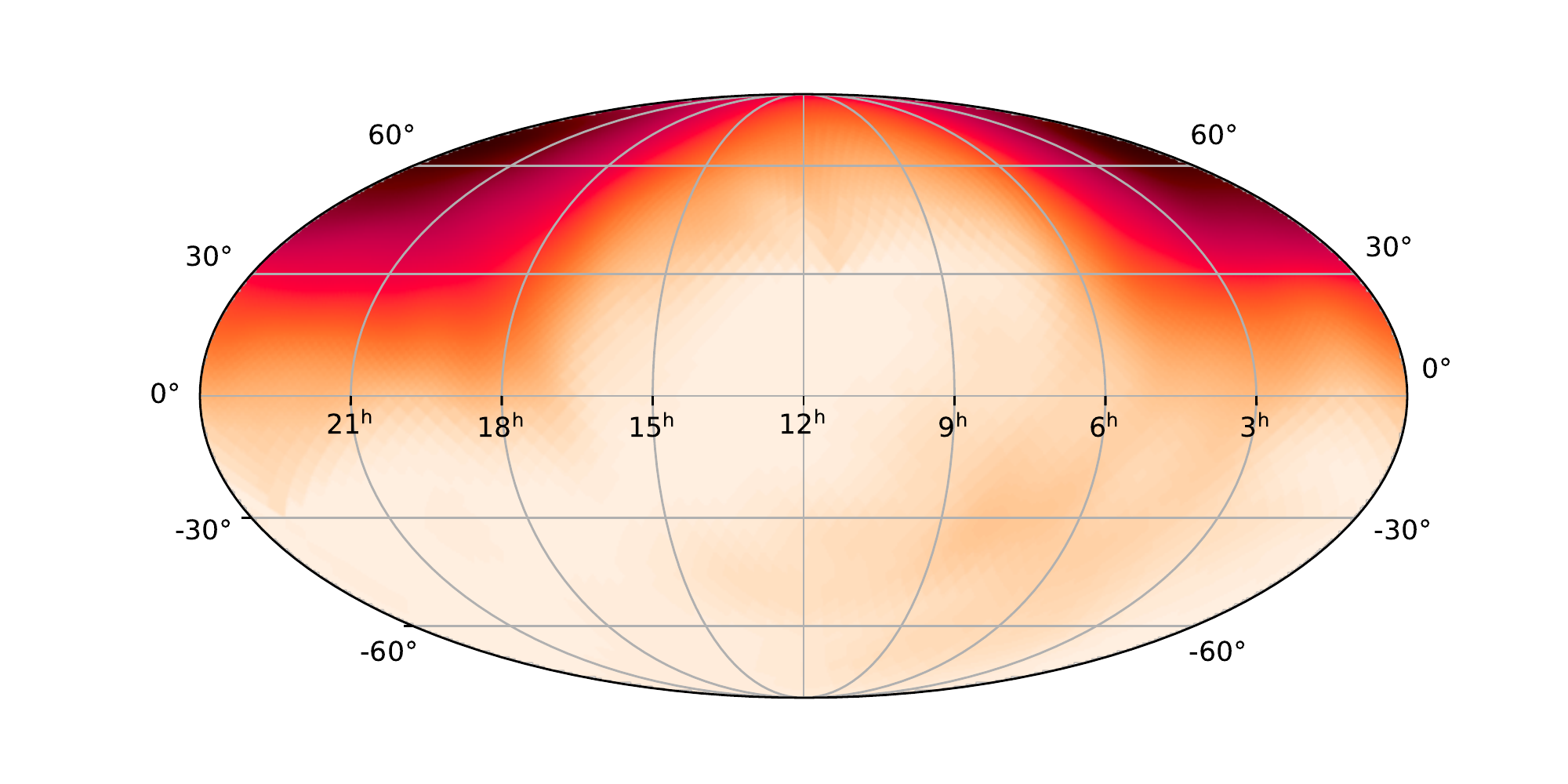}}{GW190727\_060333}
}%
\subfloat{\centering
    \stackunder{\includegraphics[width=0.20\textwidth]{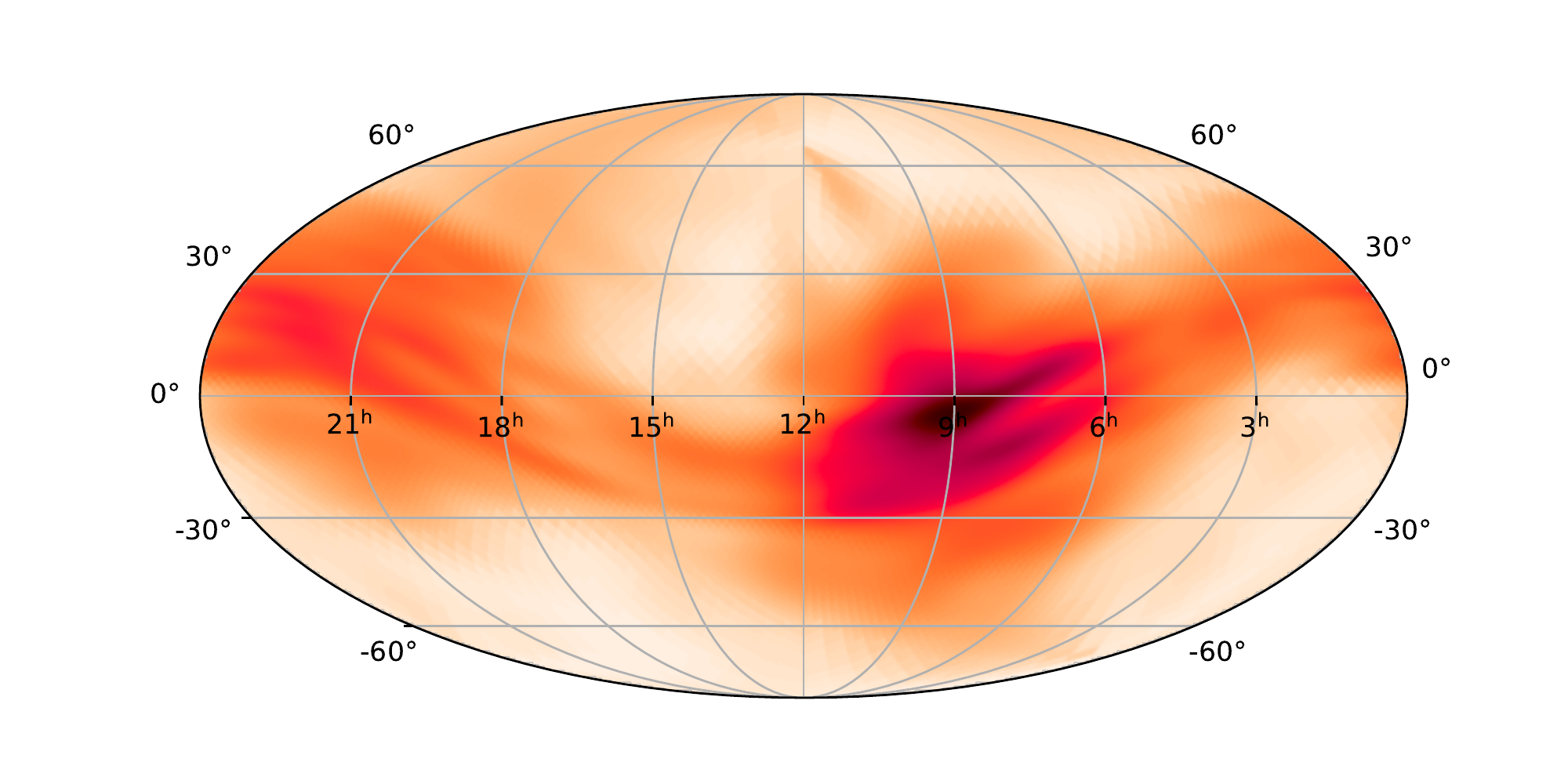}}{GW190728\_064510}
}%
\subfloat{\centering
    \stackunder{\includegraphics[width=0.20\textwidth]{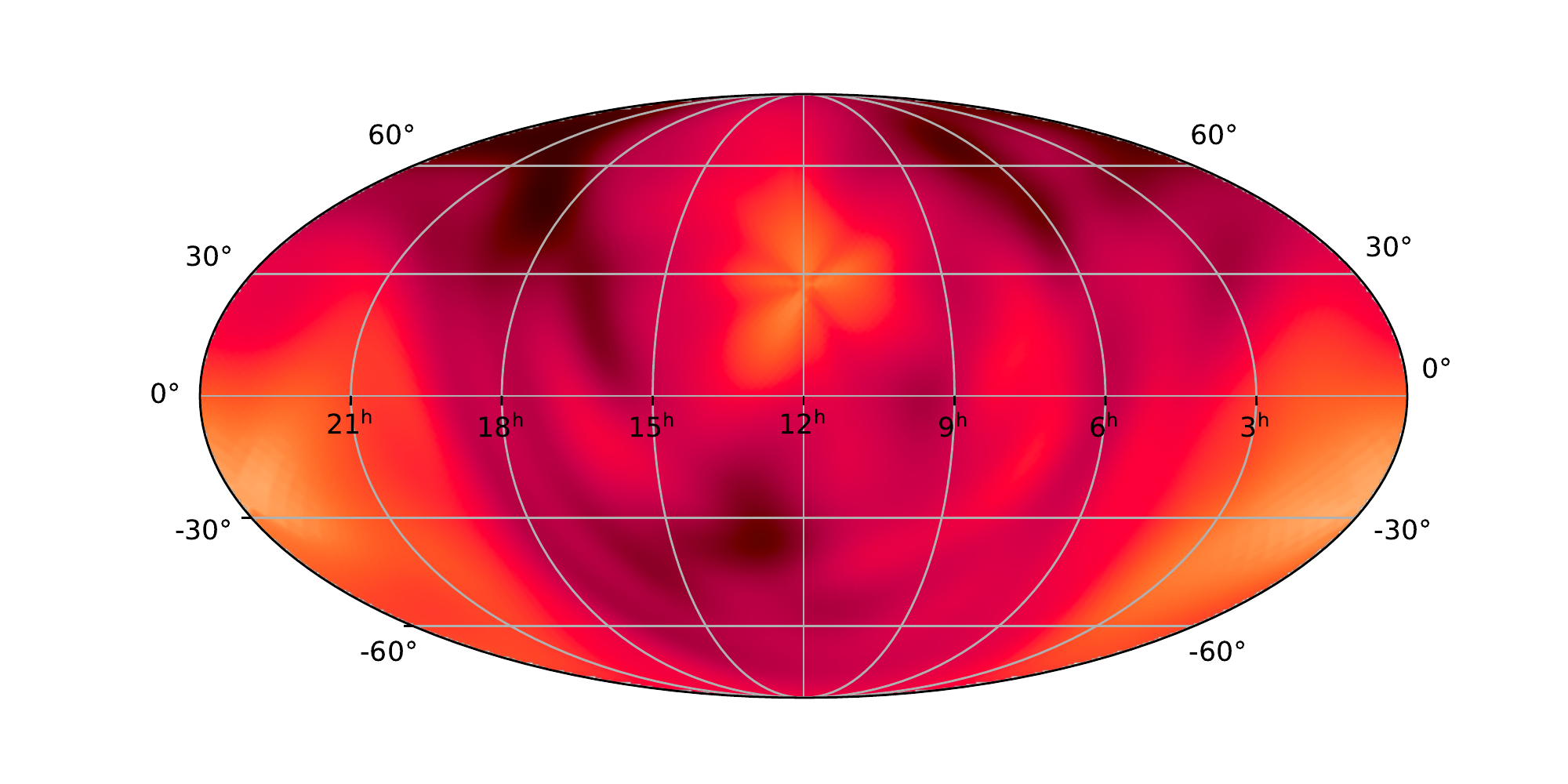}}{GW190731\_140936}
}\\
\subfloat{\centering
    \stackunder{\includegraphics[width=0.20\textwidth]{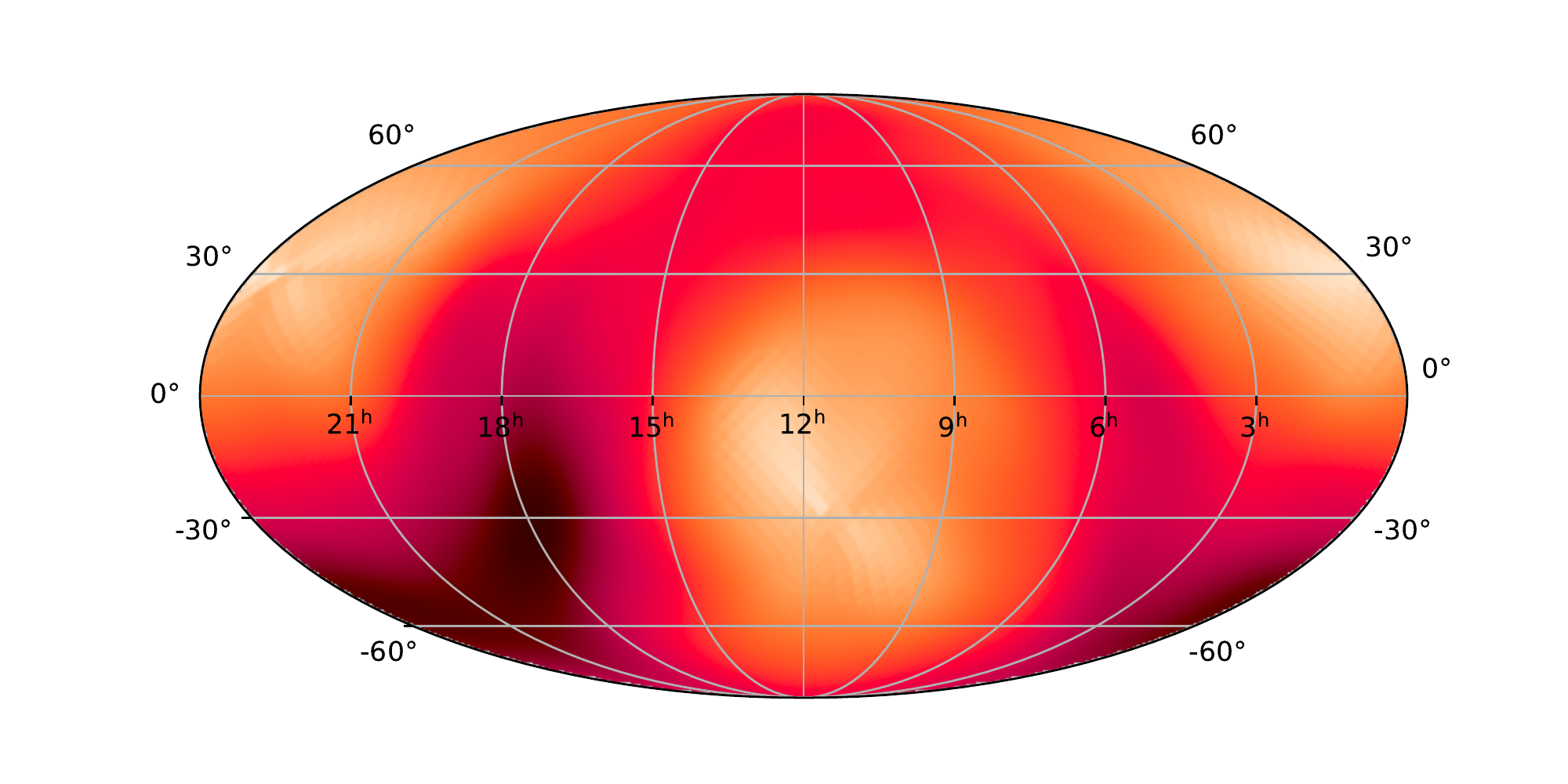}}{GW190803\_022701}
}%
\subfloat{\centering
    \stackunder{\includegraphics[width=0.20\textwidth]{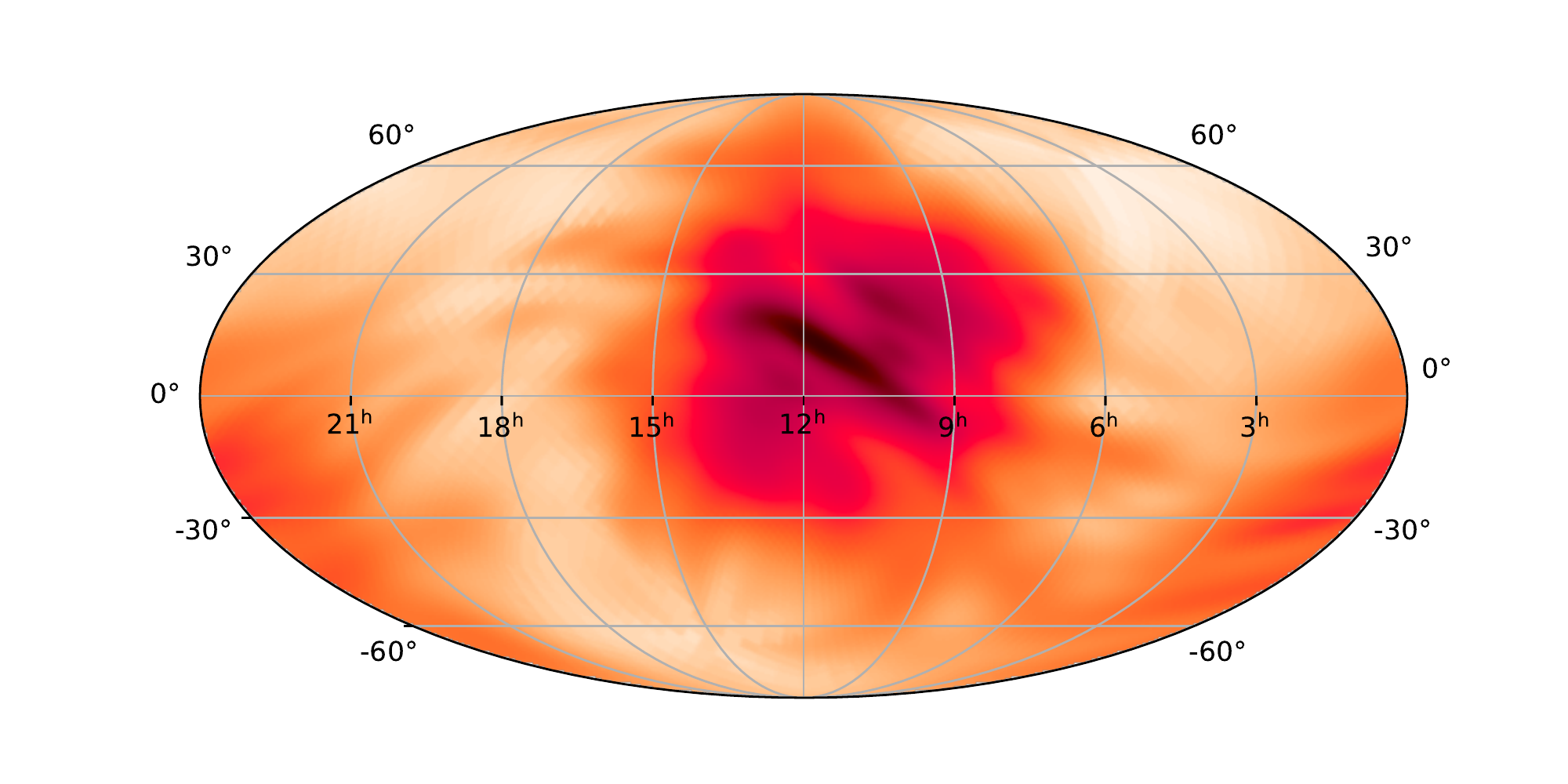}}{GW190805\_211137}
}%
\subfloat{\centering
    \stackunder{\includegraphics[width=0.20\textwidth]{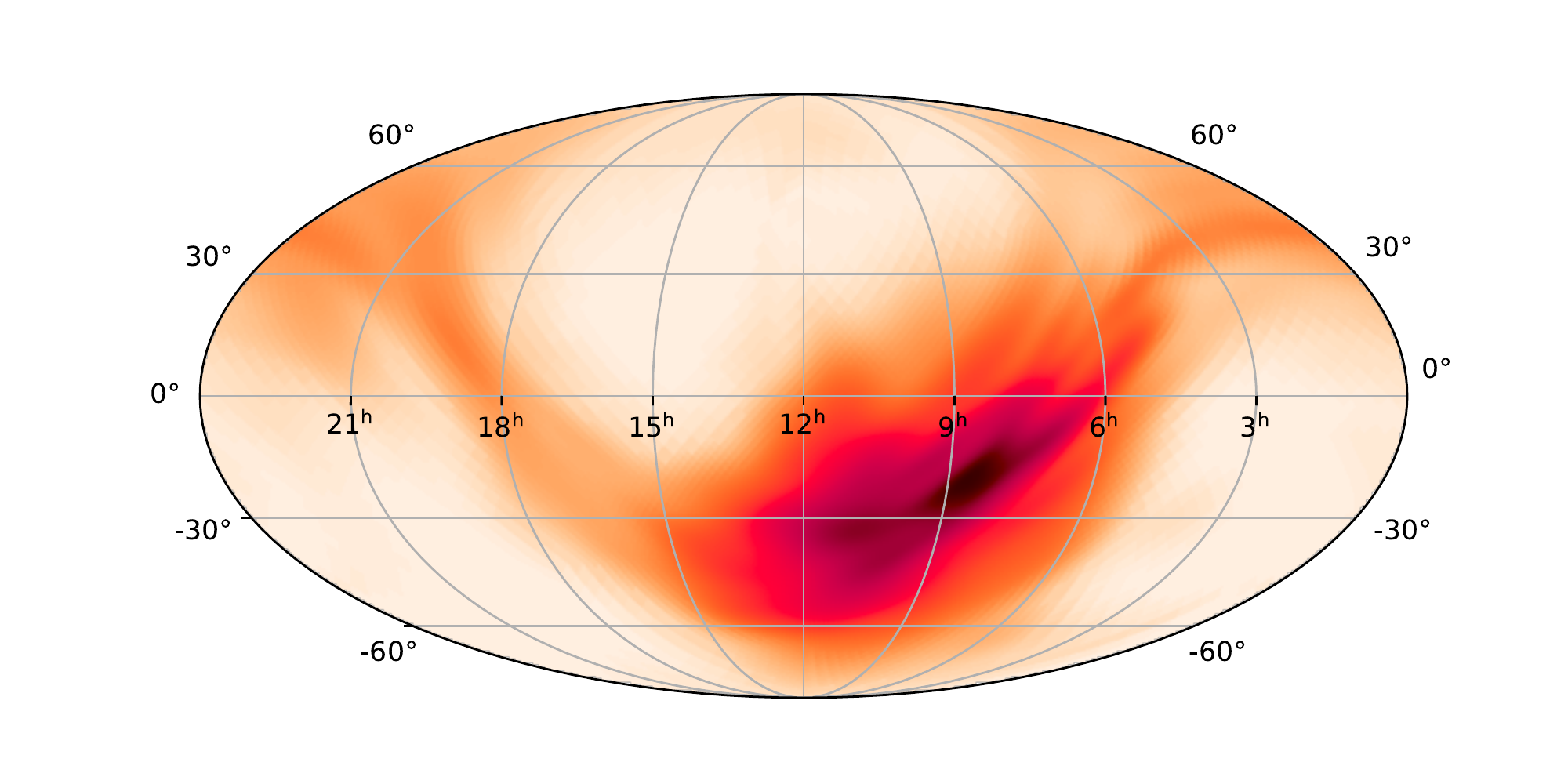}}{GW190828\_063405}
}%
\subfloat{\centering
    \stackunder{\includegraphics[width=0.20\textwidth]{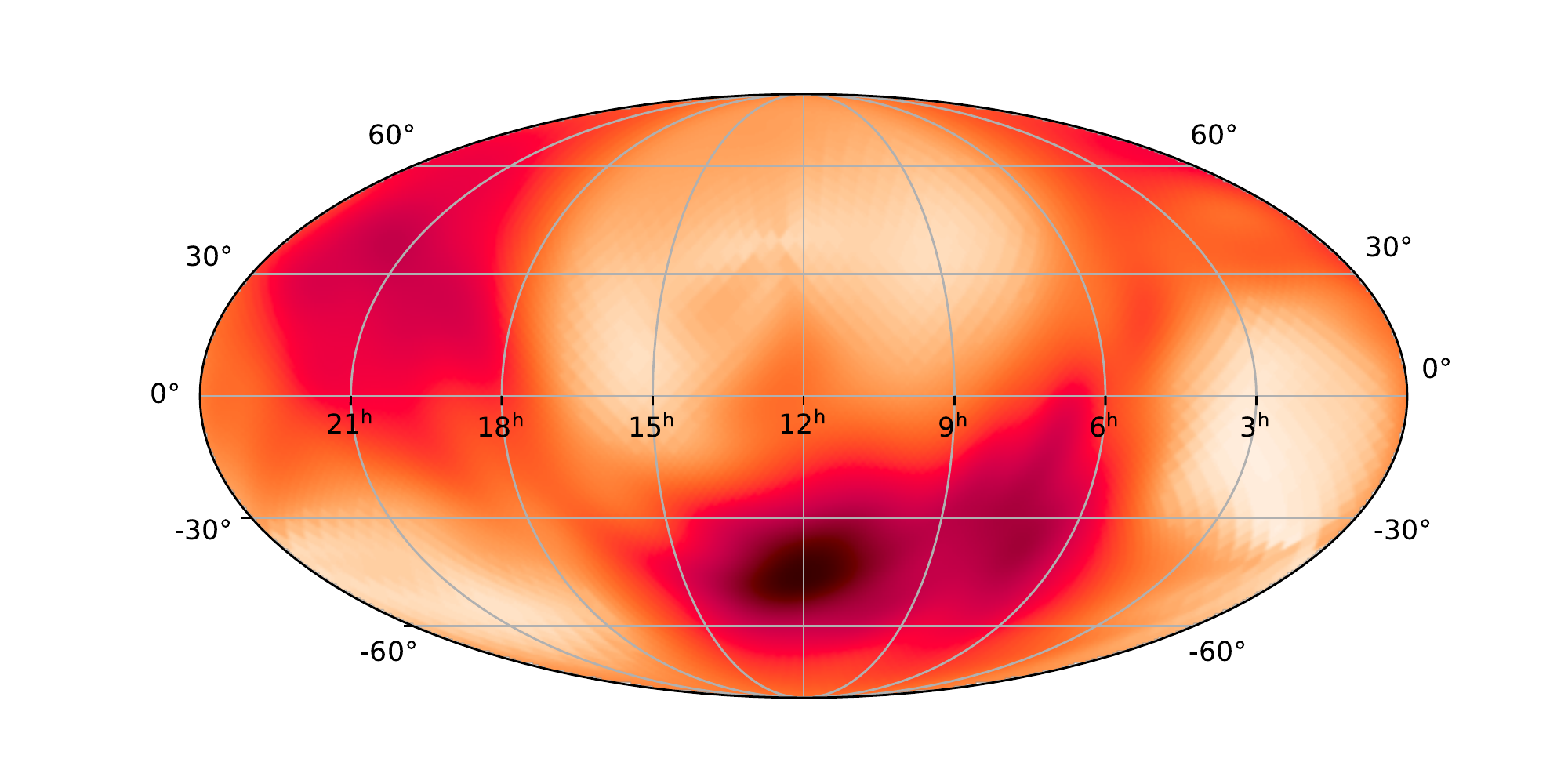}}{GW190828\_065509}
}%
\subfloat{\centering
    \stackunder{\includegraphics[width=0.20\textwidth]{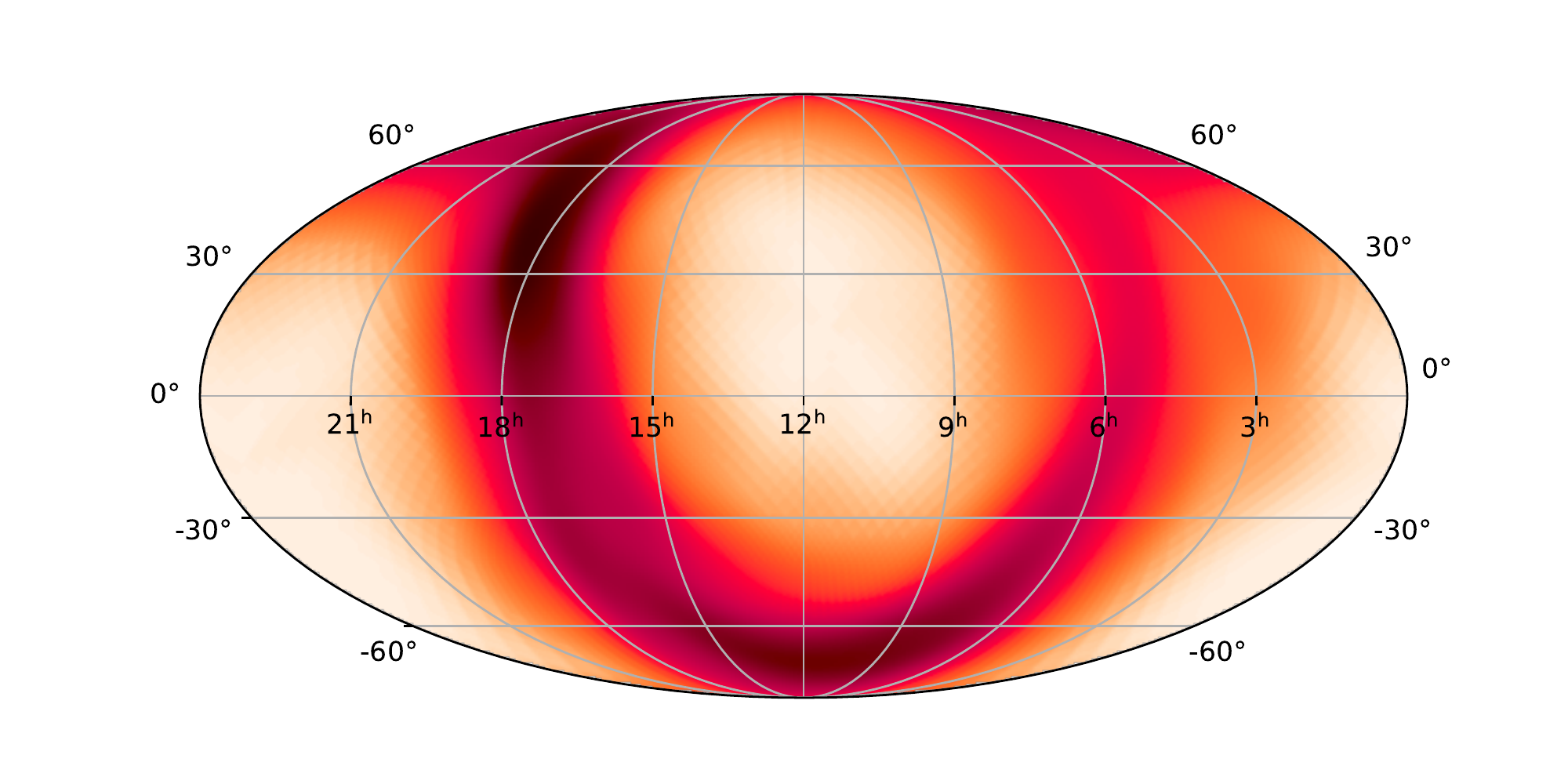}}{GW190910\_112807}
}\\
\subfloat{\centering
    \stackunder{\includegraphics[width=0.20\textwidth]{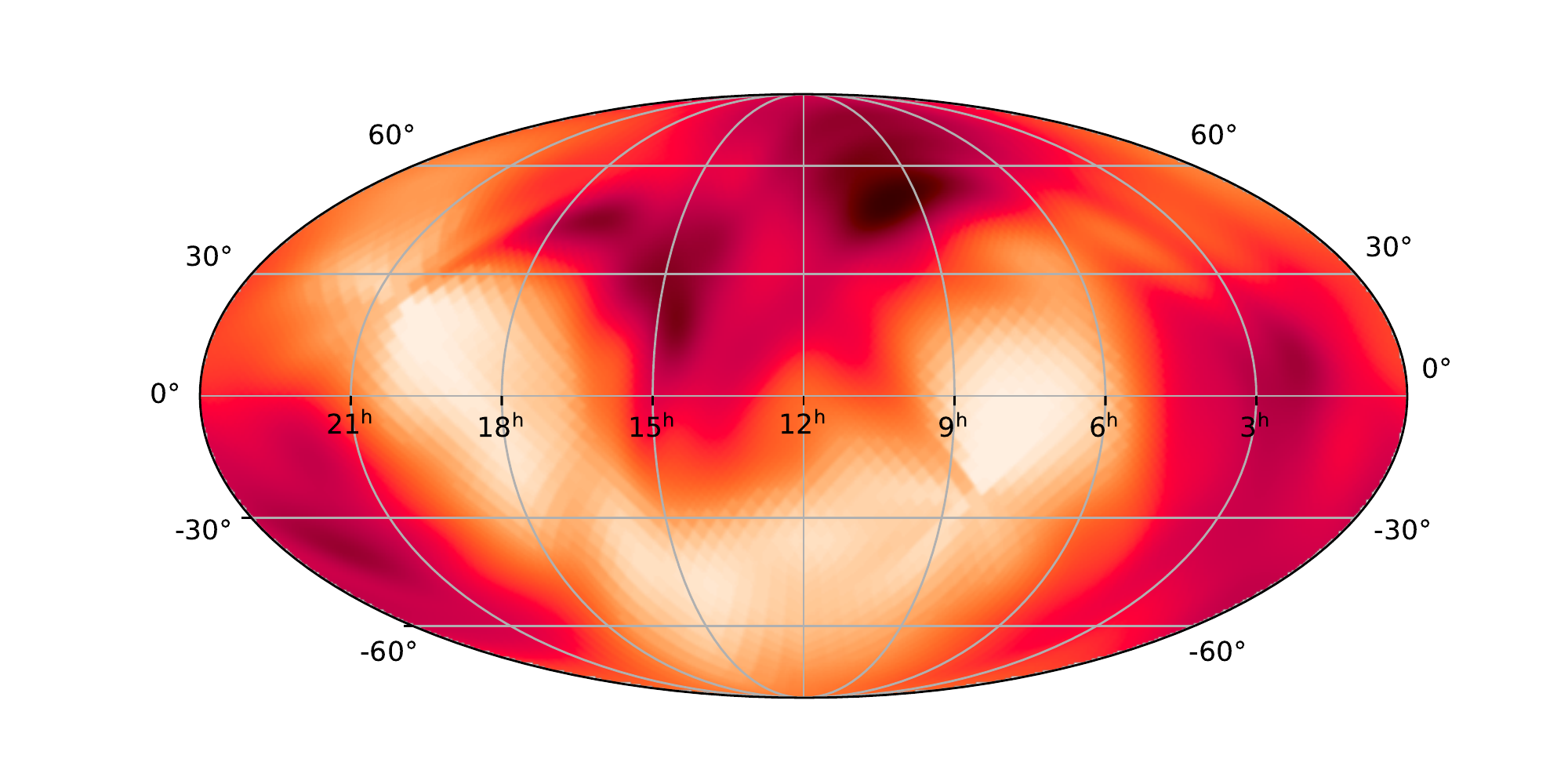}}{GW190915\_235702}
}%
\subfloat{\centering
    \stackunder{\includegraphics[width=0.20\textwidth]{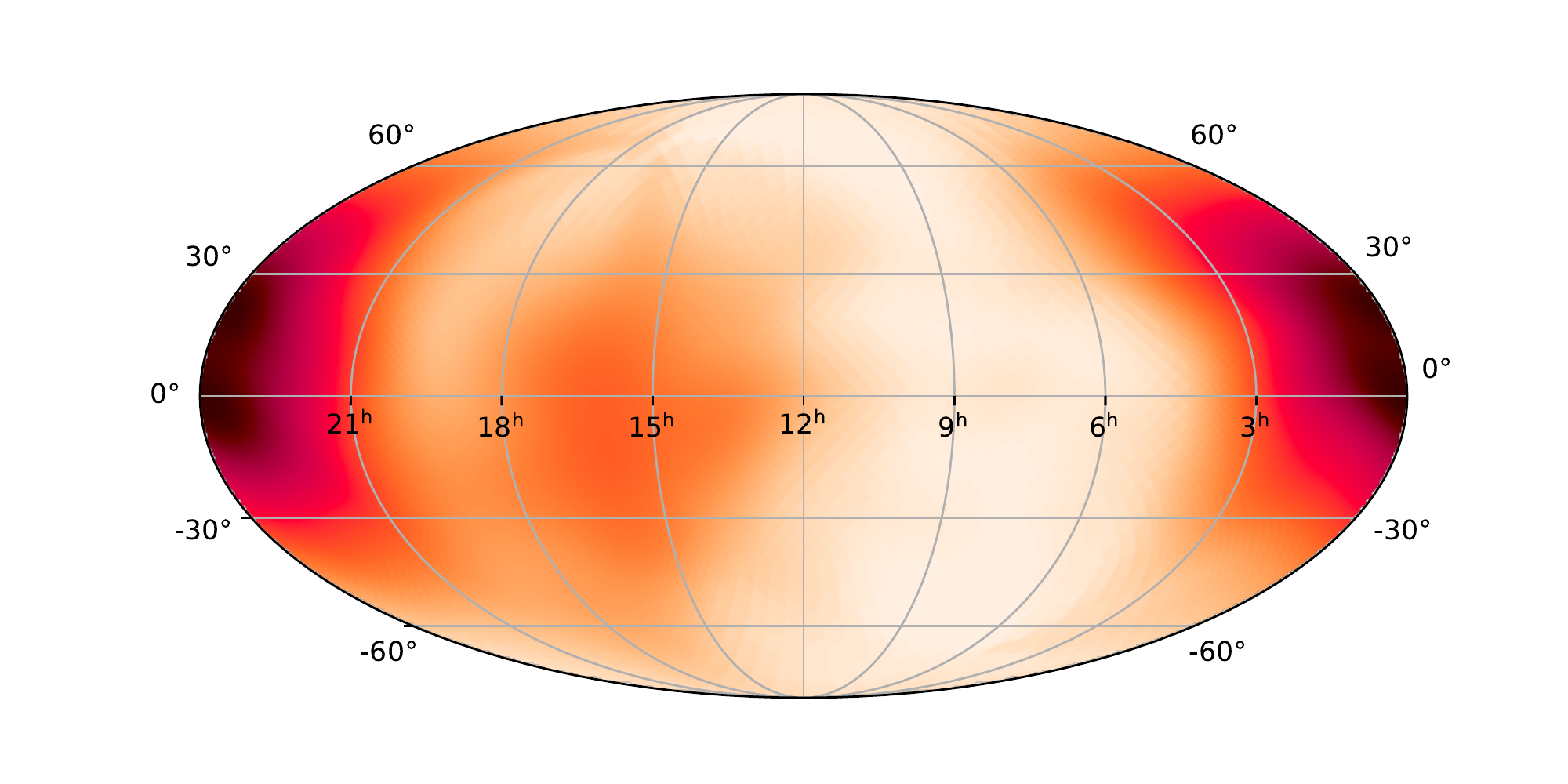}}{GW190925\_232845}
}%
\subfloat{\centering
    \stackunder{\includegraphics[width=0.20\textwidth]{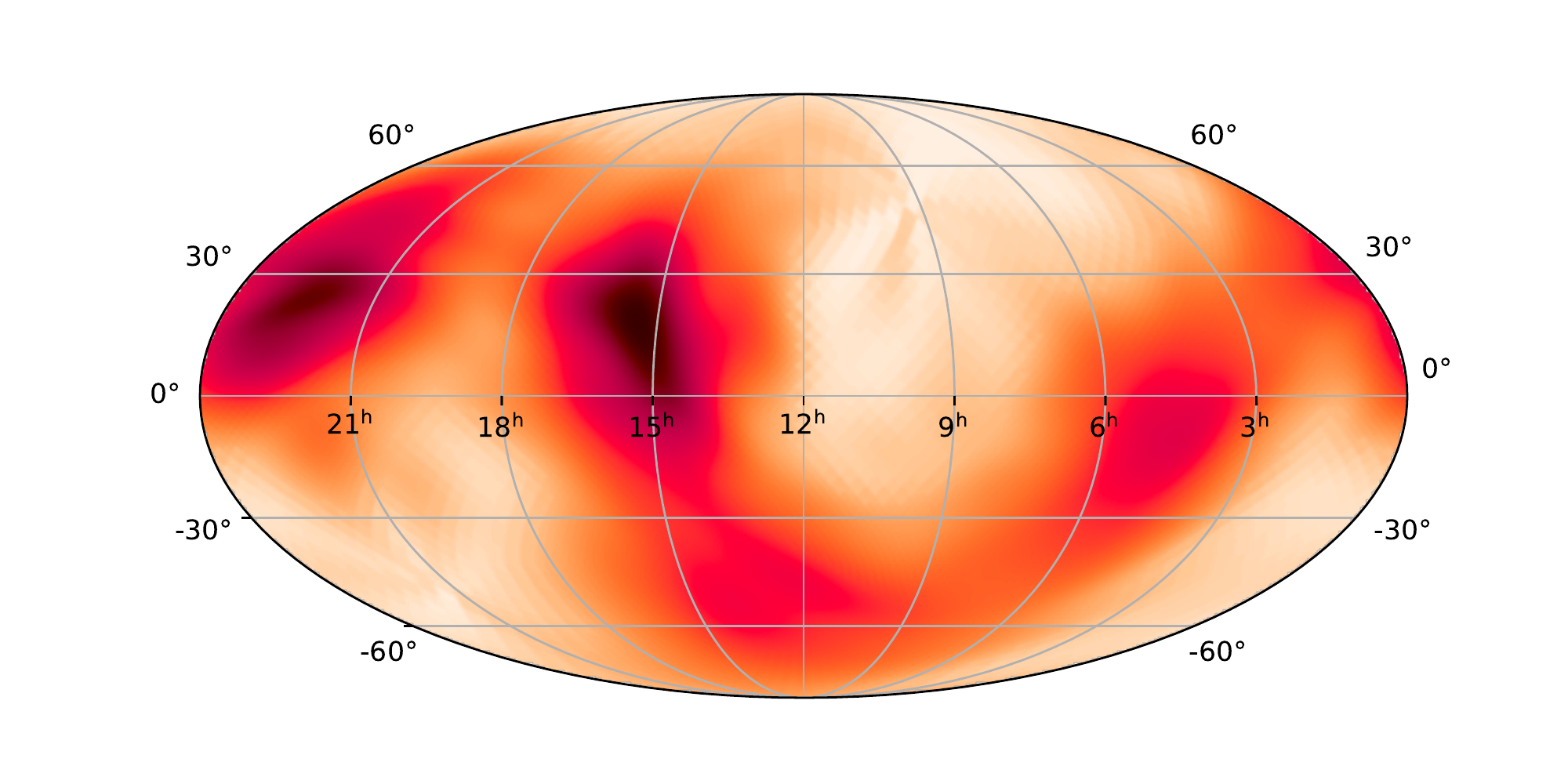}}{GW190929\_012149}
}%
\subfloat{\centering
    \stackunder{\includegraphics[width=0.20\textwidth]{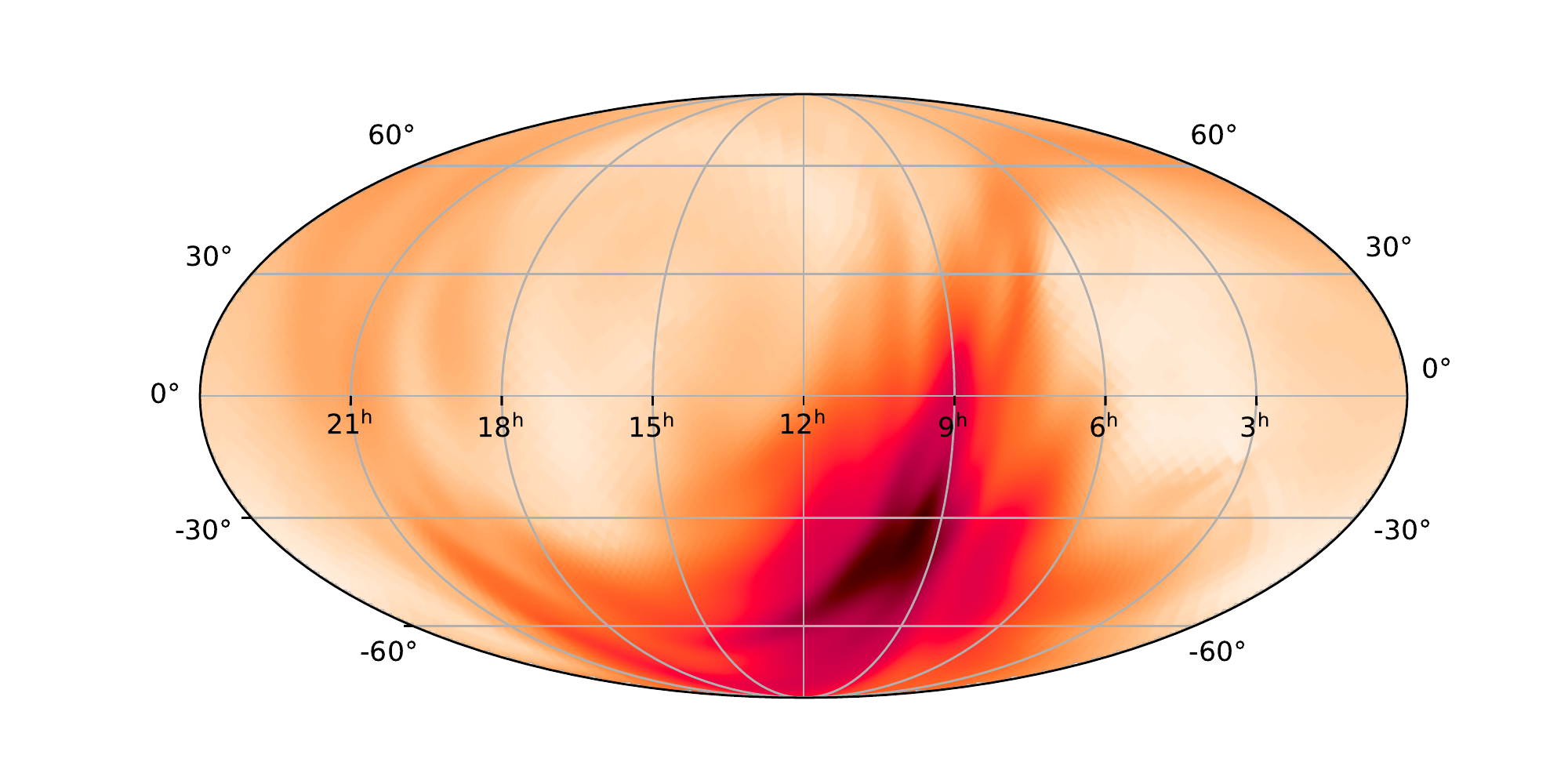}}{GW190930\_133541}
}%
\subfloat{\centering
    \stackunder{\includegraphics[width=0.20\textwidth]{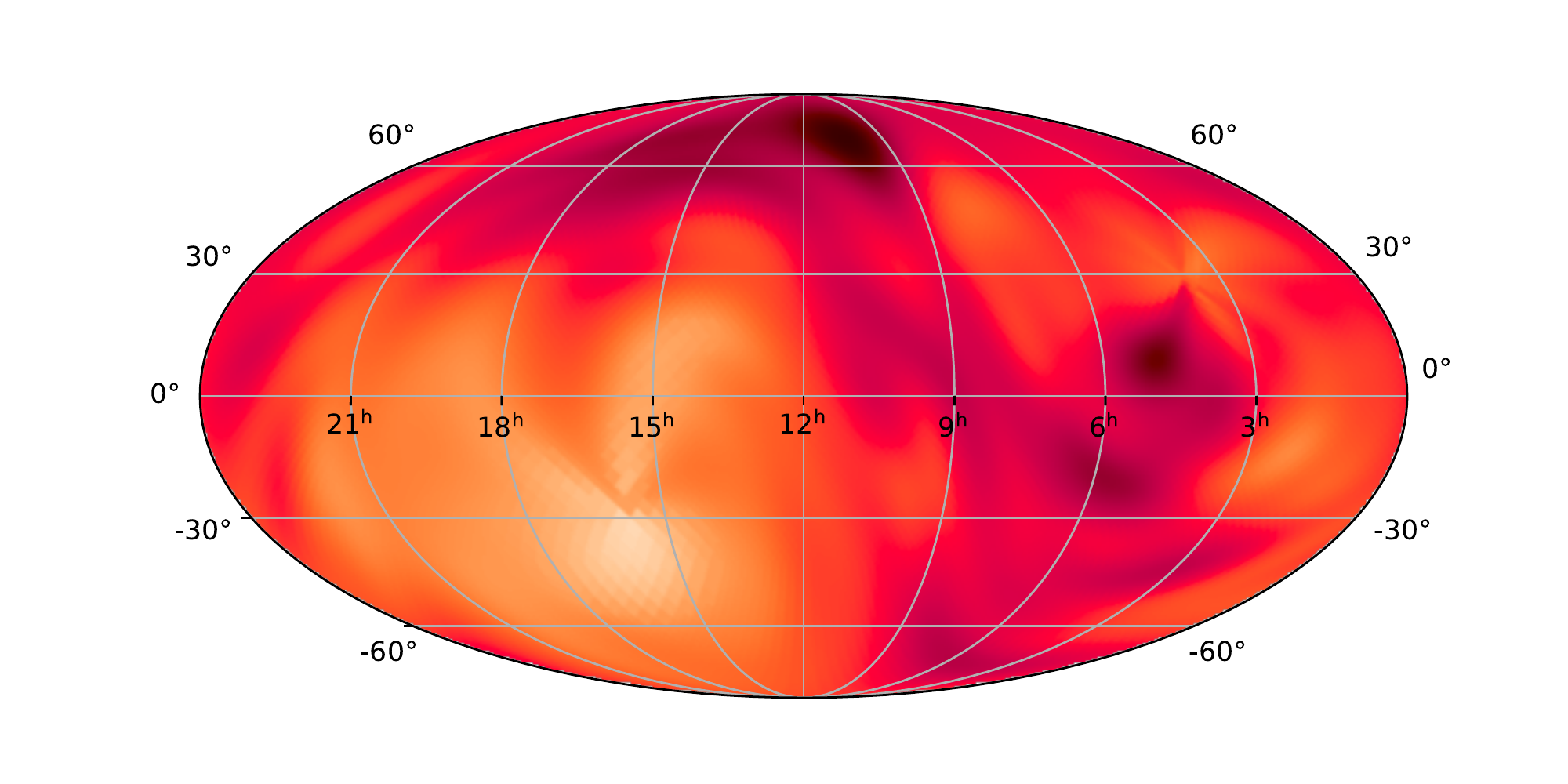}}{GW191103\_012549}
}\\
\subfloat{\centering
    \stackunder{\includegraphics[width=0.20\textwidth]{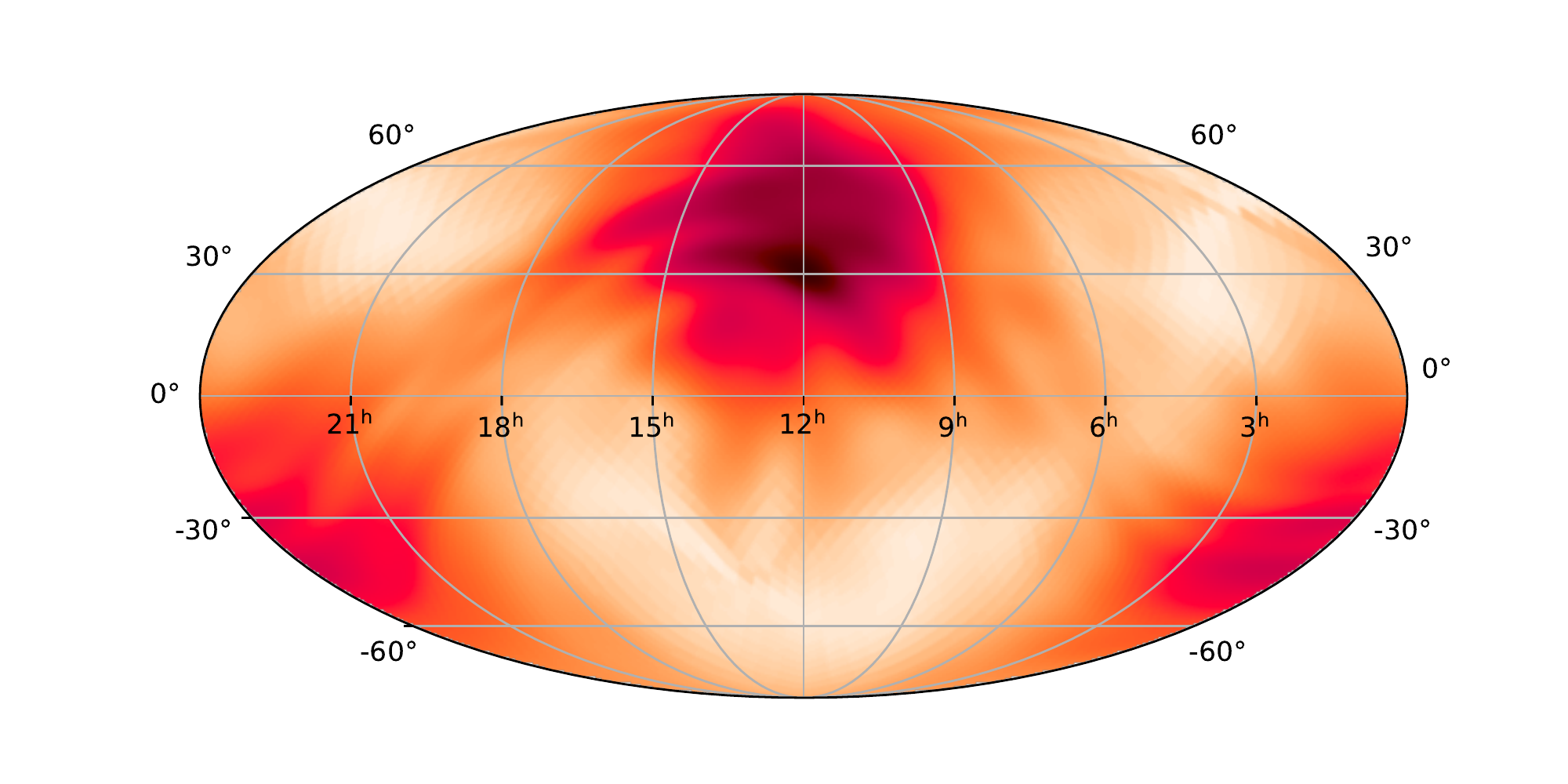}}{GW191105\_143521}
}%
\subfloat{\centering
    \stackunder{\includegraphics[width=0.20\textwidth]{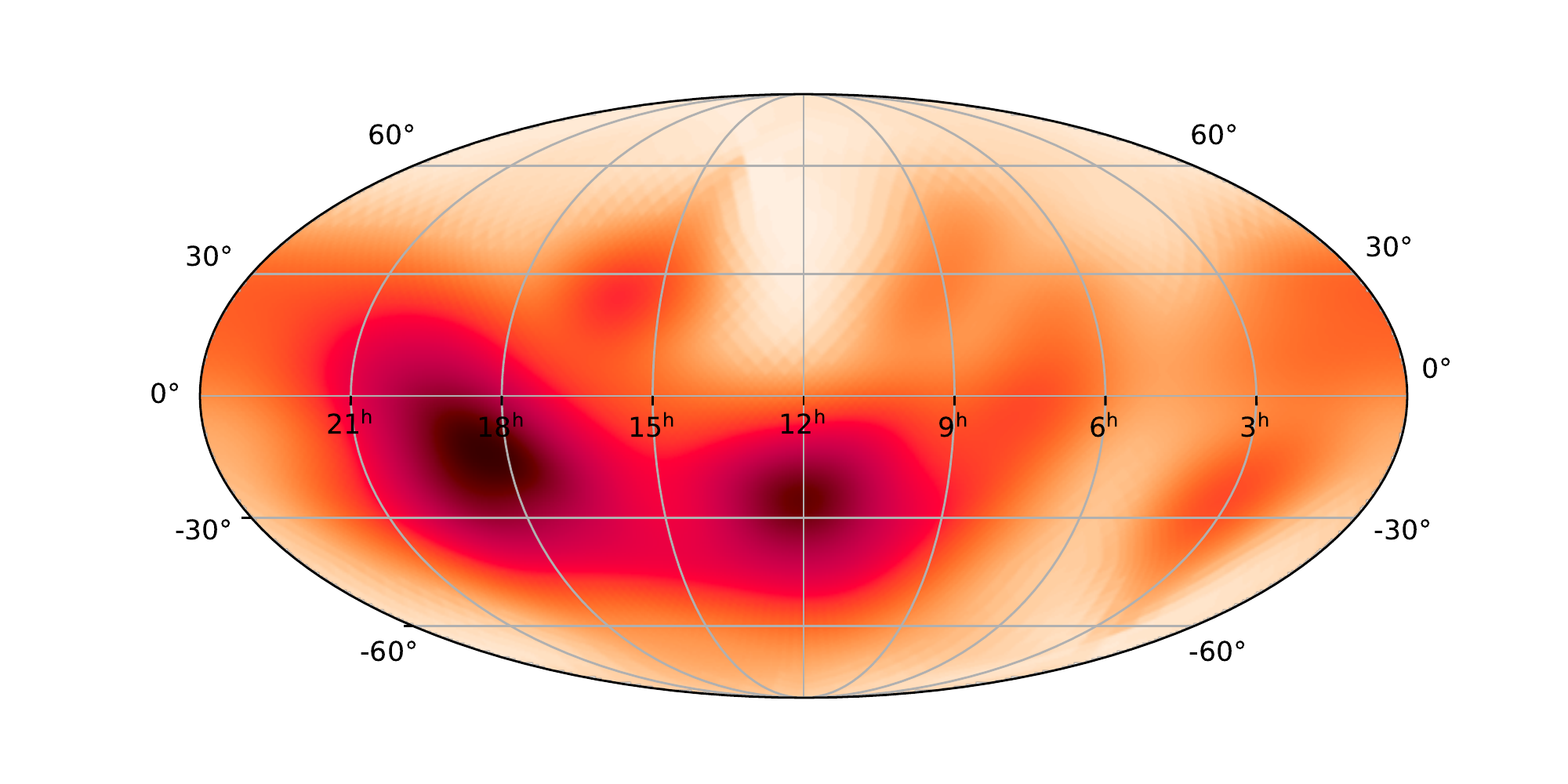}}{GW191109\_010717}
}%
\subfloat{\centering
    \stackunder{\includegraphics[width=0.20\textwidth]{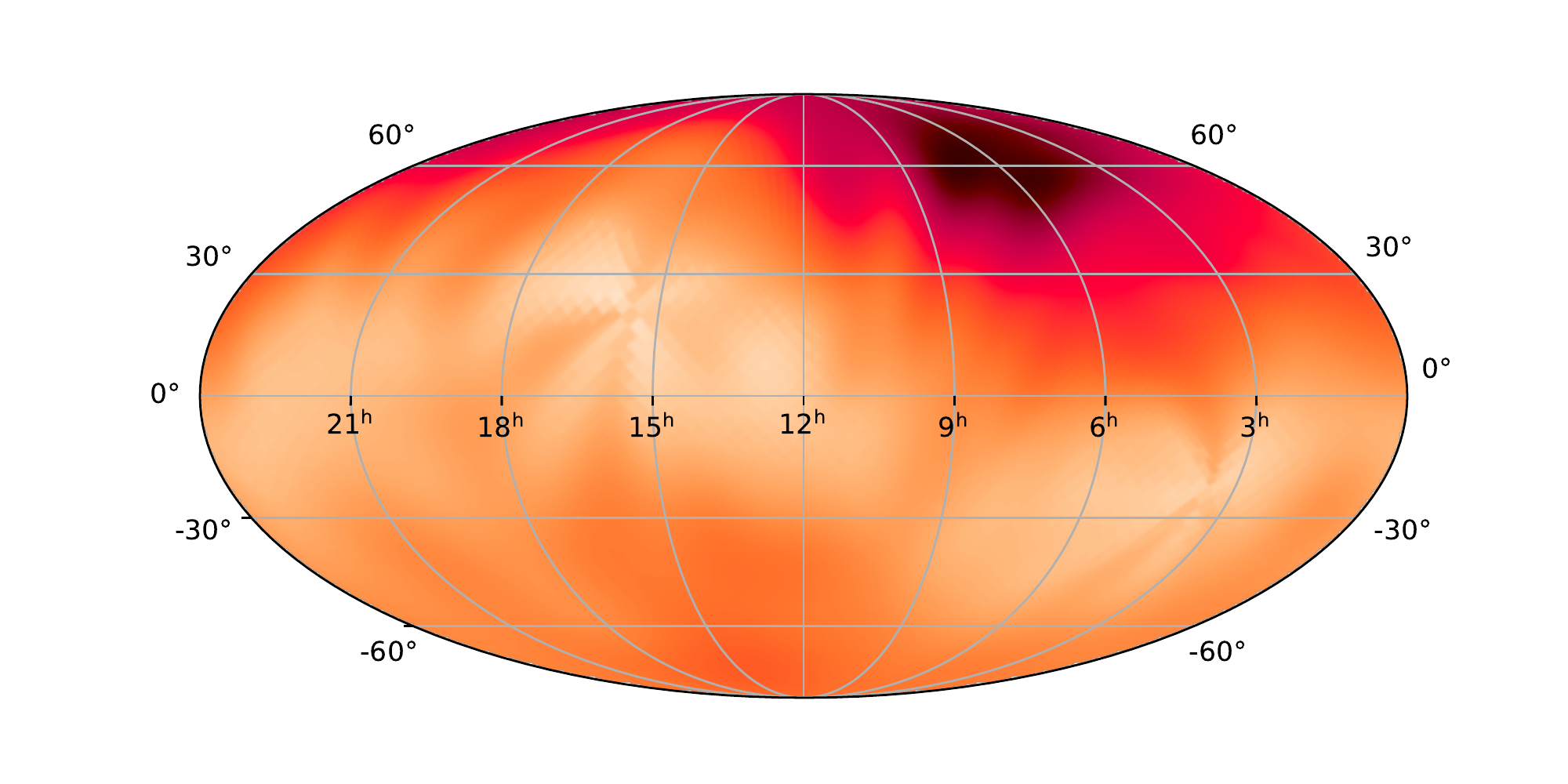}}{GW191127\_050227}
}%
\subfloat{\centering
    \stackunder{\includegraphics[width=0.20\textwidth]{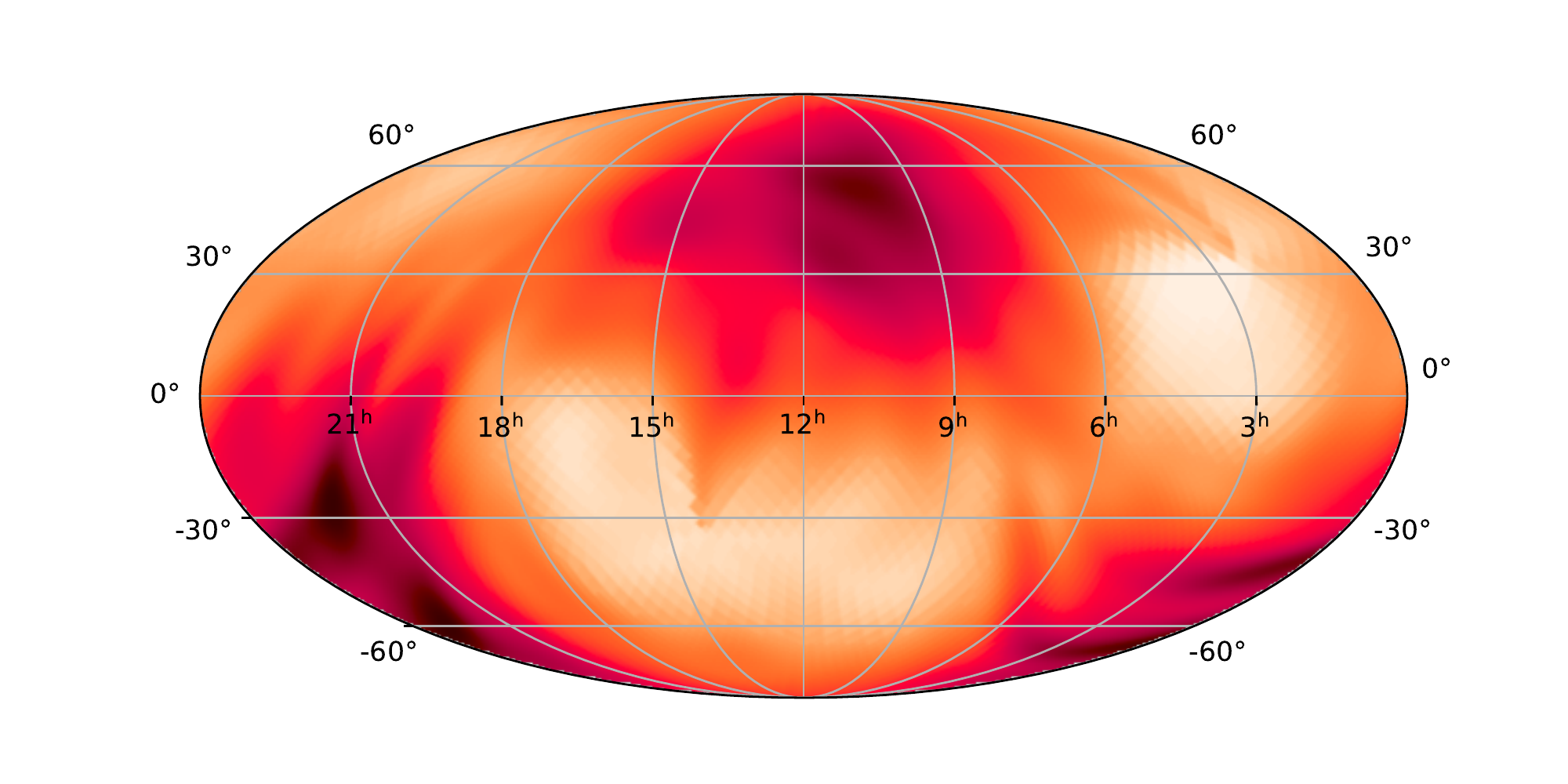}}{GW191129\_134029}
}%
\subfloat{\centering
    \stackunder{\includegraphics[width=0.20\textwidth]{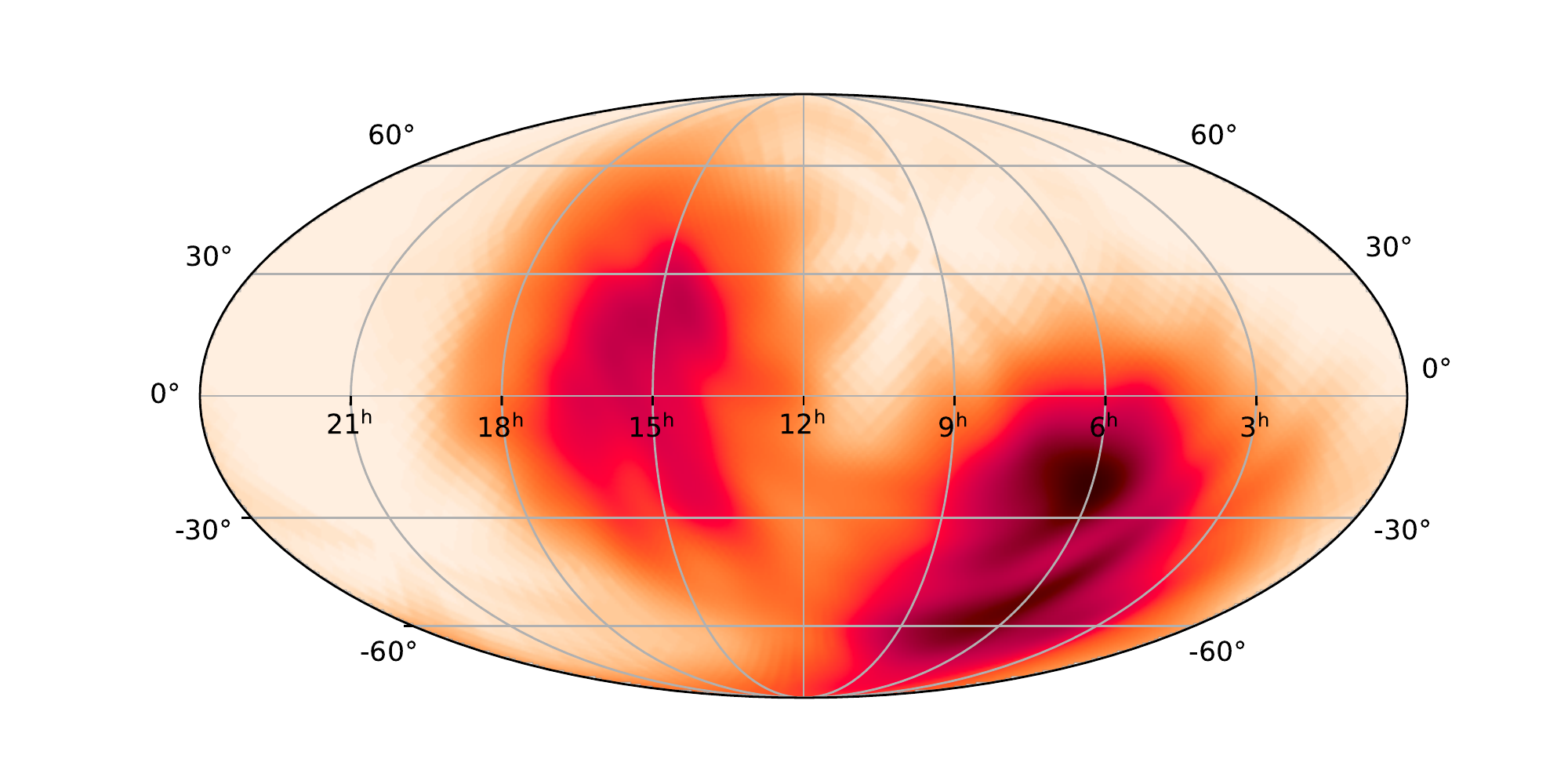}}{GW191204\_171526}
}\\
\subfloat{\centering
    \stackunder{\includegraphics[width=0.20\textwidth]{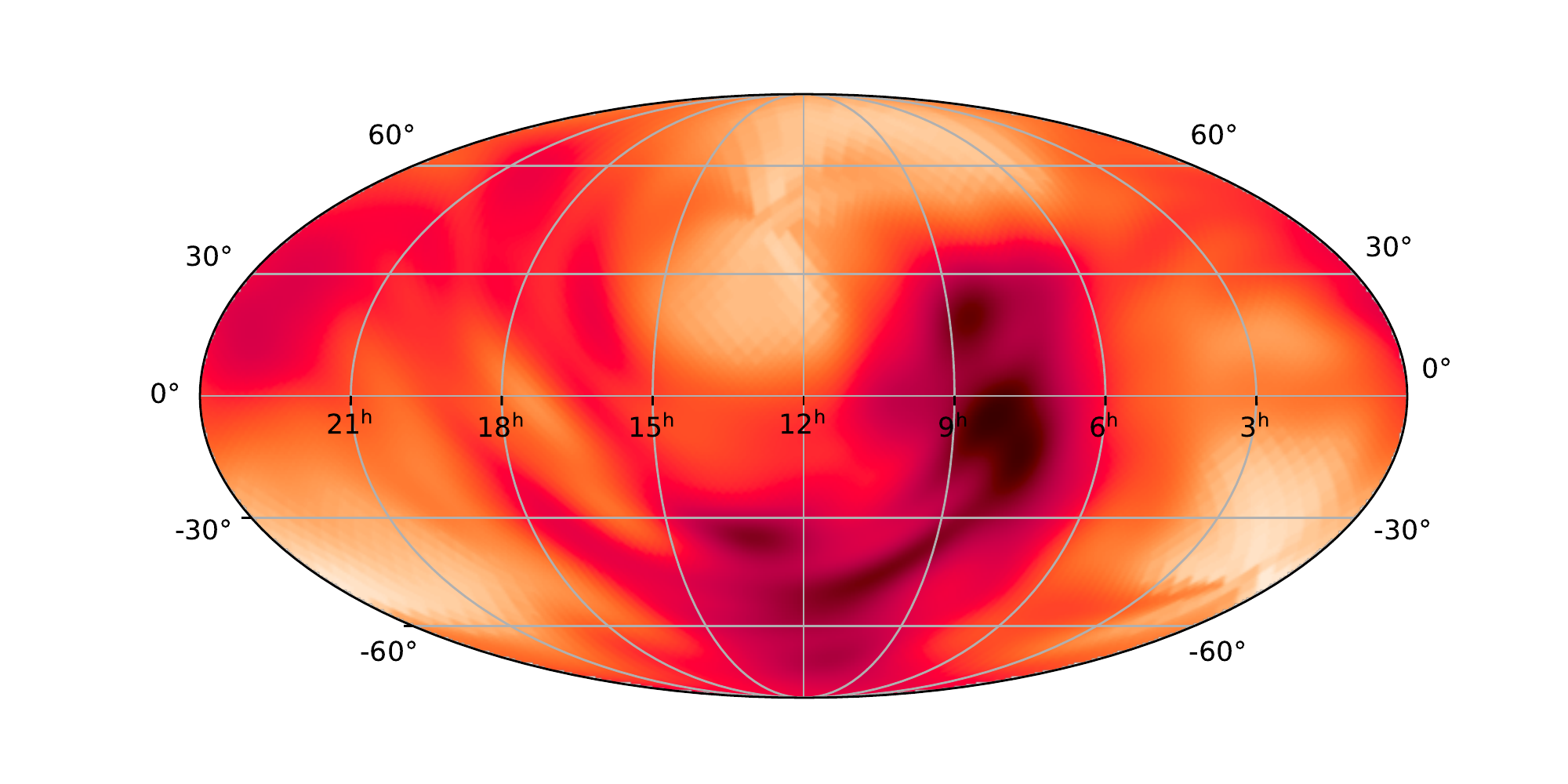}}{GW191215\_223052}
}%
\subfloat{\centering
    \stackunder{\includegraphics[width=0.20\textwidth]{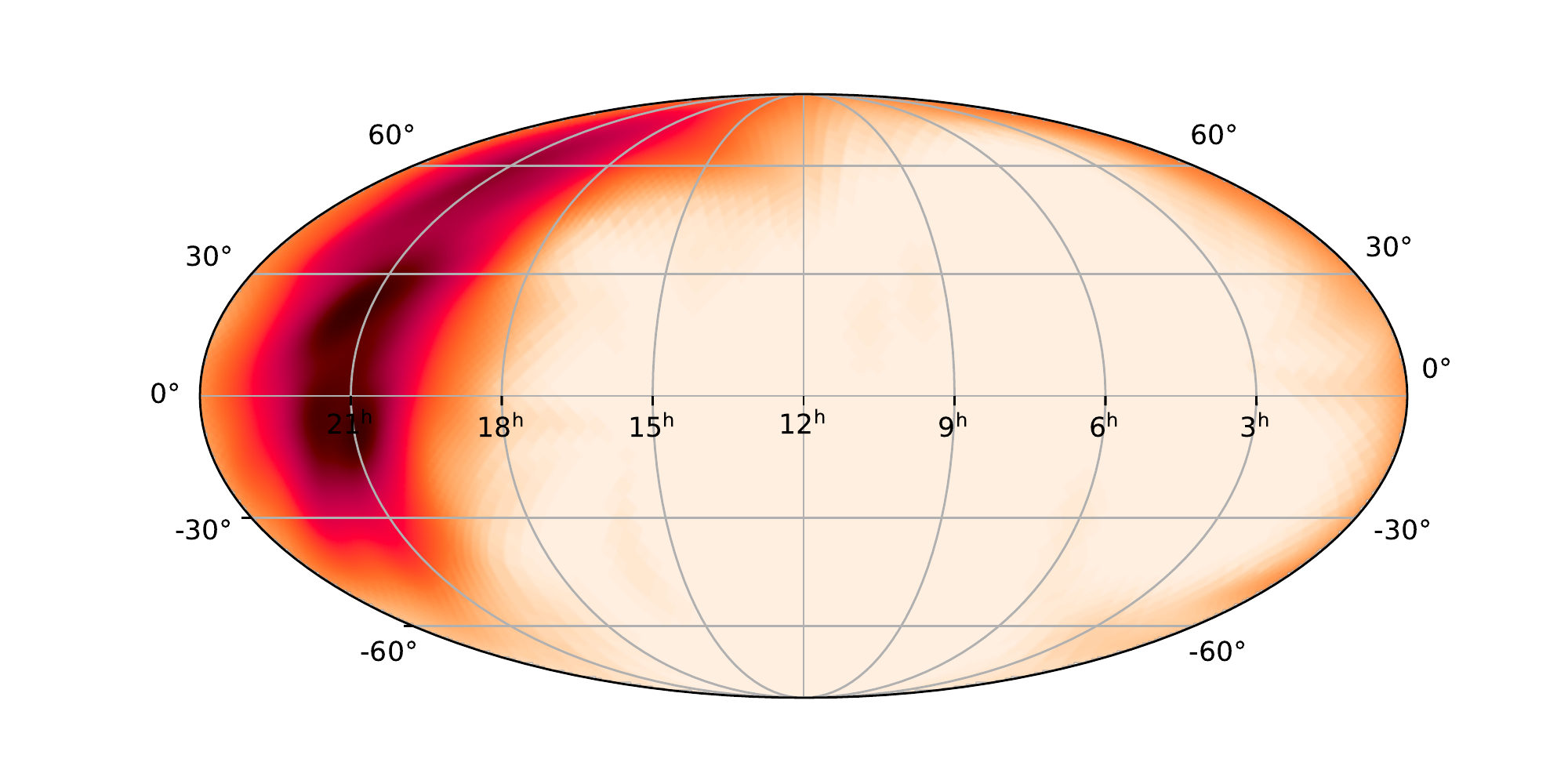}}{GW191216\_213338}
}%
\subfloat{\centering
    \stackunder{\includegraphics[width=0.20\textwidth]{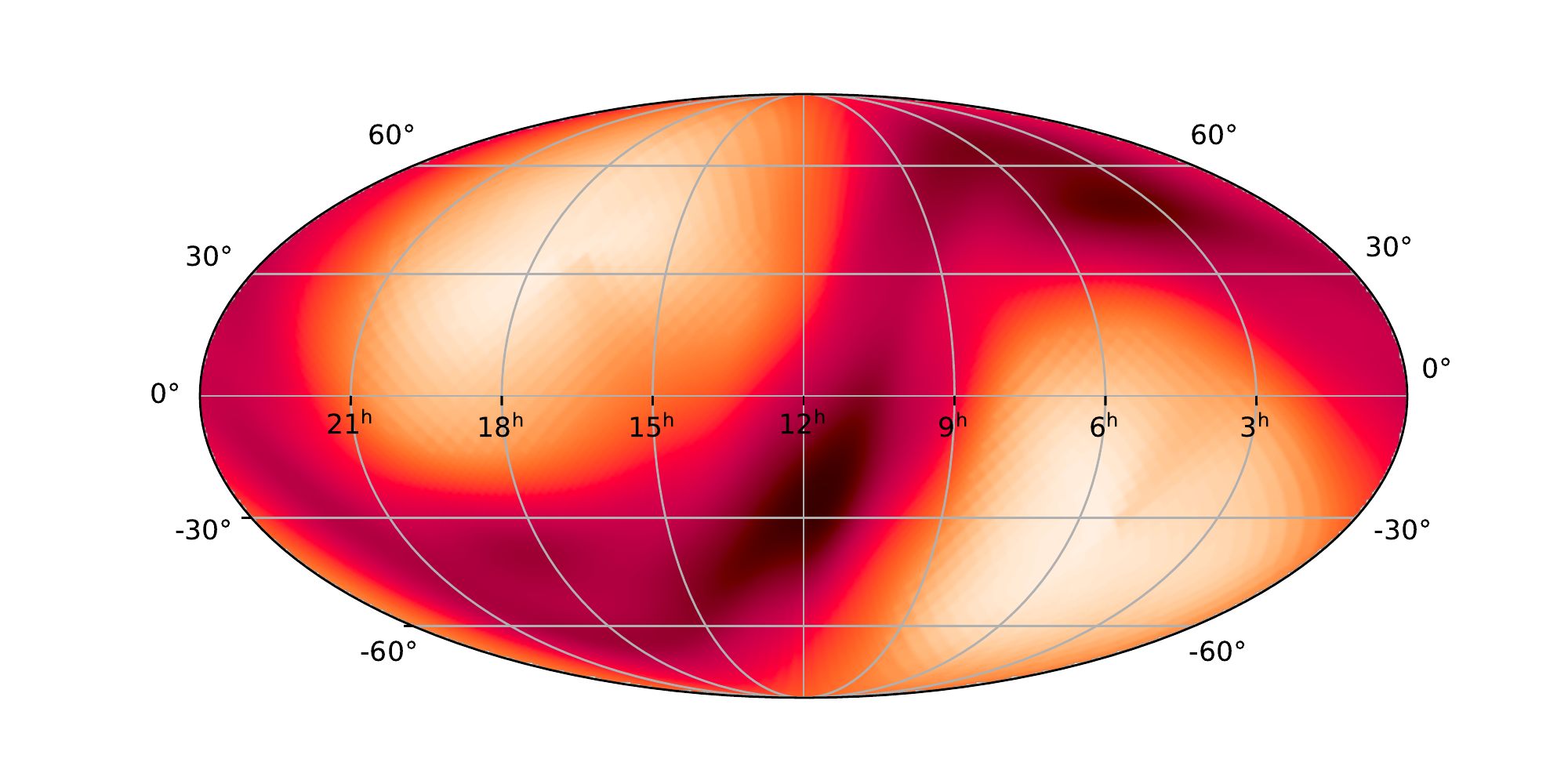}}{GW191222\_033537}
}%
\subfloat{\centering
    \stackunder{\includegraphics[width=0.20\textwidth]{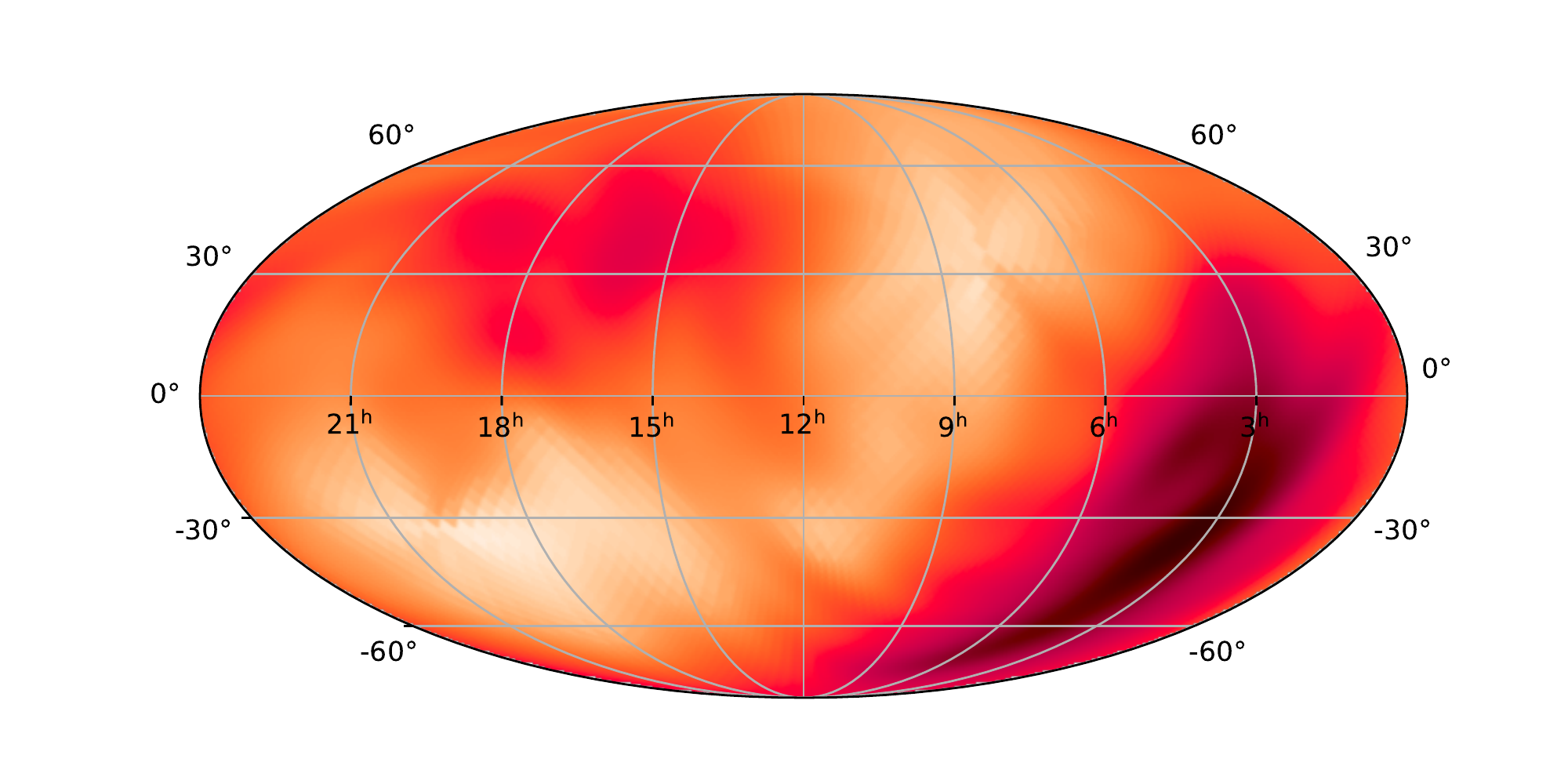}}{GW191230\_180458}
}%
\subfloat{\centering
    \stackunder{\includegraphics[width=0.20\textwidth]{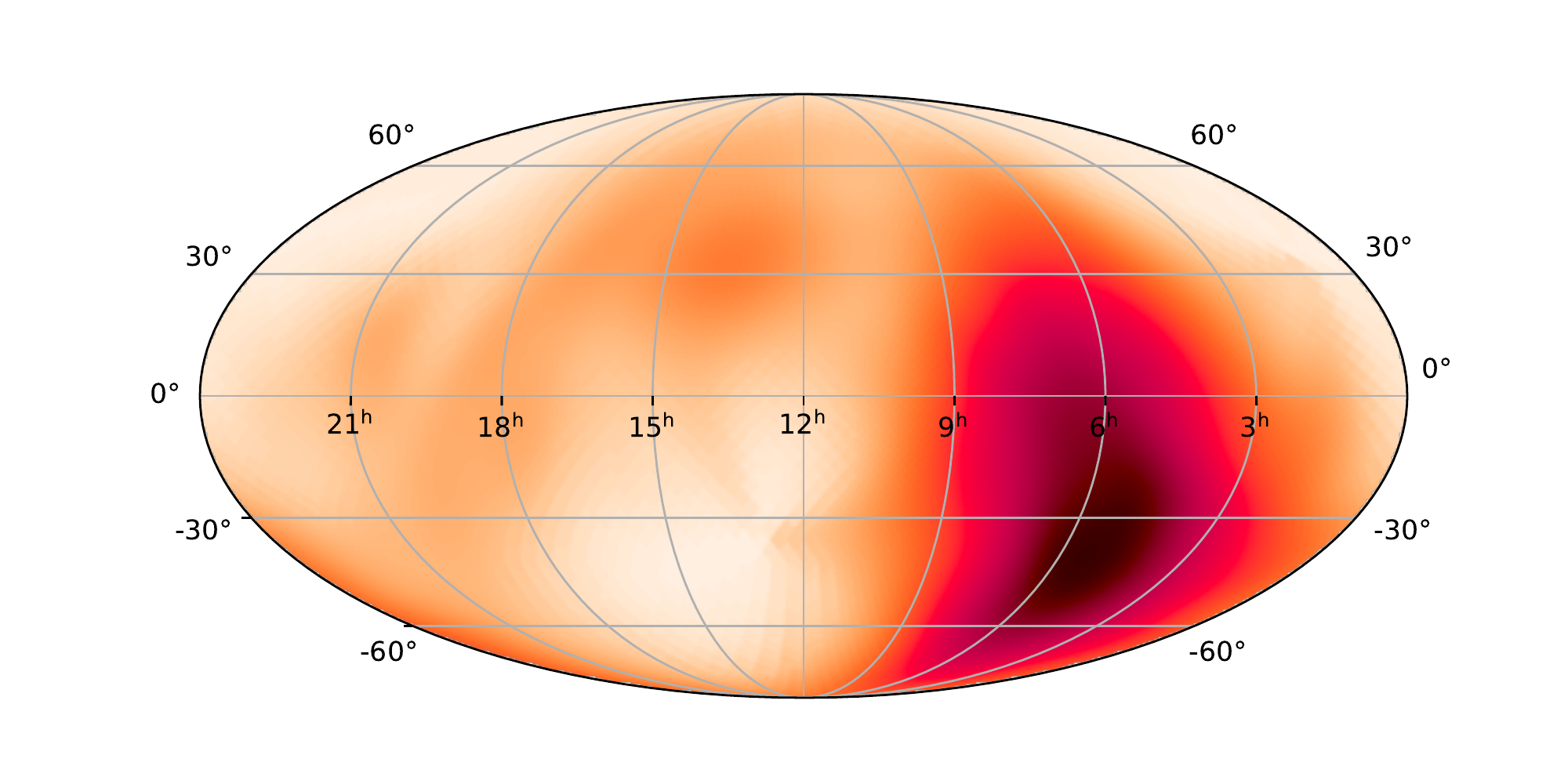}}{GW200112\_155838}
}\\
\caption{Measurements of the total angular momentum direction, $\hat{J}$, for
the events in our set, in a Mollweide projection of Earth-centric Celestial coordinates; darker color represents higher probability density for that direction in space.}
\label{fig:skymaps-1}
\end{figure*}%
\begin{figure*}\ContinuedFloat
\centering
\subfloat{\centering
    \stackunder{\includegraphics[width=0.20\textwidth]{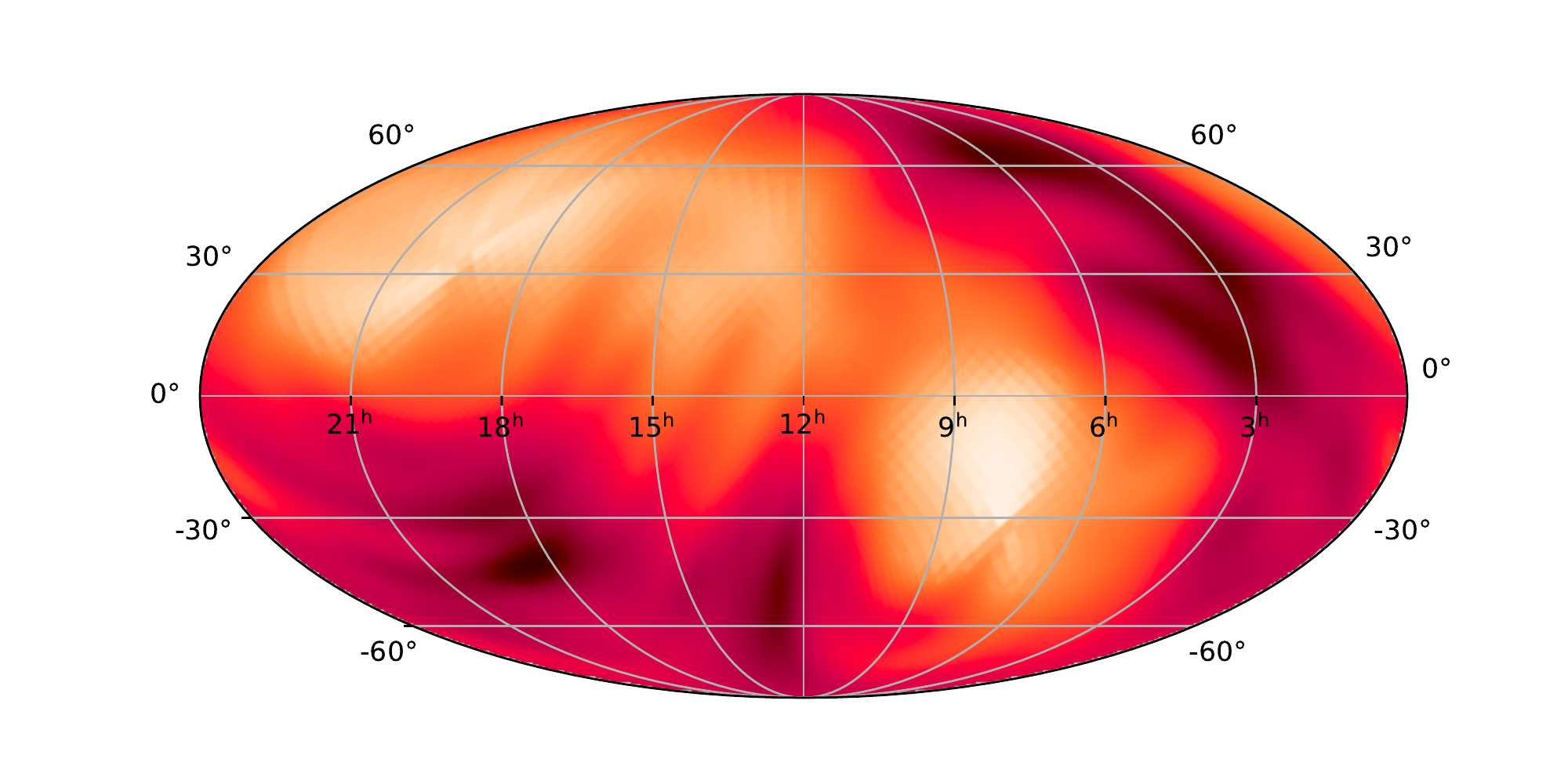}}{GW200128\_022011}
}%
\subfloat{\centering
    \stackunder{\includegraphics[width=0.20\textwidth]{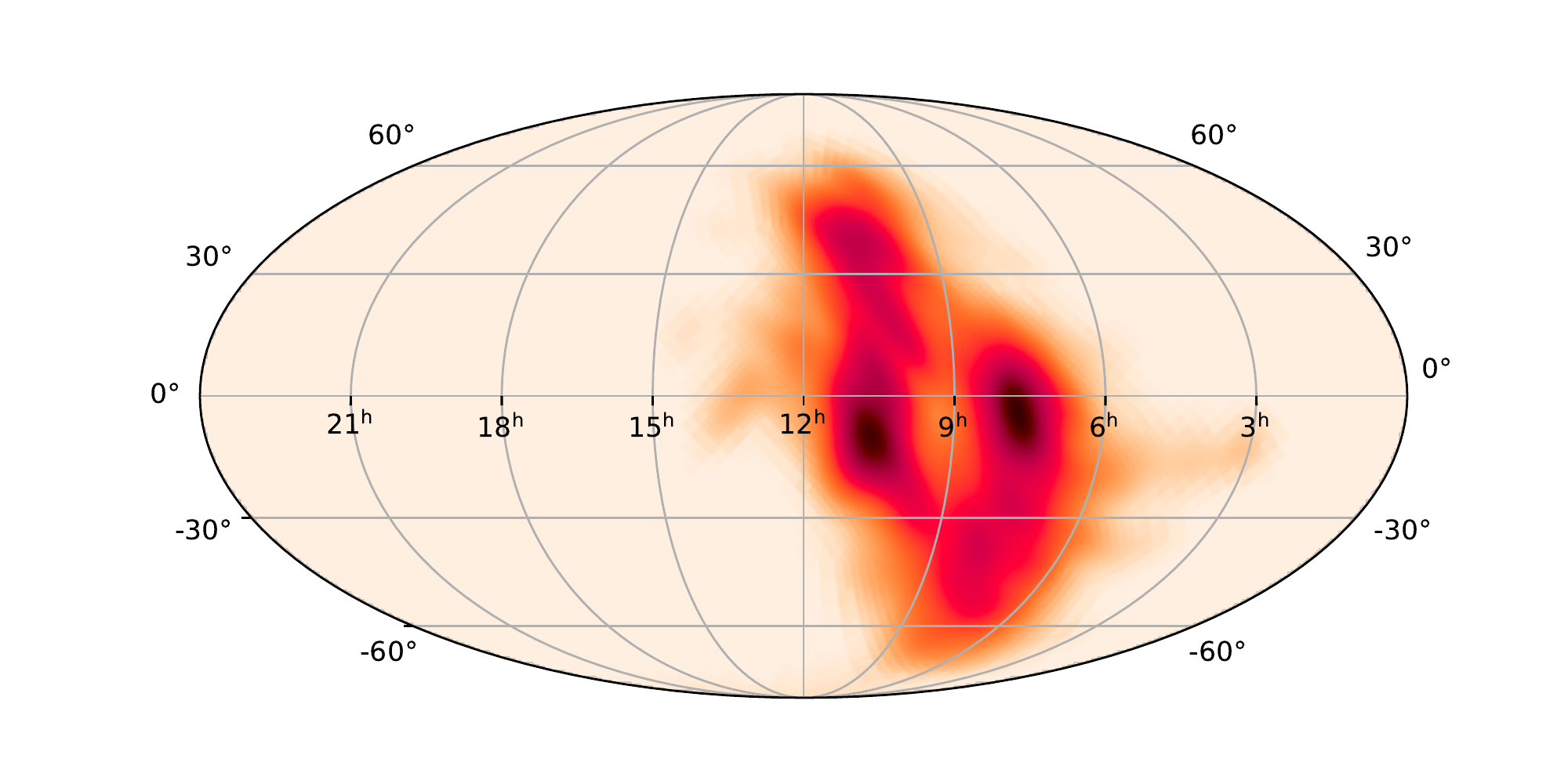}}{GW200129\_065458}
}%
\subfloat{\centering
    \stackunder{\includegraphics[width=0.20\textwidth]{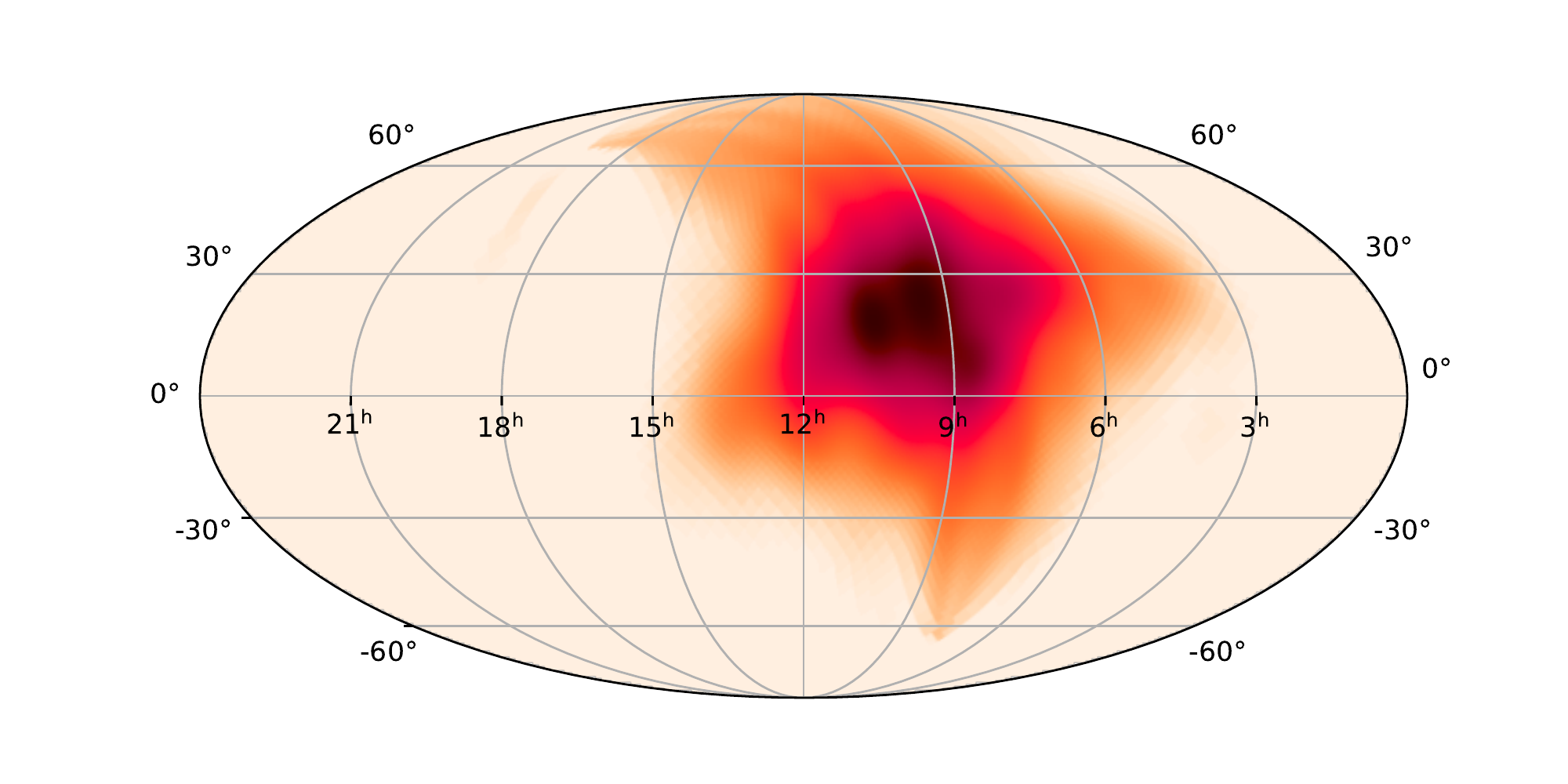}}{GW200202\_154313}
}%
\subfloat{\centering
    \stackunder{\includegraphics[width=0.20\textwidth]{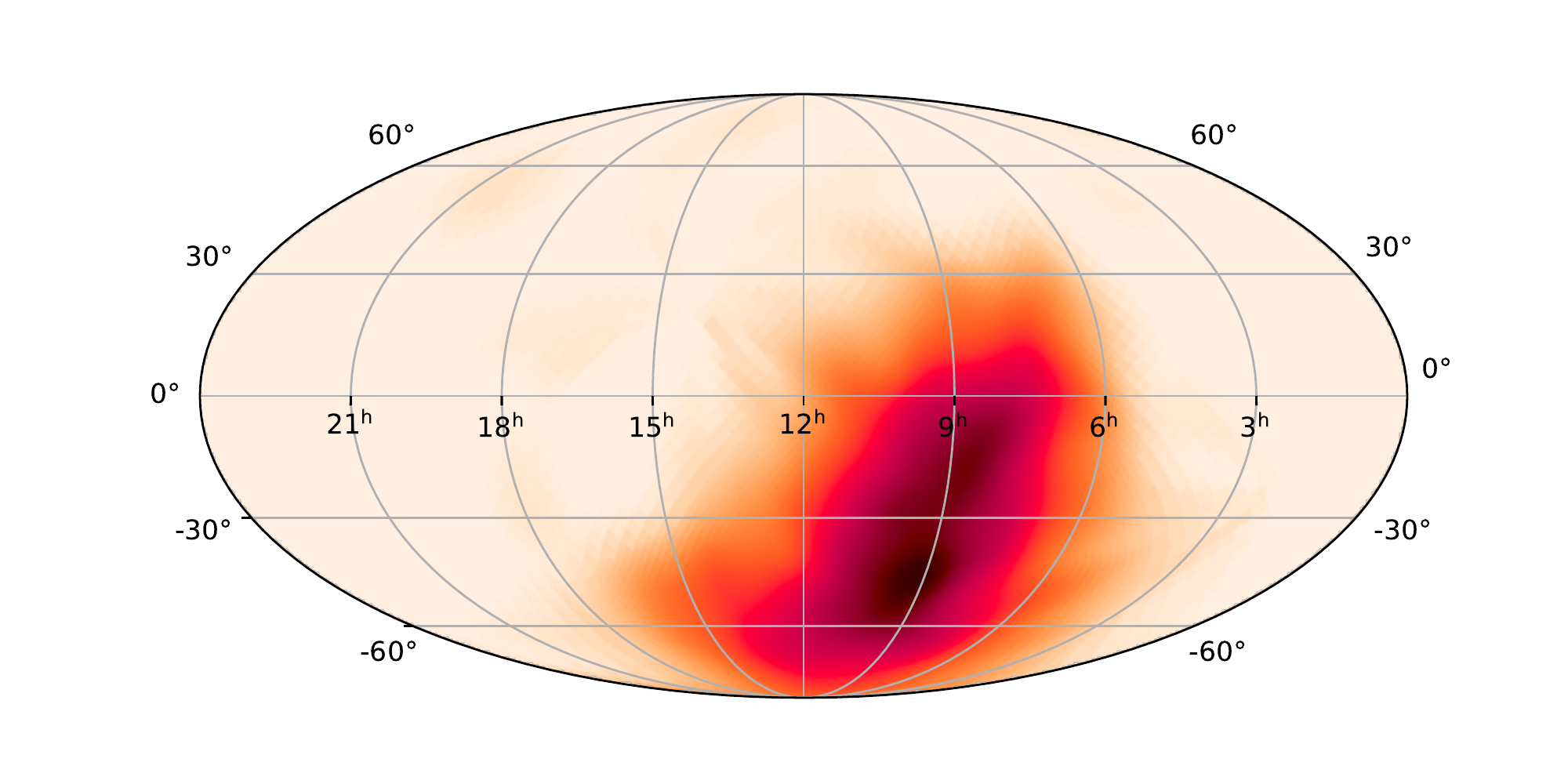}}{GW200208\_130117}
}%
\subfloat{\centering
    \stackunder{\includegraphics[width=0.20\textwidth]{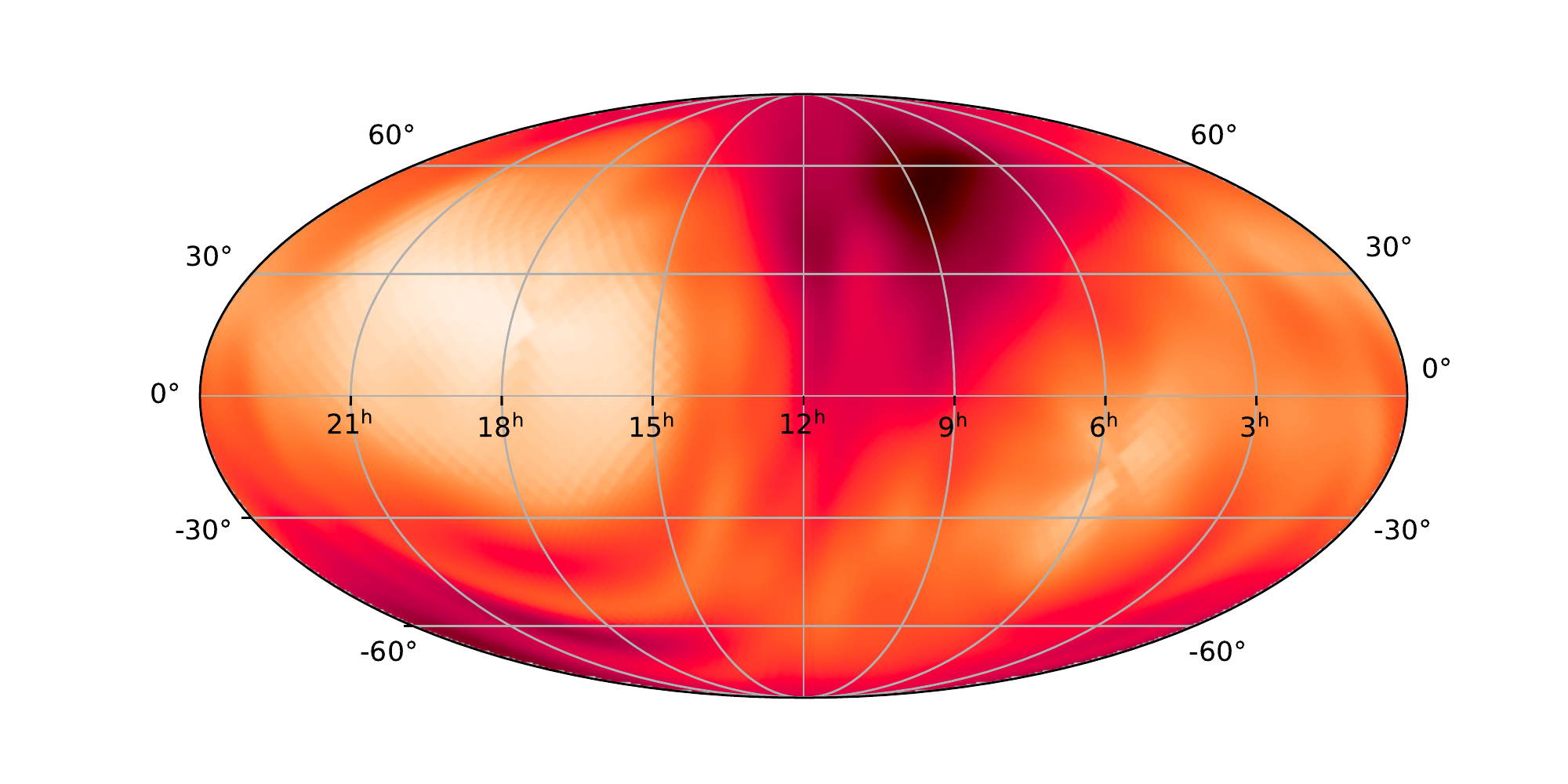}}{GW200209\_085452}
}\\
\subfloat{\centering
    \stackunder{\includegraphics[width=0.20\textwidth]{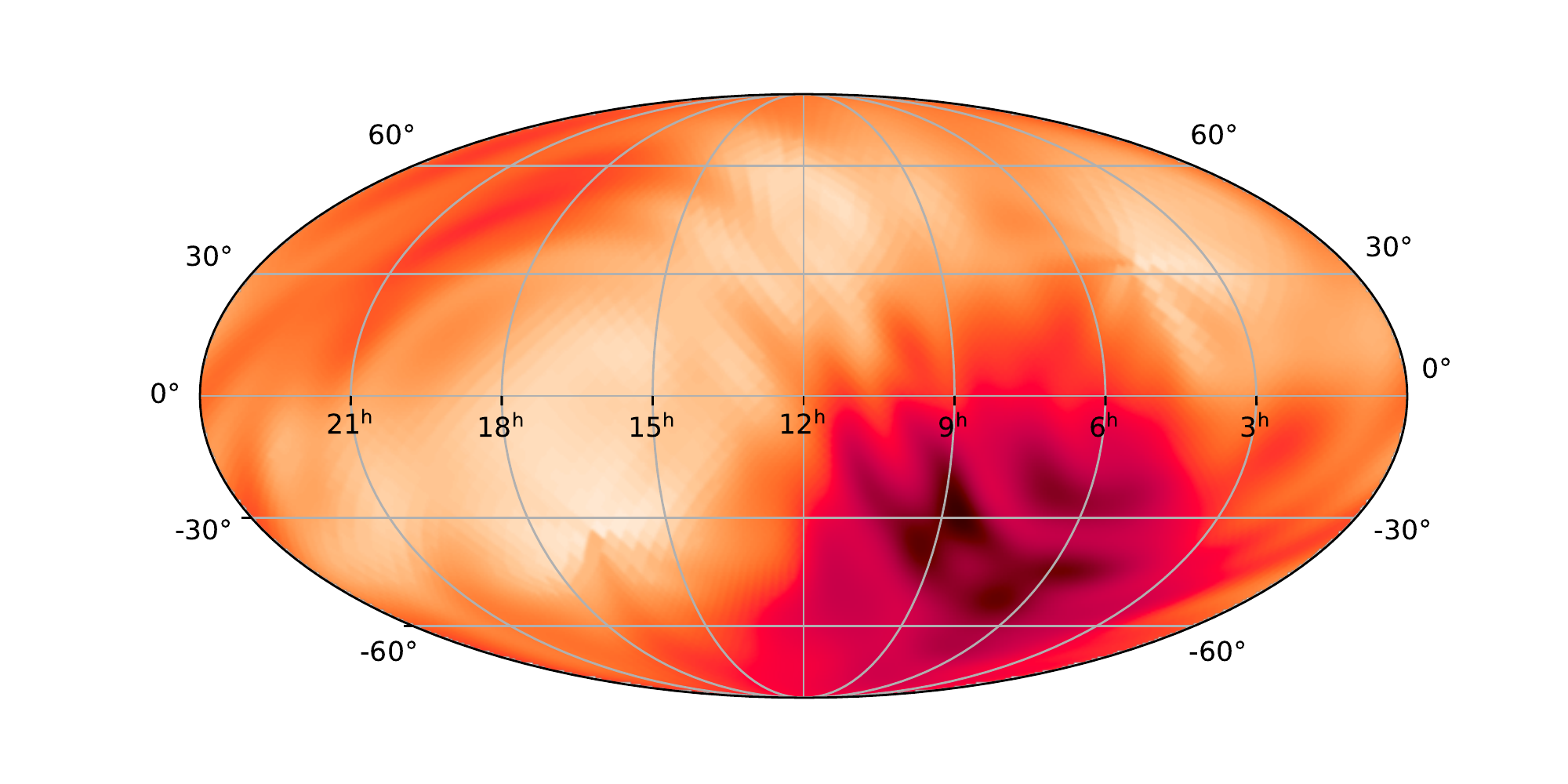}}{GW200216\_220804}
}%
\subfloat{\centering
    \stackunder{\includegraphics[width=0.20\textwidth]{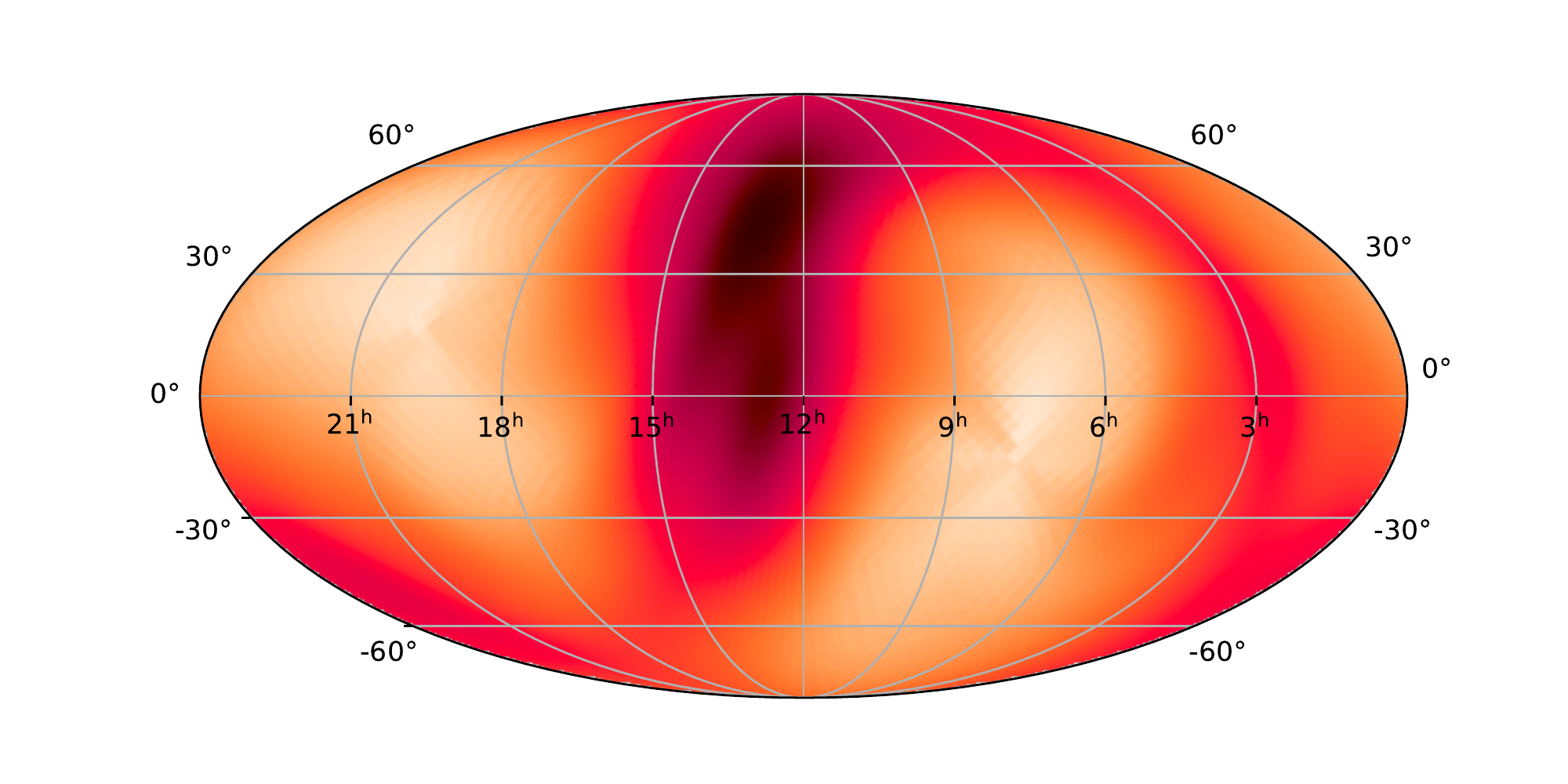}}{GW200219\_094415}
}%
\subfloat{\centering
    \stackunder{\includegraphics[width=0.20\textwidth]{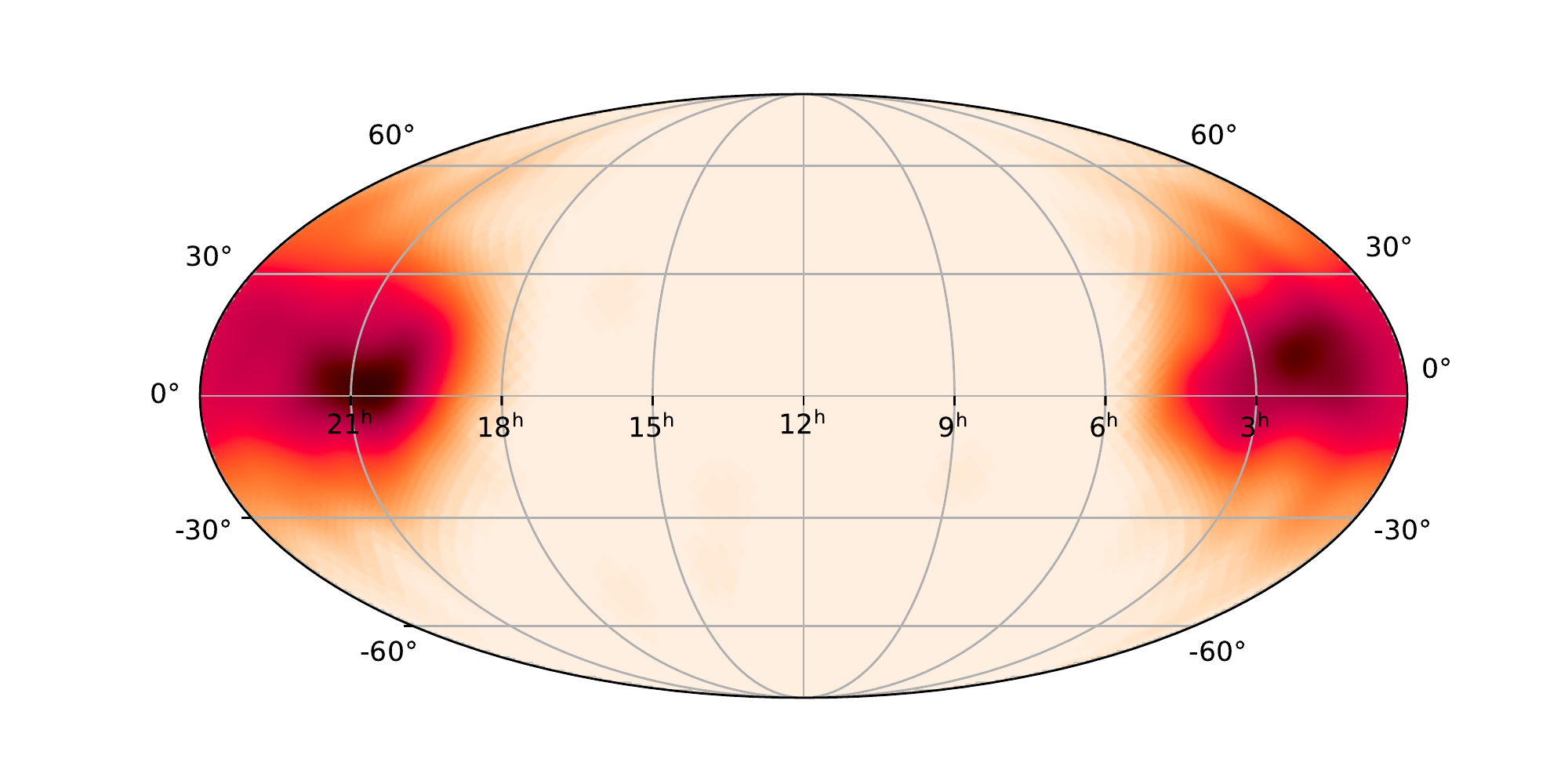}}{GW200224\_222234}
}%
\subfloat{\centering
    \stackunder{\includegraphics[width=0.20\textwidth]{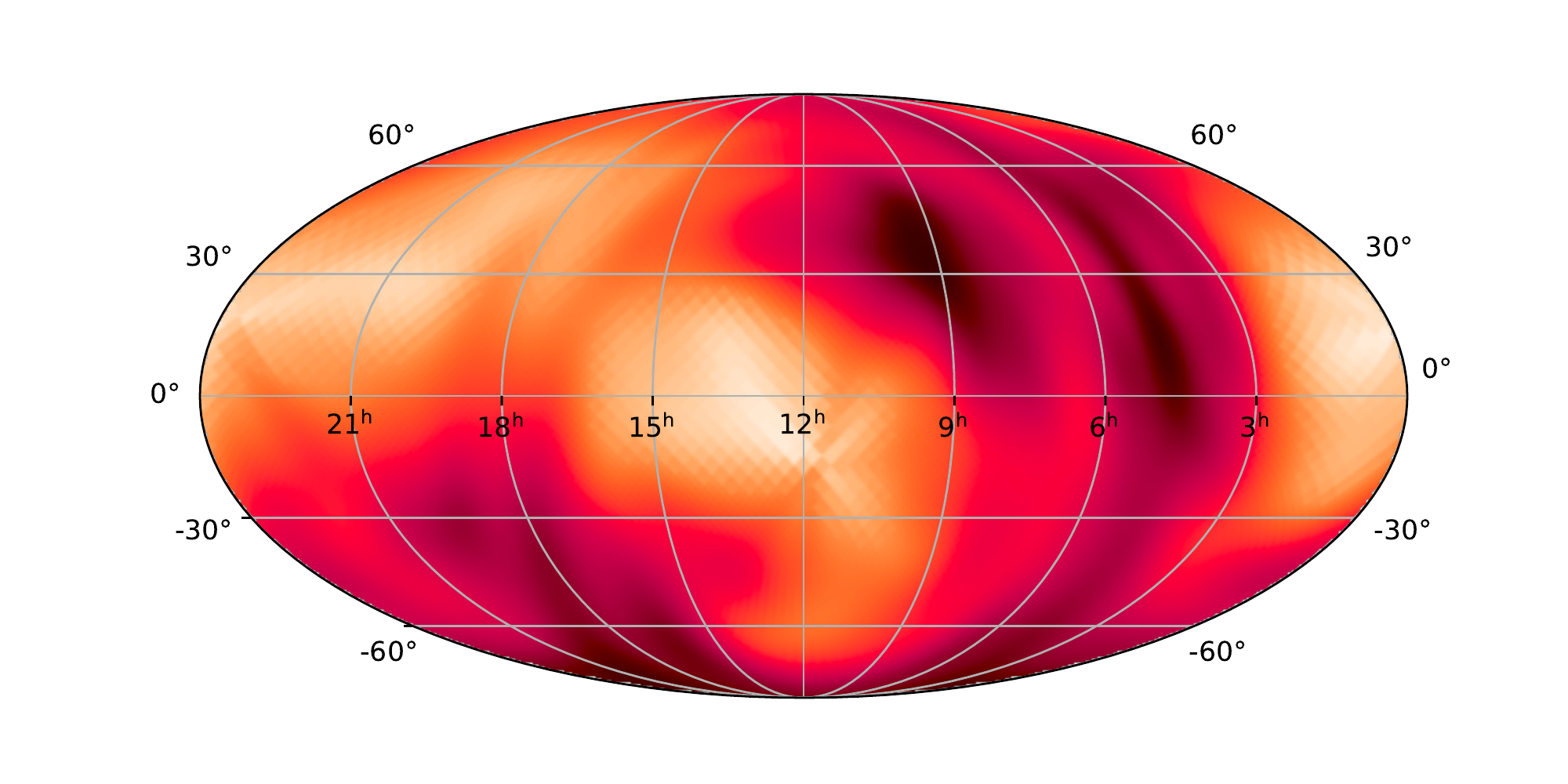}}{GW200225\_060421}
}%
\subfloat{\centering
    \stackunder{\includegraphics[width=0.20\textwidth]{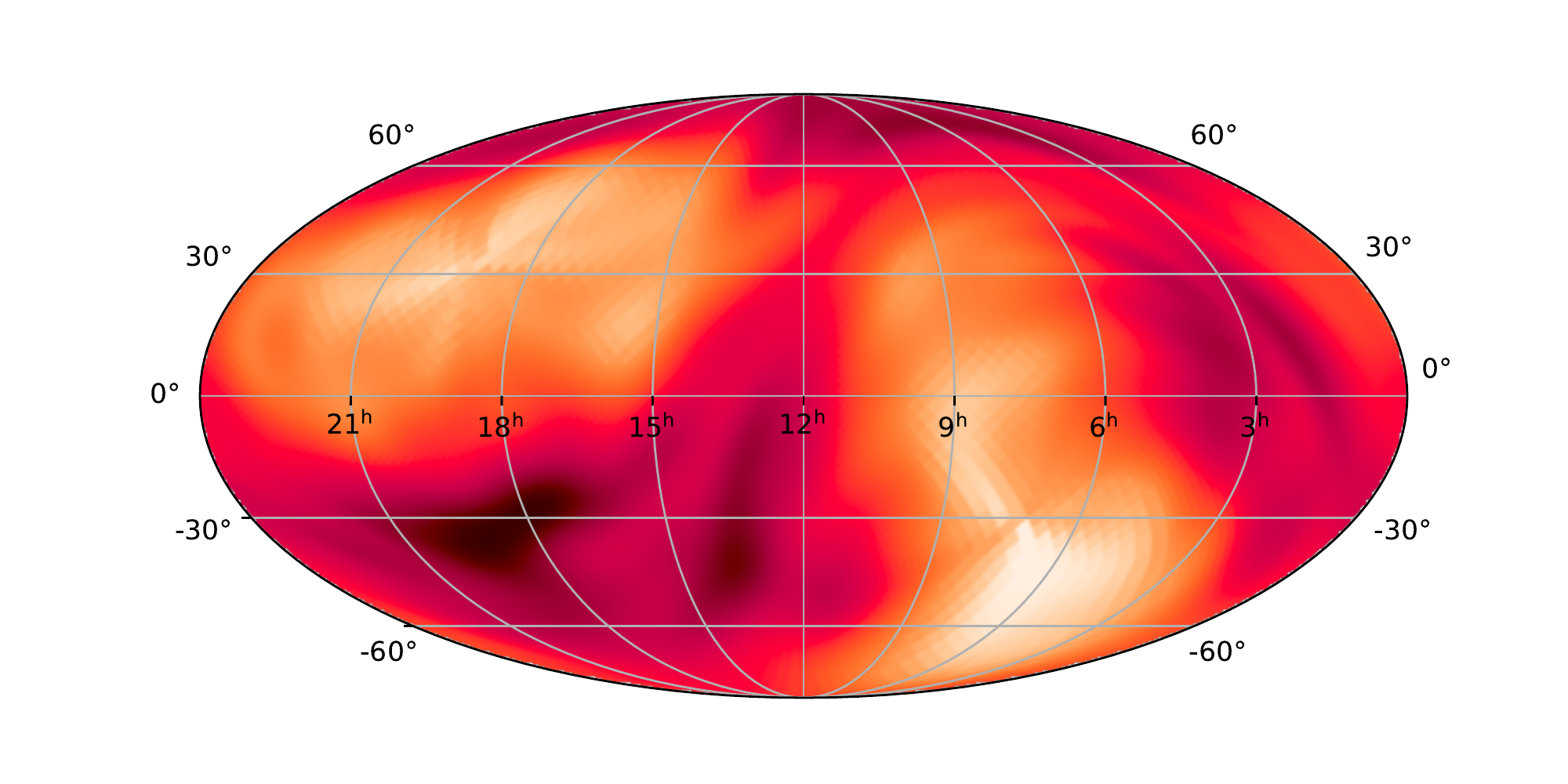}}{GW200302\_015811}
}\\
\subfloat{\centering
    \stackunder{\includegraphics[width=0.20\textwidth]{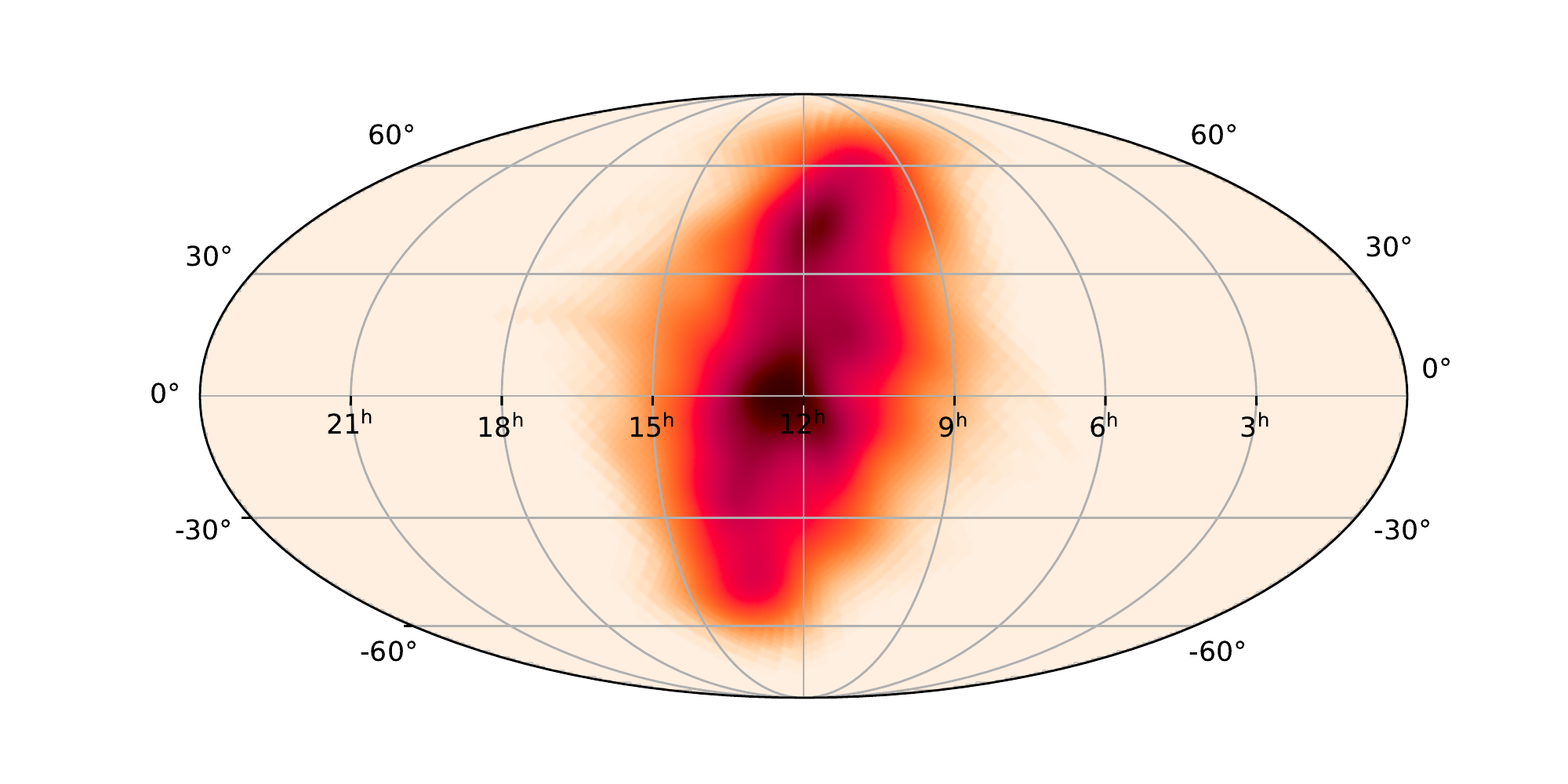}}{GW200311\_115853}
}%
\subfloat{\centering
    \stackunder{\includegraphics[width=0.20\textwidth]{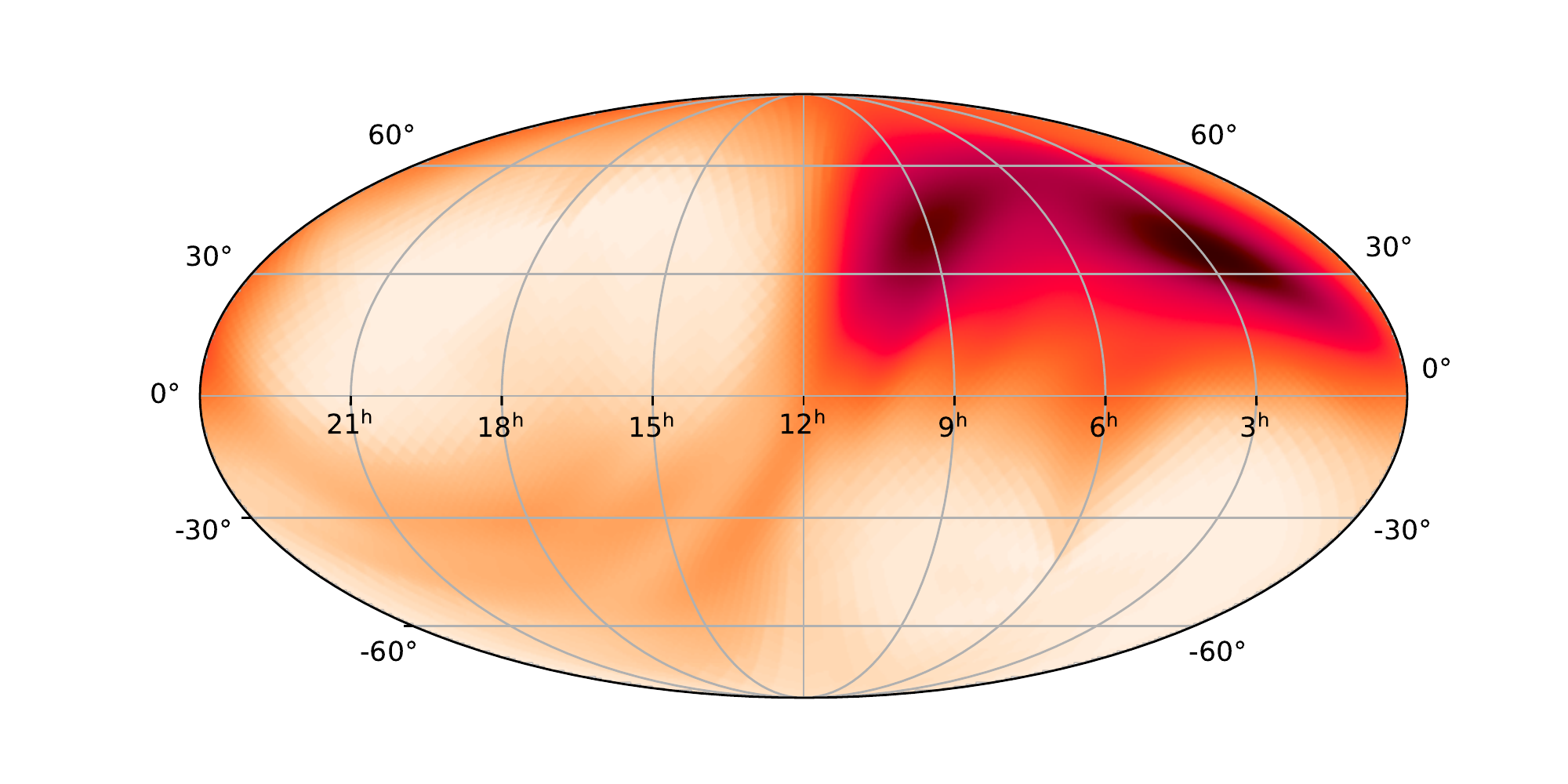}}{GW200316\_215756}
}
\caption{Measurements of the total angular momentum direction, $\hat{J}$, for
the events in our set, in a Mollweide projection of Earth-centric Celestial coordinates; darker color represents higher probability density for that direction in space (cont.).}
\label{fig:skymaps-2}
\end{figure*}\label{output/skymaps.tex}\unskip%
}

\end{document}